\documentclass[aps,prx,reprint,floatfix,footinbib,bibnotes,twocolumn,groupedaddress]{revtex4}
\usepackage[T1]{fontenc}
\usepackage[utf8]{inputenc}

\usepackage[dvipsnames]{xcolor}
\usepackage{microtype}
\usepackage{mathrsfs}
\usepackage{placeins}
\usepackage{revtex-right-top-float}
\definecolor{webblue}{rgb}{0, 0, 0.5} % less intense blue
\usepackage[bookmarks=true,
bookmarksnumbered=false, % true means bookmarks in
bookmarksopen=false, % true means only level 1
colorlinks=true,
linkcolor=webblue]{hyperref}
\hypersetup{colorlinks=true,urlcolor=webblue,citecolor=blue}
\usepackage{graphicx}
\usepackage{subfigure}
\usepackage{url}
\usepackage{hyperref}
\usepackage{inconsolata}
\usepackage{amsmath}
\allowdisplaybreaks
\usepackage[capitalize]{cleveref}
\usepackage{braket}
\usepackage{outlines}
\usepackage{enumitem}
\usepackage{soul}
\usepackage{color}

\usepackage{bm}

\usepackage{siunitx,booktabs}
\usepackage{multirow}
\usepackage{amssymb}

\usepackage{scalerel}
\usepackage{relsize}

\usepackage{soul}
\usepackage{tabularx}

\usepackage{amsmath}
\usepackage{amssymb}
\usepackage{bbm}
\usepackage{braket}
\usepackage{xcolor}

\usepackage{comment}
\usepackage{pifont}
\hypersetup{
  pdftitle={Real-Space Imaging of Band Topology via Wavefunction Zeros},
  pdfauthor={Julian Ingham and Raquel Queiroz}
}

\begin{document}

\title{Real-Space Imaging of Band Topology via Wavefunction Zeros}

\author{Julian Ingham}
\email[]{ji2322@columbia.edu}
\author{Raquel Queiroz}
\email[]{raquel.queiroz@columbia.edu}

\affiliation{Department of Physics, Columbia University, New York, NY, 10027, USA}

\date{\today}

\begin{abstract}
We prove that the wavefunction of a crystal at a high-symmetry momentum, $\Psi_{\bm{k}_*}(\bm{r})$, has symmetry-enforced zeros at certain positions in the unit cell, using a new invariant fixed uniquely by the symmorphic symmetry representation of the wavefunction. This allows one to infer the topology of an electronic band by probing zeros of the charge density, and in turn to connect scanning tunnelling microscopy  to the group representation theory of bandstructure. We apply the theorem to 1H transition metal dichalcogenides, where it detects the obstructed atomic limit of WSe$_2$, the Haldane model, where it detects the Chern number modulo three, and the Bernevig--Hughes--Zhang model, where it detects the $\mathbb{Z}_2$ index. In addition, the zeros have important consequences for interaction effects: in kagome metals, they fix the sublattice structure of Van Hove wavefunctions, and in twisted bilayer graphene, they explain the qualitative interaction-induced reshaping of the flat bands.
\end{abstract}

\maketitle

    \textit{Introduction}---Band topology has revealed phases of matter whose electronic structure is expressed through robust conduction at edges and surfaces, and quantised or otherwise unconventional electronic responses. Electronic structure is usually described in terms of Bloch bands in momentum space, yet an equally fundamental description is possible in real space, where bands are built from Wannier orbitals~\cite{Marzari2012}. This viewpoint underlies the theory of topological quantum chemistry, which connects the type and positions of orbitals to the global topology of bands across the Brillouin zone \cite{Zak1980SymmetrySpecification, Zak1981, MichelZak1999Connectivity, Bradlyn2017, Cano2018, CanoBradlyn2021}. In real space, topologically trivial bands admit symmetry-respecting localised Wannier functions, whereas topological bands are characterised by a Wannier obstruction or, in the `obstructed atomic' case, Wannier orbitals displaced away from the atoms to empty sites in the unit cell \cite{Resta1992, KingSmithVanderbilt1993, Brouder2007, po2017symmetry, Po2018}.
    
    This real-space formulation hints that band topology should leave distinct signatures in the spatial organisation of a quantum state within the unit cell. Scanning tunnelling microscopy (STM) is especially well suited to explore this, since within the Tersoff--Hamann picture, the tunnelling current probes the local density of states $\text{LDOS}(e\text{V},\bm{r})=\sum_{\bm{k}}|\Psi_{\bm{k}}(\bm{r})|^2\,\delta(e\text{V}-\varepsilon_{\bm{k}})$ with atomic spatial resolution \cite{TersoffHamann1983,TersoffHamann1985}. Yet, the quantities typically used to recognise topology are either indirect transport signatures, or momentum-space data sensitive to the phase of the wavefunction --- such as Berry phases or symmetry characters. A local density probe seems to discard much of this information, retaining only $|\Psi_{\bm{k}}(\bm{r})|^2$.  As the bias $e$V moves through a band, STM images how charge density reorganises within the unit cell, but without measuring the phase of the wavefunction. A fundamental question is whether such a phaseless real-space image can nevertheless reveal the topology of the band.
    
    Recent experiments present a concrete clue \cite{Holbrook2026, Calugaru2026}: in monolayer WSe$_2$, the charge density is centred between the atoms near the valence band maximum at $K$, but shifts onto the atomic sites at $\Gamma$ \cite{Holbrook2026}. This shift was shown to imply that WSe$_2$ is an obstructed atomic insulator \cite{Benalcazar2019,Cano2022TopologyInvisible,Gao2022Unconventional,Xu2024FillingEnforced}. These developments suggest that the distribution of charge density in Bloch wavefunctions $|\Psi_{\bm{k}}(\bm{r})|^2 $ contains symmetry information that survives the loss of phase.
    
        \begin{figure}[t]
        \centering
        \includegraphics[width=0.98\linewidth]{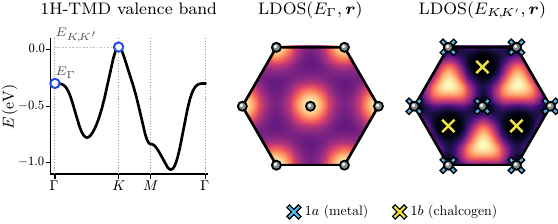}
        \vspace{-0.2cm}
        \caption{{\textbf{Imaging the band topology of TMDs.}} Left: valence-band dispersion of 1H-TMDs. The band transforms with $C_3$ eigenvalues $1$ and $e^{2\pi i/3}$ at $\Gamma$ and $K$, indicating a topologically obstructed atomic insulator (OAI). Middle: $\mathrm{LDOS}(E_\Gamma, \bm{r})$; Right: $\mathrm{LDOS}(E_{K\!,\!K'}, \bm{r})$. At $K$, symmetry-enforced zeros appear at the $1a$ and $1b$ positions (crosses); no zeros appear at $\Gamma$. The change in LDOS between these two energies is a real-space signature of an OAI.}
        \label{fig:tmd}
        \vspace{-0.5cm}
    \end{figure}

    This raises a natural question: how general is the connection between real-space STM images and band topology? Is the shift of spectral weight between sites, bonds, and hollow positions a material-specific feature, or the manifestation of a universal symmetry principle? The central claim of this paper is that there is a group-theoretic answer: the position of the charge within the unit cell is fixed by band representation theory. We derive a new real-space invariant [Eq. \eqref{indicator}] which proves that at high-symmetry momenta, Bloch wavefunctions must vanish at a set of positions we term the \emph{dark set} of the band. The dark set appears as symmetry-enforced dark spots in the LDOS whenever the bias resolves the corresponding high-symmetry states.

    We prove that for a fixed high-symmetry momentum $\bm{k}_*$, the zeros in the Bloch function $\Psi_{\bm{k}_*}(\bm{r})$ are fixed by the visible part of its irreducible representation (irrep); nonsymmorphic symmetry data are invisible to local density probes. In symmorphic crystals, \textit{the dark set uniquely determines the representation content of a band}, up to complex-conjugate irrep pairs at time-reversal invariant momenta that no density measurement can distinguish. Symmetry indicators allow the inference of topological invariants from this data, such as the Chern number or $\mathbb{Z}_2$ invariant \cite{FuKane2007, fang2012bulk, po2017symmetry, Slager2013SpaceGroupClassification, Kruthoff2017BandCombinatorics, Song2018QuantitativeMappings, Khalaf2018SymmetryIndicators}.

    Motivated by possible applications to STM, we focus on two-dimensional systems. We derive the general dark set selection rule, tabulate the symmetry-enforced dark sets across all 17 two-dimensional wallpaper groups, and apply the formalism to representative topological bands. For 1H transition metal dichalcogenides (TMDs), it diagnoses the obstructed atomic limit of the valence band \cite{Holbrook2026, Calugaru2026}. For the Haldane \cite{Haldane1988} and BHZ \cite{Bernevig2006} models, we show how STM can diagnose a nonzero Chern number and $\mathbb{Z}_2$ index. The framework developed here suggests a new form of topological band spectroscopy, by which topology can be determined from the intracell structure of the bulk wavefunction.
    
    The consequences of these zeros extend beyond STM, as the structure of charge density within the unit cell determines how electrons interact with one another. A symmetry-enforced zero can suppress certain interaction channels, or allow selected states to avoid an interaction-generated potential. On the kagome lattice, we show the zeros explain the sublattice texture of Van Hove wavefunctions --- which promotes long-range Coulomb interactions known to stabilise exotic loop current orders. In twisted bilayer graphene (TBG), we show that the zeros explain the qualitative reshaping of the narrow bands by interactions: the Bloch states at $\Gamma$ have a zero which allows them to avoid the Hartree potential created by the states at $K$. Thus, dark sets are not only STM signatures of topology; they are real-space constraints on the interacting physics of topological bands.

    \textit{Derivation}---Define a space-group element $h=\{\mathcal{R}| \bm{t}\}$, comprising a proper rotation or reflection $\mathcal{R}$ followed by a translation by $\bm{t}$. We define a high-symmetry momentum $\bm{k}_*$ and its little group $G_{\bm{k}_*}$, the subgroup of the crystal symmetries which leaves $\bm{k}_*$ invariant modulo a reciprocal lattice vector, $G_{\bm{k}_*}=\left\{h: \mathcal{R} \bm{k}_*=\bm{k}_*+\bm{G}\right\}$. We also define a high-symmetry real-space position $\bm{r}_*$ and the associated site symmetry group $G_{\bm{r}_*}$, i.e. those symmetries which leave $\bm{r}_*$ invariant modulo a lattice vector, $G_{\bm{r}_*}=\{h: \mathcal{R}\bm{r}_*+\bm{t}=\bm{r}_*\}$. For such $h$, we denote $\bm{t}_h(\bm{r}_*)=(1-\mathcal{R}) \bm{r}_*$.

    An electronic band has a little group representation at $\bm{k}_*$, defined via matrices $\mathcal{D}_{\rho\bm{k}_*}(h)$ whose action encodes how the wavefunction transforms under the symmetry operation $h$ (see the Appendix for a basic theoretical introduction), i.e. $h$ acts as a unitary transformation $U_h$
    \begin{align}
    \label{symmetry_op}
    U_h \Psi_{\bm{k}_{*}, a}=e^{-i \bm{k}_{*} \cdot \bm{t}} \left[\mathcal{D}_{\rho\bm{k}_{*}}(h)\right]_{ab} \Psi_{\bm{k}_{*},b}
    \end{align}
    where $\rho$ denotes the irrep, and $a,b$ index the wavefunction components (summation over $b$ is implicit).

    Our goal is to find selection rules which force the amplitude of the wavefunction to vanish at certain points depending on its symmetry. Intuitively, these are positions in real space where the phases of nearby orbitals interfere destructively, as determined by their orbital character and positions. The single-orbital limit is familiar: the nodes of a spherical harmonic identify its angular momentum; the dark set generalises this from a single orbital to a crystal.

    Consider the symmetry operations $h$ contained in both the site symmetry group $G_{\bm{r}_*}$ and the little group $G_{\bm{k}_*}$; we denote this intersection $H_{\bm{k}_*\bm{r}_*} = G_{\bm{r}_*}\cap G_{\bm{k}_*}$. Evaluating Eq. \eqref{symmetry_op} at a position left invariant by the symmetry, $\bm{r}_*=\mathcal{R}\bm{r}_*+\bm{t}$, we conclude
    \begin{align}
        \Psi_{\bm{k}_*}(\bm{r}_*)=e^{-i\bm{k}_*\cdot\bm{t}} \, \mathcal{D}_{\rho\bm{k}_*}(h)\,\Psi_{\bm{k}_*}(\bm{r}_*)
    \end{align}
    Hence, \textit{either} $e^{-i\bm{k}_*\cdot\bm{t}} \mathcal{D}_{\rho\bm{k}_*}(h)$ leaves the wavefunction invariant, \textit{or else} $\Psi_{\bm{k}_*}(\bm{r}_*)=0$.

    To identify the vector space for which the matrix $e^{-i\bm{k}_*\cdot\bm{t}} \mathcal{D}_{\rho\bm{k}_*}(h)$ acts as a projector, we sum over the elements of $H_{\bm{k}_*\bm{r}_*}$,
    \begin{align}
        \mathcal{P}_{\bm{k}_*,\bm{r}_*} = \frac{1}{|H_{\bm{k}_*\bm{r}_*}|}\sum_{h \in H_{\bm{k}_*\bm{r}_*}} e^{-i\bm{k}_*\cdot\bm{t}_h(\bm{r}_*)} \, \mathcal{D}_{\rho\bm{k}_*}(h)
    \end{align}
    This object, known as a Reynolds operator \cite{DerksenKemper2015, Serre1977}, projects onto the subspace left invariant by $e^{-i\bm{k}_*\cdot\bm{t}} \mathcal{D}_{\rho\bm{k}_*}(h)$ for all $h$. Taking the trace, we find that
    \begin{align}
    \label{indicator}
    \boxed{m_{\bm{k}_*,\bm{r}_*} =\frac{1}{|H_{\bm{k}_*\bm{r}_*}|} \sum_{h \in H_{\bm{k}_*\bm{r}_*}} e^{-i \bm{k}_* \cdot \bm{t}_h(\bm{r}_*)} \,\chi_{\rho\bm{k}_*}(h)}
    \end{align}
    must be nonzero, else the wavefunction vanishes at $\bm{r}_*$, where we introduced the characters of the irreducible representation $\rho$, $\chi_{\rho\bm{k}_*}(h) = \text{Tr}[\mathcal{D}_{\rho\bm{k}_*}(h)]$.
    
    This formula determines the symmetry-enforced nodes in a Bloch wavefunction. The set of high-symmetry positions with $m_{\bm{k}_*,\bm{r}_*}=0$ is the \textit{dark set of the irrep} $\rho$: the orbital phases encoded in $\chi_{\rho\bm{k}_*}(h)$, dictated by the symmetry representation, interfere destructively and force $\Psi_{\bm{k}_*}(\bm{r}_*)=0$. The positions of the dark set are gauge invariant: redefinition of the rotation axes, coordinate origin, or orbital embedding does not change the physical locations where $\Psi_{\bm{k}}(\bm{r})$ is forced to vanish (SM Sec.~\ref{gauge}). The phase of the wavefunction, by contrast, is gauge dependent and experimentally inaccessible.  Away from a high-symmetry momentum $\bm{k}_*$, the wavefunction does not transform as a single irrep; moving through the Brillouin zone, its charge density evolves between different irreps with different patterns of zeros. This enforces an energy dependence of the charge density across the band. 

    The differential current $dI/dV$ measured in STM is proportional to the LDOS \cite{Coleman2015}; if the bias is chosen to intersect the band at a single high-symmetry momentum $\bm{k}_*$, then one can image $|\Psi_{\bm{k}_*}(\bm{r})|^2$ and hence probe the existence of the symmetry-enforced zeros. The selection rule acts irrep by irrep at $\bm{k}_*$; when an energy window contains several irreps, the only exact zeros are those shared by every contributing irrep. Two physical conditions control the visibility of these zeros. First, if a band is composed of tightly localised orbitals with minimal support off the atomic sites, then a significant fraction of the unit cell will appear dark simply because the density of states is small, as opposed to being strictly zero due to the selection rule. To discern regions which are forced-dark, the constituent orbitals must have considerable overlap with their neighbours, such that the density of states will be large unless a zero is required by symmetry. Second, certain high-symmetry momenta will not be isolated energetically, but as we will show in model calculations, they can still dominate the LDOS at a fixed energy, allowing inference of the dark spots. In practice, the diagnostic is the position and symmetry of the LDOS minima.

    The question naturally arises: at a given high-symmetry momentum $\bm{k}_*$, for a single isolated irrep $\rho$, does observation of the dark set suffice to determine the irrep of the wavefunction? The answer is the following:

    \noindent\fbox{\parbox{\dimexpr\columnwidth-2\fboxsep-2\fboxrule\relax}{%
    \textbf{Theorem.} Probed over the high-symmetry positions of the unit cell, the symmetry-forced dark set of Eq.~\eqref{indicator} determines $\rho$ \emph{exactly} up to complex conjugation $\rho\leftrightarrow\bar{\rho}$ at time-reversal invariant momenta (TRIM), and up to the character data of nonsymmorphic symmetries.}}

    We tabulate the dark sets for all irreps in the 17 wallpaper groups in SM Sec.~\ref{supp-wallpaper}, verifying the theorem constructively; a non-constructive proof is given in SM Sec.~\ref{sm:general-proof}. Conjugate pairs $E_\pm$ appear with identical dark sets at $\Gamma$ in $p3$, $p4$, and $p6$, and at $M$ in $p4$; in systems with (time-reversal or) mirror symmetries, $\rho$ and $\bar{\rho}$ form degenerate components $E_\pm$ of a single (co)irrep, meaning the only remaining ambiguity in such systems is that due to nonsymmorphic symmetries.

    \textit{Applications: Transition Metal Dichalcogenides---}The first example we will consider is the semiconducting 1H transition metal dichalcogenides (1H-TMDs), a prototypical example of which is WSe$_2$. The space group of this family is $P\bar{6}m2$, and the point group $D_{3h}$. Denoting the position $(a,b) = a\bm{R}_1+b\bm{R}_2$ where $\bm{R}_1=(1,0)$ and $\bm{R}_2=(-\tfrac{1}{2},\tfrac{\sqrt{3}}{2})$, we consider the high-symmetry positions: the 1a position at the origin of the unit cell $(0,0)$, the 1b position at $(\tfrac{1}{3},\tfrac{2}{3})$, and the 1c position at $(\tfrac{2}{3},\tfrac{1}{3})$. 1H-TMDs exhibit an atomically obstructed valence band, for which localised Wannier functions cannot be constructed if one insists on centring them at the atomic sites. This follows from the mismatch of the $C_3$ irreps at $\Gamma$ and $K$: $\mathcal{D}_{\Gamma}(C_3) = 1$ and $\mathcal{D}_{\tau K}(C_3) =e^{2\pi \tau i/3}$ where $\tau=\pm$.

    Consider first the $\tau K$ point, at which the symmetry group is $C_{3h}$. At a $C_3$ centre defined by $\bm{r}=C_3\bm{r} +\bm{t}_{C_3}({\bm{r}})$
\begin{align}
m_{\tau\bm{K},\bm{r}}
=
\tfrac13 \left[1+\vartheta_{\tau K,\bm{r}}\,\xi_{\tau K}+
(\vartheta_{\tau K,\bm{r}}\,\xi_{\tau K})^2\right] , 
\end{align}
where $\vartheta_{\tau K,\bm{r}}=e^{-i\tau\bm{K}\cdot \bm{t}_{C_3}({\bm{r}})}$, and $\xi_{\tau K}$ is the $C_3$ eigenvalue of the Bloch state. Hence the state is bright at precisely the centre for which $\vartheta_{\tau K,\bm{r}}\,\xi_{\tau K}=1$. For the three $C_3$ centres, the phases at $K$ are $\vartheta_{K,\{1a,1b,1c\}}=\{1,\omega,\omega^2\}$. The valence band of the TMD has $\xi_\Gamma = 1$ and $\xi_{\tau K}=\omega^\tau$, so it is nodeless at $\Gamma$ and bright only at $1c$ at $K$ (Fig. \ref{fig:tmd}); the calculation is elaborated in the Appendix. An unobstructed valence band would have $\xi_{\Gamma}=\omega^\pm$, whose dark set includes all three $C_3$ centres (see SM Sec. \ref{sm:tmd}).

    \begin{figure}[t]
        \centering
        \includegraphics[width=\linewidth]{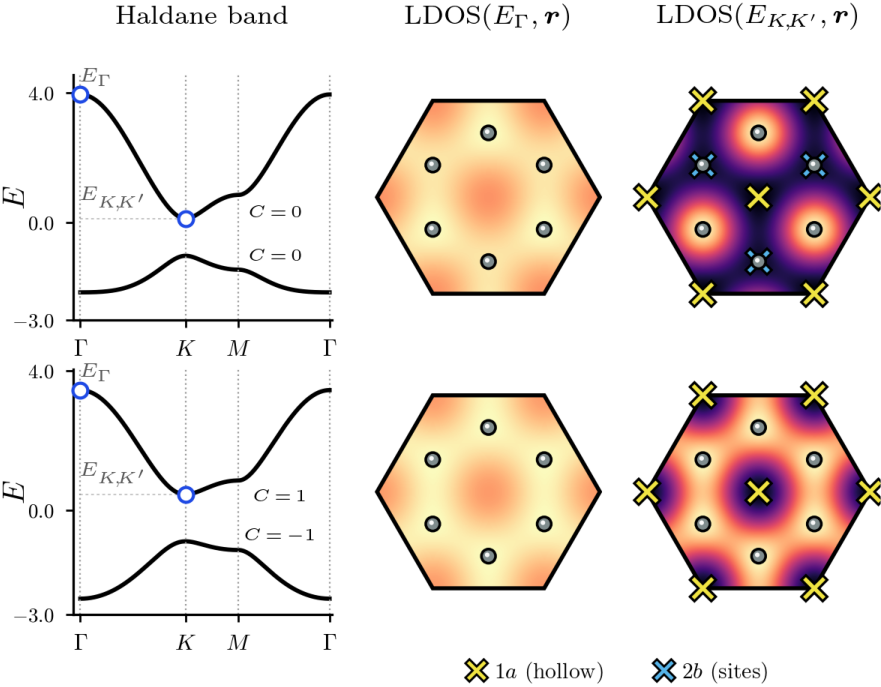}
        \caption{\textbf{Imaging the Chern number of the Haldane model.} Top row: trivial phase, $C=0$; bottom row: Chern phase, $|C|=1$. In the trivial phase, the $K$/$K'$ states have zeros on the same sublattice. In the Chern phase the two valleys have zeros (crosses) on opposite sublattices, leaving only the hexagon centre dark in the total LDOS. The distinct dark sets allow inference of the Chern number modulo three.}
        \label{fig:haldane}
        \vspace{-0.5cm}
    \end{figure}
    
    \textit{Haldane model---}The next example we consider is the Haldane model of a honeycomb lattice Chern insulator \cite{Haldane1988}. For any $C_3$-symmetric two-dimensional insulator, the Chern number mod 3 is fixed by the $C_3$ eigenvalues at the $C_3$-invariant momenta $\xi_{\Gamma}, \xi_{K}, \xi_{K^{\prime}}$; for a single occupied band, 
    $e^{2\pi i C/3}=\xi_{\Gamma} \xi_K \xi_{K^{\prime}}$ \cite{fang2012bulk}.
    For a honeycomb lattice of $s$ orbitals, $\xi_{\Gamma}=1$, and so it remains to determine the characters at the $K/K'$ points to infer the Chern number mod 3. We find (Appendix) that $\xi_K=e^{\pm 2\pi i/3}$ implies the wavefunction at $K$ is dark on the A/B sublattice, while $\xi_{K'}=e^{\pm 2\pi i/3}$ implies the wavefunction at $K'$ is dark on the B/A sublattice. Hence, a single centre bright at both $K$ and $K'$ implies $C=0$, whereas two centres bright, each originating from $K$ and $K'$, imply $|C|= 1$. We demonstrate this with numerical LDOS simulations for $|C|=0,1$ (Fig.~\ref{fig:haldane} top, bottom). The key observation is that for $C=0$, one sublattice is dark at $K$/$K'$; for $C=\pm 1$, only the origin is dark. When $K$ and $K'$ are not degenerate, the selection rule is even more transparent: in the trivial case only one sublattice is bright as the bias is varied, whereas in the topological case a second sublattice suddenly turns bright when the bias energy intersects the second valley (SM Sec. \ref{sm:generalised}).

    Prior theory work on indenene \cite{bauernfeind2021design} has proposed a connection between the characters at $K$ and the charge density --- specifically, that $s$ orbitals at the honeycomb positions produce the same maxima of charge density near $K$ as $p_{x,y}$ orbitals at the triangular position \cite{Eck2022}. Our formalism makes rigorous those results.

    \textit{Bernevig-Hughes-Zhang model}---The next example we consider is the BHZ model, a minimal example of a square lattice $\mathbb{Z}_2$ topological insulator \cite{Bernevig2006}. The topologically nontrivial phase is characterised by a finite spin Hall conductivity, and is diagnosed by a $\mathbb{Z}_2$ invariant. The presence of time-reversal symmetry ensures the Chern number vanishes, yet a $\mathbb{Z}_2$ topology remains, characterised by the Kane--Mele invariant $\nu$ \cite{KaneMele2005Z2,FuKane2006}. For a single occupied Kramers pair such as the lower band of the BHZ model, inversion symmetry implies $(-1)^\nu=\zeta_\Gamma \zeta_X \zeta_Y \zeta_M$, where $\zeta$ are the inversion eigenvalues at the TRIM \cite{FuKane2007}. The resulting dark sets are calculated in the Appendix.

\begin{figure}
        \centering
        \includegraphics[width=\linewidth]{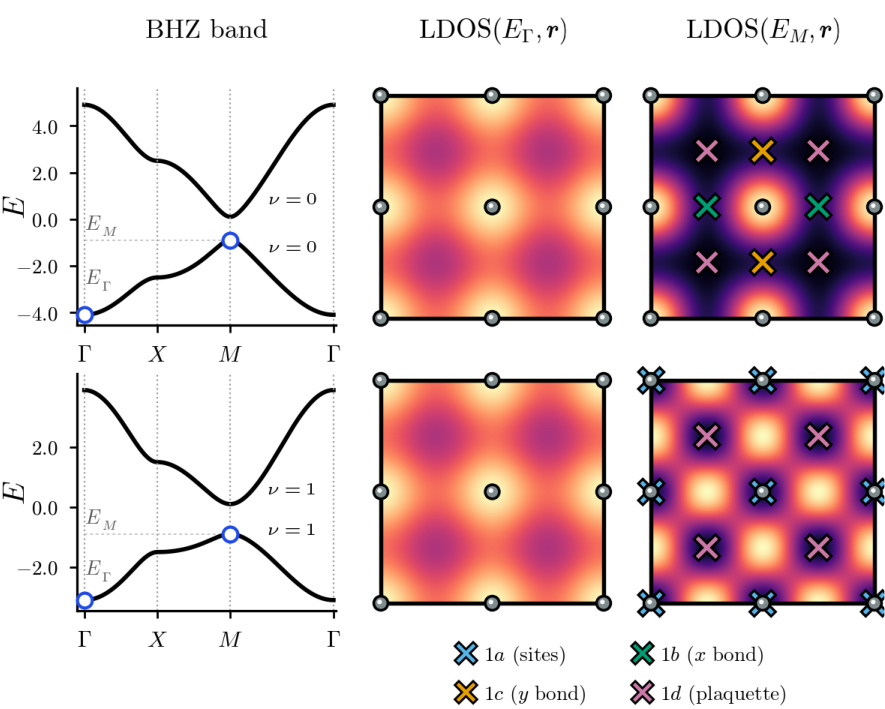}
        \caption{\textbf{Imaging the $\mathbb{Z}_2$ invariant of the BHZ model.}  Top row: trivial phase, $\nu=0$; bottom row: quantum spin Hall phase, $\nu=1$. In the trivial phase the bonds and plaquette centre have zeros (crosses) at $M$; in the quantum spin Hall phase the bonds are bright. The Kane-Mele $\mathbb{Z}_2$ invariant can be diagnosed simply by whether the bonds have the same ($\nu=1$) or opposite ($\nu=0$) contrast at $\Gamma$ and $M$.}
        \label{fig:bhz}
        \vspace{-0.5cm}
    \end{figure}

    Let $\bm{r}_*$ be an inversion centre, i.e. $I\bm{r}_*=\bm{r}_*$. Then, $\bm{t}_{I}(\bm{r}_*)=2\bm{r}_*$, and if no other symmetries are present, $m_{\bm{k}_* \bm{r}_*}=1+e^{-i\bm{k}_*\cdot \bm{t}_I}\zeta_I(\bm{k}_*)$. A simple rule follows: the number of zeros mod 2 across all TRIM, at an inversion centre \textit{with no other symmetry}, equals the $\mathbb{Z}_2$ invariant (SM Sec. \ref{sm:kane-mele}). With $C_4$ symmetry, the rule becomes simpler: the bonds have the same (opposite) contrast if $\nu=1$ ($\nu=0$). We demonstrate this with LDOS simulations for the BHZ valence band in Fig.~\ref{fig:bhz}: in the topological case (bottom) the bonds are bright at both $\Gamma$ and $M$, yet in the trivial case (top) they are dark at $M$.

    \textit{Kagome Van Hove wavefunctions}---In the following examples, we move beyond the application of STM and explore the consequences the dark set has when interactions are included. Intuitively, the arrangement of charge density within the unit cell should strongly influence how electrons interact. A paradigmatic example is the kagome lattice --- the wavefunctions near the Van Hove singularities (VHS) at the $M$-point exhibit an unusual structure known as the \textit{sublattice interference effect}, whereby the wavefunction resides on only one or two of the three sublattices  \cite{Kiesel2012a, Kiesel2013}. We show how this sublattice texture originates from the particular band representations at the $M$-point, i.e. the dark set of $\Psi_{\bm{M}_i}(\bm{r})$.

\begin{figure}[t]
    \centering
    \includegraphics[width=\linewidth]{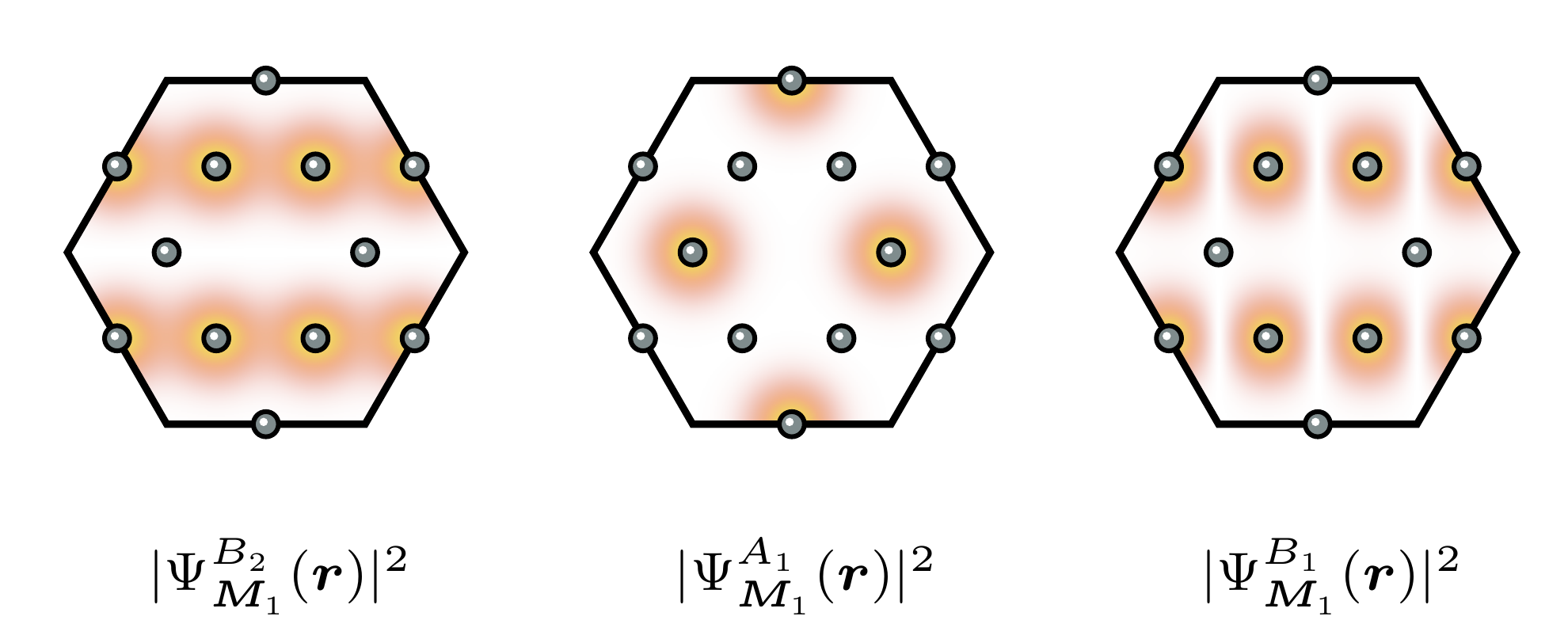}
    \vspace{-0.5cm}
    \caption{{\textbf{Kagome Van Hove dark sets.}} The wavefunctions
    at the VHS of an $s$ orbital kagome lattice correspond to irreps
    $A_1$ and $B_{1,2}$. At the momentum $M_1$, the $A_1$ irrep is
    forced-dark on $B$ and $C$, while the $B_{1,2}$ irreps are forced
    dark on $A$.}
    \label{fig:sli}
    \vspace{-0.5cm}
    \end{figure}

    The three sublattices are located at the $3c$ Wyckoff positions of the group $p6mm$, inducing the $M$-point irreps $A_1 \oplus B_1 \oplus B_2$ (see SM Sec. \ref{sm:kagome}). Calculation of the dark sets reveals that at $M_1$, the $A_1$ wavefunction vanishes on both the $B$ and $C$ sublattices, while the $B_{1,2}$ wavefunctions vanish on the $A$ sublattice (Fig. \ref{fig:sli}); cyclic permutation gives the result at the other two $M$ points. In the nearest-neighbour kagome model, the middle (so-called $p$-type) Van Hove state is the $A_1$ irrep, while the lower (so-called $m$-type) Van Hove state is the $B_{2}$ irrep; the flat band at $M$ transforms as $B_1$, and is distinguished from $B_2$ by a node on the bonds. Therefore, the sublattice texture of the kagome Van Hove states is a consequence of the dark sets of the $M$-point irreps $A_1$ and $B_2$.

    Hence, the dark set gives insight into the spatial structure of wavefunctions, which in turn has striking consequences for interaction effects \cite{Kiesel2012a, Kiesel2013, christensen2022loop, scammell2023chiral, ingham2024theory, ingham2025vestigial, dong2023loop, tazai2022mechanism, tazai2023charge, li2024intertwined, Profe2024, Wu2023SLI}. At the $p$-type VHS, different $M$-points live on strictly distinct sublattices --- scattering between $M$-points therefore has zero projection onto onsite interactions, enhancing the effect of longer-range couplings which stabilise exotic loop current orders. This example highlights how symmetry-enforced zeros can produce fascinating correlation effects.

    \textit{Twisted Bilayer Graphene}---A second example of this is found in magic-angle TBG, where Coulomb interactions strongly reshape the nominally flat bands in a highly momentum-selective way \cite{Xiao2026QTMFlatBands}; the dark set explains the origin of this selectivity. Computation of the dark sets (SM Sec. \ref{sm:tbg}) reveals that the states near $K$ have no zeros common to both layers, whereas the states at $\Gamma$ have a $C_3$-enforced zero at the AA stacking regions (Fig. \ref{fig:tbg}). This zero and its connection to the Chern number have previously been observed by Ledwith et al. \cite{ledwith2025nonlocal}, who have shown that its existence prevents the formation of a fully gapped Mott state, producing a so-called `Mott semimetal'. The zero has profound consequences for how the bands are renormalised \cite{GuineaWalet2018, RademakerAbaninMellado2019, Lewandowski2021FillingDependent, Kwan2025MeanFieldGuide}: the Hartree potential is strong at the AA regions, due to the concentrated density from the $K$ states, which consequently shift upwards in energy sharply. However, the $\Gamma$ states have a zero at the AA regions, allowing them to avoid the Hartree potential, and so do not shift significantly. It is worth emphasising that this zero is a non-perturbative consequence of representation theory, and cannot be removed by interactions unless they close the gap to remote bands or break symmetries. The difference in symmetry irreps at $\Gamma$ and $K$ therefore explains the qualitative reshaping of the flat bands.

    \textit{Discussion}---The examples above show that symmetry-enforced zeros give a local way to connect momentum-space irreps to real-space wavefunction structure, in the spirit of real-space invariants \cite{Xu2021RealSpaceInvariants}. A classic momentum-space counterpart is Kohmoto's result that the Hall conductance is determined by the total vorticity of wavefunction zeros across the Brillouin zone \cite{Kohmoto1985}. The dark set interchanges momentum and position in this construction: at fixed $\bm{k}_*$, the irrep pins zeros to definite locations in real space. Previous works have also considered formulations of topology in real space using the projected position operator or Hamiltonian, retaining phase information absent from the charge density \cite{Kitaev2006,BiancoResta2011,HastingsLoring2010,Prodan2009,Loring2015,LoringSchulzBaldes2020,ProdanSchulzBaldes2016}.
    
    As the bias moves through a band, the LDOS samples different regions of the Brillouin zone and can track how spectral weight is reorganised among sites, bonds and hollow positions. Such bias-dependent textures are a real-space window onto the momentum dependence of Bloch wavefunctions, and hence onto the wavefunction data from which quantum geometry is built \cite{VermaQueiroz2026, YuBernevig2025, TormaPeotta2022}. STM studies of kagome metal CoSn have visualised energy-dependent redistribution of intracell spectral weight between flat and dispersive bands \cite{chen2023visualizing}; the symmetry-enforced part of such a texture is the dark set structure developed here.

    Spectral geometry asks the question ``\textit{Can one hear the shape of a drum?}''; can the shape of a vibrating surface be inferred from its sound \cite{kac1966can}? This paper poses a variation: ``\textit{Can one see the band topology of a wavefunction in its charge density?}'' Na\"ively, the wavefunction amplitude should be agnostic to the phase information from which band topology is built. But the amplitude of a complex function is fixed by its phase near vortices or domain walls, where it is forced to vanish. The dark set makes this intuition precise, demonstrating that crystal symmetry converts band topology into exact zeros of the wavefunction. These zeros are the crystalline analogue of atomic nodes; Bloch states can be viewed as many-atom molecular orbitals, and the dark set comprises their generalised nodes. This represents a surprising result in group representation theory, and might be of broader relevance to the study of irreducible representations of finite groups.
    
    The dark set allows symmetry-indicated topology to be read from the organisation of charge within the unit cell, but also determines the form factors through which electrons interact. In kagome metals, $M$ states near a $p$-type VHS avoid each other by living on different sublattices, promoting the relevance of long-range Coulomb interactions, known to stabilise loop current orders \cite{Kiesel2012a, Kiesel2013, christensen2022loop, scammell2023chiral, ingham2024theory, ingham2025vestigial, dong2023loop, tazai2022mechanism, tazai2023charge, li2024intertwined, Profe2024, Wu2023SLI}. Similarly, the $\Gamma$ and $K$ states in TBG live in different regions of the unit cell as a consequence of their $C_3$ irreps, causing the $\Gamma$ states to avoid the Hartree potential from the $K$ states \cite{Kwan2025MeanFieldGuide}. Thus, the same real-space structure that makes band topology visible can decide which interaction channels survive in topological materials. Wavefunction zeros therefore provide a bridge between three views of quantum materials that are usually separated: momentum-space topology, real-space spectroscopy, and correlated electronic order.

\begin{righttopfigure}[t!]
    \centering
    \includegraphics[width=0.95\linewidth]{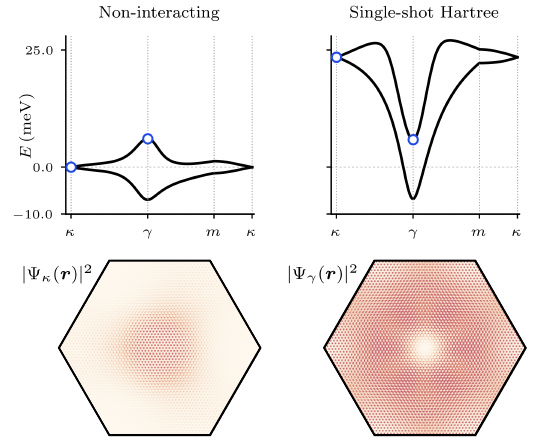}
    \caption{{\textbf{Band renormalisation in moiré graphene.}} Top: non-interacting and Hartree-renormalised narrow bands of TBG near $\theta=1.05^\circ$. The $\Gamma$ states (bottom right)
    are forced-dark at the AA-stacked regions, where the $K$ states (bottom left) are bright.
    The Hartree potential is strongest at the AA regions, strongly increasing the energy at $K$, but not at $\Gamma$ where the
    wavefunction avoids the Hartree potential.}
    \label{fig:tbg}
\end{righttopfigure}

    \FloatBarrier
\section*{Acknowledgements}
\vspace{-0.5cm}

JI thanks Owen Howell and Harley Scammell for interesting discussions, as well as Madisen Holbrook, Jim Hone, and Abhay Pasupathy for the experimental collaboration which inspired this project. This work is supported by the Sloan Foundation, and the NSF MRSEC program at Columbia University through the Center for Precision-Assembled Quantum Materials (DMR-2011738).

\bibliography{arxiv_refs}

\newpage \newpage

\begin{center}
\textbf{\large Appendix}
\end{center}

\setcounter{equation}{0}
\setcounter{table}{0}
\setcounter{section}{0}
\setcounter{figure}{0}
%\makeatletter
\renewcommand{\theequation}{E\arabic{equation}}
\renewcommand{\thefigure}{E\arabic{figure}}
\renewcommand{\thesection}{E\arabic{section}}
%\renewcommand{\citenumfont}[1]{S#1}
%\begin{widetext}
\vspace{-0.6cm}

\section*{Introduction to band representation theory}
\label{em:background}
\vspace{-0.2cm}

A symmetry $g$ acts on a wavefunction $|\alpha\rangle$ via
\begin{align*}
\hat{U}_g |\alpha\rangle
=
\sum_\beta D_{\beta\alpha}(g)|\beta\rangle .
\end{align*}
An \textit{irreducible representation} (irrep) is the ``smallest symmetry-closed space of states''; no change of basis can split it into smaller independent spaces which are each closed under all symmetries --- equivalently, $D_{\beta\alpha}(g)$ cannot be block-diagonalised. A familiar example is the atomic $s,p,d,...$ orbitals: an $s$ orbital is invariant under rotations, the $p$ orbitals form a three-dimensional space mixed by rotations, and the $d$ orbitals form a five-dimensional space. In group theory language, each such symmetry type is an irrep. 

In a crystal, the continuous rotation group is reduced to a \textit{point group}, so the symmetry labels become coarser than atomic angular momentum labels --- e.g. on a triangular lattice, angular momentum is only conserved mod 3, and so an orbital transforming trivially under $C_3$ may contain both $s$- and $f$-wave components. The general statement is that functions can be classified as irreps of the point group, which are often denoted with \textit{Mulliken notation} using a letter and subscript --- e.g. $A_{2g}$, $E_{1u}$.

Bloch states can be understood using representation theory in both momentum and real-space; we begin by discussing the former. A \textit{space group} operation $g=\{R_g|\bm{t}\}$ sends a Bloch state at crystal momentum $\bm{k}$ to $R_g\bm{k}$. Whereas in a translationally invariant system the only momentum invariant under rotations is $\bm{k}=0$, in a crystal, momenta differing by a reciprocal lattice vector are equivalent. Hence, at \textit{high-symmetry momenta} for which  $R_g\bm{k}_*=\bm{k}_*+\bm{G}$, the wavefunction can be simultaneously diagonalised with the symmetry $g$, and the Bloch eigenstates transform as irreps of the symmetries which leave $\bm{k}_*$ invariant --- referred to as the \textit{little group}, $G_{\bm{k}_*}$,
\begin{align*}
\hat{U}_g|\psi_{\bm{k}_*,a}\rangle
=
e^{-i \bm{k}_{*} \cdot \bm{t}} \sum_b
D^\rho_{ba}(g)|\psi_{\bm{k}_*,b}\rangle,
\qquad
 g\in G_{\bm{k}_*}.
\end{align*}
The label $\rho$ specifies the type of irrep; irreps can be fully classified by the \textit{characters} $\chi_\rho(g)=\mathrm{Tr}\,D^\rho(g)$ of the symmetry matrices which act on them.

Band-representation theory relates these momentum-space irreps to the real-space irreps and locations of the atomic orbitals which give rise to them. A \textit{Wyckoff orbit} is a set of \textit{Wyckoff positions} $\bm{r}_*$ which map to themselves under a subset of the symmetries, referred to as the \textit{site-symmetry group} $G_{\bm{r}_*}$. A local orbital placed at $\bm{r}_*$ therefore transforms as an irrep of $G_{\bm{r}_*}$ --- e.g. an $s$ orbital at a $C_3$ centre transforms trivially, whereas a chiral $p_+$ orbital transforms with a factor $e^{2\pi i/3}$.

From orbitals $|\phi_{\bm{R},\alpha}\rangle$ one forms Bloch states,
\begin{align}
|\psi_{\bm{k},\alpha}\rangle
=
\sum_{\bm{R}}
e^{i\bm{k}\cdot(\bm{R}+\bm{r}_\alpha)}
|\phi_{\bm{R},\alpha}\rangle,
\label{bloch_induced}
\end{align}
where $\bm{r}_\alpha$ is the position of the orbital within the unit cell. At $\bm{k}_*$, a symmetry operation $g\in G_{\bm{k}_*}$ acts on Eq. \eqref{bloch_induced} through two ingredients. First, it rotates or reflects the local orbital $|\phi_{\bm{R},\alpha}\rangle$ according to its $G_{\bm{r}_*}$ irrep. Second, it may move the orbital to an equivalent position in a neighbouring unit cell, contributing a Bloch phase via $e^{i\bm{k}_*\cdot(\bm{R}+\bm{r}_\alpha)}$. This allows one to determine the $G_{\bm{k}_*}$ irrep implied by Eq. \eqref{bloch_induced} when $\bm{k}=\bm{k}_*$.

To take a simple example, consider $s$ orbitals on two honeycomb sublattices located at positions $\bm{r}_\pm$. The $s$ orbitals transform trivially under $C_3$, i.e. transform trivially under the site symmetry group. However, under a $C_3$ rotation about the hexagon centre, the Bloch sum acquires a phase $e^{i\bm{k}_*\cdot (1-C_3)\bm{r}_\pm}$. At $K$, the phase is $e^{i\bm{K}\cdot (1-C_3)\bm{r}_\pm}=\omega^\pm$. Hence the two $s$ orbital Bloch sums transform at $K$ as the one-dimensional $C_3$ irreps with characters $\omega$ and $\omega^2$. At the opposite valley $\bm{K}'=-\bm{K}$ the phases are complex conjugated, while at $\Gamma$ both phases are unity. Thus, when only $C_3$ is present, the induced band representation has irrep content $\Gamma:1\oplus1$, $K:\omega\oplus\omega^2$, and $K':\omega^2\oplus\omega$. Another example --- of $s$ orbitals on a kagome lattice --- is detailed in SM Sec. \ref{sm:kag-induction}.

This is the organising principle behind topological quantum chemistry and band-representation theory \cite{Bradlyn2017}. A set of localised Wannier orbitals generates a constrained pattern of irreps at high-symmetry momenta such as $\Gamma$, $K$, $M$. Comparing the irrep content of a group of bands with the catalogue of band representations gives information about their parent real-space description.

The irreps at high-symmetry momenta imply constraints on the ability to construct the band as a Fourier transform of some local orbital as in Eq. \eqref{bloch_induced}. If the high-symmetry irreps of a group of bands cannot be obtained from symmetric exponentially localised Wannier orbitals, the bands have a \textit{Wannier obstruction}, and are otherwise said to possess an \textit{atomic limit}. If they can be obtained from localised Wannier orbitals only by placing the Wannier centres at positions displaced from the atoms, the bands have an \textit{obstructed atomic limit}.

\textit{Symmetry-based indicators} compare the observed irreps at high-symmetry momenta with those obtained from atomic limits, allowing inference of topological invariants such as the Chern number --- e.g. in a $C_3$ symmetric two-dimensional insulator with a single occupied band,
\begin{align*}
e^{2\pi i C/3}
=
\xi_\Gamma \xi_K \xi_{K'},
\end{align*}
where $\xi_{\bm{k}_*}$ is the $C_3$ eigenvalue of the occupied band at $\bm{k}_*$; thus $\xi_{\bm{k}_*}$ determine $C$ mod 3. For an inversion-symmetric time-reversal invariant insulator with one occupied Kramers pair, the inversion eigenvalues $\zeta_{\bm{k}_*}$ at the time-reversal invariant momenta (TRIM) give
\begin{align*}
(-1)^\nu
=
\prod_{\bm{k}_*\in\mathrm{TRIM}}
\zeta_{\bm{k}_*},
\end{align*}
where $\nu$ is the Kane-Mele $\mathbbm{Z}_2$ index. The main text proves that momentum space irreps can be determined from \textit{zeros of the wavefunction}, thus allowing inference of topological invariants via symmetry-based indicators.

\section*{Dark sets of 1H-TMDs}
\label{em:tmd}

Let $\omega=e^{2\pi i/3}$ and denote the three $C_3$
centres by $1a=(0,0)$, $1b=(\tfrac13,\tfrac23)$, and
$1c=(\tfrac23, \tfrac13)$.  The states retained in the three-band model of 1H-TMDs are even under the horizontal mirror, so the $C_{3h}$ projector reduces to its $C_3$ part. At $K$, the local phases $(e^{-i\bm{K}\cdot \bm{t}_{C_3}(1a)},e^{-i\bm{K}\cdot \bm{t}_{C_3}(1b)},e^{-i\bm{K}\cdot \bm{t}_{C_3}(1c)})$ are
$(1,\omega,\omega^2)$; at $K'$ they are the complex conjugates
$(1,\omega^2,\omega)$.  For a state with $C_3$
eigenvalue $\xi_K$ at the $K$-point,
\begin{align}
m_{K,{\bm{r}}}
=
\tfrac13
\sum_{n=0}^2
\left(\vartheta_{K,\bm{r}}\,\xi_K\right)^n
\end{align}
Hence, $\bm{r}$ is symmetry-allowed precisely when
$\vartheta_{K,\bm{r}}\,\xi_K=1$.

The TMD valence band has $C_3$ eigenvalues
$\xi_\Gamma=1$, $\xi_K=\omega$, and
$\xi_{K'}=\omega^2$.  Therefore
\begin{align}
D_\Gamma&=\varnothing,
&
D_K&=\{1a,1b\},
&
D_{K'}&=\{1a,1b\}.
\end{align}
Thus the valley states are allowed only at the hollow
centre $1c$, whereas the $\Gamma$ state has no
symmetry-enforced zero at the three $C_3$ centres.

For comparison, a chiral $E'$ doublet at $\Gamma$ would have
\begin{align}
D_\Gamma(E')=\{1a,1b,1c\}.
\end{align}

\section*{Dark sets of the Haldane model}

Let $O$ be the hexagon centre and $A,B$ the two honeycomb sublattice centres.  Their local $C_3$ phases at $K$ $(e^{-i\bm{K}\cdot \bm{t}_{C_3}(O)},e^{-i\bm{K}\cdot \bm{t}_{C_3}(A)},e^{-i\bm{K}\cdot \bm{t}_{C_3}(B)})$ are $(1,\omega,\omega^2)$; at $K'$ the phases are $(1,\omega^2,\omega)$.

As in the TMD case, a valley state with $C_3$ eigenvalue $\xi$ has $\bm{r}$ symmetry-allowed precisely when $\vartheta_{K,\bm{r}}\,\xi=1$. The dark sets are therefore
\begin{gather}
D_K(1)=\{A,B\},\,
D_K(\omega)=D_{K'}(\omega^2)=\{O,A\},
\nonumber \\
D_{K'}(1)=\{A,B\},\,
D_{K'}(\omega)=D_K(\omega^2)=\{O,B\},
\end{gather}
Hence, a honeycomb $s$ orbital band has no symmetry-enforced $C_3$-centre zero at $\Gamma$. At $K$, the hexagon centre is dark and exactly one atomic sublattice is bright.  If the same sublattice is bright at $K$ and $K'$, the total LDOS is dark on the opposite sublattice, and $O$.  If different sublattices are bright at the two valleys, their total LDOS has only $O$ dark.

\section*{Dark sets of the BHZ model}

We evaluate the four Wyckoff positions $\bm{r}_{1a}=(0,0)$, $\bm{r}_{1b}=(\tfrac12,\tfrac12)$, $\bm{r}_{2c_x}=(\tfrac12,0)$, and $\bm{r}_{2c_y}=(0,\tfrac12)$ at $\Gamma=(0,0)$, $X=(\pi,0)$, $Y=(0,\pi)$, and $M=(\pi,\pi)$. The points $1a$ and $1b$ are $C_4$ centres --- namely, they contain $C_4$ in their site symmetry group --- at $\Gamma$ and $M$, while $2c_x$ and $2c_y$ are only $C_2$ centres. When $C_4$ symmetry is broken, the notation changes as $2c_x\rightarrow 1b$, $2c_y\rightarrow 1c$, $1b\rightarrow 1d$.

For a Kramers pair with inversion eigenvalue $\zeta(\Lambda)$,
\begin{align}
m_{\bm{\Lambda},\bm{r}_i} = 1+e^{-2i\Lambda\cdot \bm{r}_i}\zeta(\Lambda)
\end{align}
at the $C_2$ centres $\bm{r}_i$, which are therefore dark when $e^{-2i\bm{\Lambda}\cdot \bm{r}}\zeta(\Lambda)=-1$.  In the BHZ model, $\zeta(\Gamma)=\operatorname{sgn}(m+2)$, $\zeta(X)=\zeta(Y)=\operatorname{sgn}(m)$, and $\zeta(M)=\operatorname{sgn}(m-2)$.  The $C_2$ phases $(e^{-2i\bm{\Lambda}\cdot\bm{r}_{1a}}, e^{-2i\bm{\Lambda}\cdot\bm{r}_{2c_x}}, e^{-2i\bm{\Lambda}\cdot\bm{r}_{2c_y}}, e^{-2i\bm{\Lambda}\cdot\bm{r}_{1b}})$ at $X,Y,M$ are $(1,-1,1,-1)$, $(1,1,-1,-1)$, and $(1,-1,-1,1)$.

At the $C_4$ centres $1a$ and $1b$, the selection rule depends on the $C_4$ characters.  If the two members of the Kramers pair have $C_4$ eigenvalues $\chi_1,\chi_2$, then
\begin{align}
m_{\Lambda,r}
=
\tfrac{1}{4}\sum_{a=1}^{2}
\sum_{j=0}^{3}
(
\varphi_{\bm{\Lambda},\bm{r}}\,\chi_a
)^j.
\end{align}
where $\varphi_{\Lambda,\bm{r}}=e^{-i\bm{\Lambda}\cdot(1-C_4)\bm{r}}$. The $s$-like pair has $(\chi_1,\chi_2)=(1,1)$, while the $p_\pm$ pair has $(\chi_1,\chi_2)=(i,-i)$. We find $\varphi_{\Gamma, 1a}=\varphi_{\Gamma, 1b}=1$, while $\varphi_{M,1a}=1$ and $\varphi_{M,1b}=-1$.  Therefore the $C_4$ centres are dark at $\Gamma$ for the $p_\pm$ irrep; at $M$, the plaquette centre $1b$ is always dark, and the site centre $1a$ is dark only for the $p_\pm$ irrep.

Combining the $C_2$ and $C_4$ projectors gives
\begin{align}
D_\Gamma
&=
\begin{cases}
\varnothing, & m>-2,\\
\{1a,1b,2c_x,2c_y\}, & m<-2,
\end{cases}
\\
D_M
&=
\begin{cases}
\{1b,2c_x,2c_y\}, & m>2,\\
\{1a,1b\}, & m<2.
\end{cases}
\end{align}
At $X$/$Y$ the result is determined only by the
sign of $m$:
\begin{align}
m>0:\quad&
D_X=\{1b,2c_x\},\qquad
D_Y=\{1b,2c_y\},
\\
m<0:\quad&
D_X=\{1a,2c_y\},\qquad
D_Y=\{1a,2c_x\}.
\end{align}

Restricting to the two bond centres gives the
inversion-clean reduction used in the main text.  For
$|m|<2$, neither bond centre is dark at $\Gamma$ or
$M$.  For $m>2$, $D_\Gamma^{\rm bond}=\varnothing$
and $D_M^{\rm bond}=\{2c_x,2c_y\}$; for $m<-2$ the
two are exchanged, so
$D_\Gamma^{\rm bond}=\{2c_x,2c_y\}$ and
$D_M^{\rm bond}=\varnothing$.  At $X$ and $Y$, the
bond-centre dark sets depend only on the sign of $m$:
for $m>0$ one has
$D_X^{\rm bond}=\{2c_x\}$ and
$D_Y^{\rm bond}=\{2c_y\}$, while for $m<0$ one has
$D_X^{\rm bond}=\{2c_y\}$ and
$D_Y^{\rm bond}=\{2c_x\}$.

For the two regimes plotted in the main text, both bond centres are dark at $M$ in the trivial phase $m>2$, while neither bond centre is dark at $M$ in the quantum spin Hall phase $0<m<2$.  The site and plaquette centres are not inversion-clean probes at $\Gamma$ and $M$, because their stabiliser contains $C_4$.

\widetext
\newpage
\begin{center}
\textbf{\large Supplementary Material}
\end{center}

\vspace{-0.5cm}
\tableofcontents
\newpage
\setcounter{equation}{0}
\setcounter{table}{0}
\setcounter{section}{0}
\setcounter{figure}{0}
\makeatletter
\renewcommand{\theequation}{S\arabic{equation}}
\renewcommand{\thefigure}{S\arabic{figure}}
\renewcommand{\thesection}{S\arabic{section}}
%\renewcommand{\citenumfont}[1]{S#1}
%\begin{widetext}

\section{Dark sets in all 17 wallpaper groups}
\label{supp-wallpaper}

Here we enumerate the dark sets for all irreps in the 17 wallpaper groups: $p1$, $p2$, $pm$, $pg$, $cm$, $pmm$, $pmg$, $pgg$, $cmm$, $p4$, $p4m$, $p4g$, $p3$, $p3m1$, $p31m$, $p6$, and $p6m$. We take the high-symmetry data from the Bilbao Crystallographic Server \cite{Aroyo2006BilbaoI, Aroyo2006BilbaoII, Aroyo2014BrillouinZoneDatabase, DeLaFlor2021LayerGroups} and compute each dark set using the formula in the main text. We present tables for the 15 groups with non-vacuous dark sets. In Group No.~1 ($p1$), the only point stabiliser is $C_1$, so no symmetry-enforced dark spot is possible. In Group No.~4 ($pg$), the only non-translation point operation is a fixed-point-free glide; it cannot produce a real-space stabiliser and likewise produces no dark set rule.

Column orthogonality of characters supplies a sharp consistency check on the tables: for each probe position $\bm{r}_*$, a sum rule holds:
\begin{align}
    \sum_\rho d_\rho\, m_{\bm{k}_*,\bm{r}_*}(\rho)=|\bar{G}_{\bm{k}_*}|/|H_{\bm{k}_*,\bm{r}_*}|
\end{align}
where $m_{\bm{k}_*,\bm{r}_*}(\rho)$ is the index of the main text Eq.~\eqref{indicator} evaluated for irrep $\rho$, $d_\rho$ is the irrep dimension, and $\bar{G}_{\bm{k}_*}$ is the little co-group. Each probe is bright in exactly as many irreps, weighted by dimension, as the index of its stabiliser allows. At a $C_3$ centre at $K$, for instance, the sum equals one: exactly one of the three $C_3$ irreps is bright per centre.

A fixed-point-free nonsymmorphic operation, such as a glide reflection, cannot belong to a real-space stabiliser. Consequently, one-point real-space probes cannot resolve character data carried only by such operations; they distinguish only irreps whose restrictions to the visible, fixed-point symmetries differ. Specifically, we identify the invisible nonsymmorphic sectors for all nonsymmorphic groups as follows. In Group No.~7 ($pmg$), despite the group being nonsymmorphic, all inequivalent irreps at $\Gamma$ and $Y$ have distinct dark sets, while at $X$ and $S$ there is only a single two-dimensional projective irrep, so no further ambiguity arises. In Group No.~8 ($pgg$), the $\Gamma$-point irreps $A_1$ and $A_2$ cannot be distinguished, nor can $B_1$ and $B_2$, because the members of each pair differ only in their characters under the glide reflections. At $X$ and $Y$ there is only a single two-dimensional projective irrep, so no ambiguity arises at those high-symmetry momenta. At $M$, the four one-dimensional projective irreps form two complex-conjugate pairs, yet the two pairs have distinct dark sets, with one pair dark at the $2a$ centres and the other at the $2b$ centres. In Group No.~12 ($p4g$), all the $\Gamma$-point irreps are resolvable. At $X$ and $Y$ there is only a single two-dimensional projective irrep, while at $M$ the four one-dimensional projective irreps form two complex-conjugate pairs and the remaining irrep is two-dimensional. These three physical classes have distinct dark sets, so the $M$-point irreps are resolvable up to the complex-conjugate pairing.

Within the visible sector, all irreps at all high-symmetry points in the wallpaper groups have a unique dark set, with one systematic exception: at time-reversal invariant momenta (TRIM) --- i.e. $\bm{k}_*$ for which $2\bm{k}_*$ is a reciprocal lattice vector --- complex-conjugate irrep pairs have identical dark sets. Examples include $E_\pm$ of the little groups $C_3$, $C_4$, $C_6$. The collision of their dark sets can be understood physically: at such momenta, every realisable local phase is $\pm1$, and the pair $(\rho,\bar{\rho})$ consists of time-reversal partners with pointwise identical charge density, which no density measurement can distinguish (SM Sec.~\ref{sm:general-proof}). However, we note that in systems with time-reversal symmetry, complex-conjugate irreps are properly thought of as two components of a single irreducible corepresentation, and therefore this ambiguity is only meaningful in systems with broken time-reversal symmetry.

The figures and tables below use a common member-resolution convention. Colours identify complete wallpaper-group Wyckoff families. The pale-grey polygon marks the central primitive Wigner--Seitz cell. Repeated markers or line segments outside the grey cell are lattice-related copies included to make the periodic geometry visible. Nearby mathematical labels identify individual members only when a high-symmetry momentum splits a full Wyckoff orbit into subsets with different dark sets. Correspondingly, a table uses the unsplit family label whenever all members of that family have the same dark status, and uses member-resolved labels only when a proper subset must be distinguished. The meaning of solid and dashed symmetry lines is stated in each figure caption.

\makeatletter
\providecommand{\SOnePageStart}{%
  \clearpage
  \renewcommand{\arraystretch}{1.2}%
  \noindent\begin{minipage}[t][\textheight][s]{\textwidth}%
}
\providecommand{\SOnePageEnd}{%
  \end{minipage}%
  \par\FloatBarrier\clearpage
}
\providecommand{\SOneTableStart}{%
  \begin{center}%
  \def\@captype{table}%
}
\providecommand{\SOneTableEnd}{%
  \end{center}%
  \vfill
}
\providecommand{\SOneFigureStart}{%
  \begin{center}%
  \def\@captype{figure}%
}
\providecommand{\SOneFigureEnd}{%
  \end{center}%
}
\makeatother

\SOnePageStart
\SOneTableStart
\caption{\textbf{$p2$ (No.\ 2) Dark sets.}
Probe positions are Wyckoff centres $1a=(0,0)$, $1b=(0,\tfrac12)$, $1c=(\tfrac12,0)$, and $1d=(\tfrac12,\tfrac12)$, all with site symmetry $2$.
High-symmetry momenta are $\Gamma=(0,0)$, $X=(\pi,0)$, $Y=(0,\pi)$, and $M=(\pi,\pi)$.
Entries list the Wyckoff centres that are \emph{symmetry-forbidden (dark)} by the local projector for each little-group irrep.
All little groups are $C_2$, with one-dimensional irreps $A$ (even) and $B$ (odd).}
\label{tab:p2_dark_sets_style}
\vspace{0.1cm}
\begin{ruledtabular}
\begin{tabular}{cll} \\[-3mm]
 & \textbf{Irrep} & \textbf{Dark set} \\[1.5mm] \hline \vspace{-0.2cm} \\
\multirow{2}{*}{\rotatebox[origin=c]{90}{$\Gamma$; $C_2$}}
& $A$ & $\varnothing$ \\
& $B$ & $\{1a,\,1b,\,1c,\,1d\}$ \\[1.5mm] \hline \vspace{-0.2cm} \\
\multirow{2}{*}{\rotatebox[origin=c]{90}{$X$; $C_2$}}
& $A$ & $\{1c,\,1d\}$ \\
& $B$ & $\{1a,\,1b\}$ \\[1.5mm] \hline \vspace{-0.2cm} \\
\multirow{2}{*}{\rotatebox[origin=c]{90}{$Y$; $C_2$}}
& $A$ & $\{1b,\,1d\}$ \\
& $B$ & $\{1a,\,1c\}$ \\[1.5mm] \hline \vspace{-0.2cm} \\
\multirow{2}{*}{\rotatebox[origin=c]{90}{$M$; $C_2$}}
& $A$ & $\{1b,\,1c\}$ \\
& $B$ & $\{1a,\,1d\}$ \\[+1.5mm]
\end{tabular}
\end{ruledtabular}
\SOneTableEnd

\SOneFigureStart
    \includegraphics[width=\linewidth]{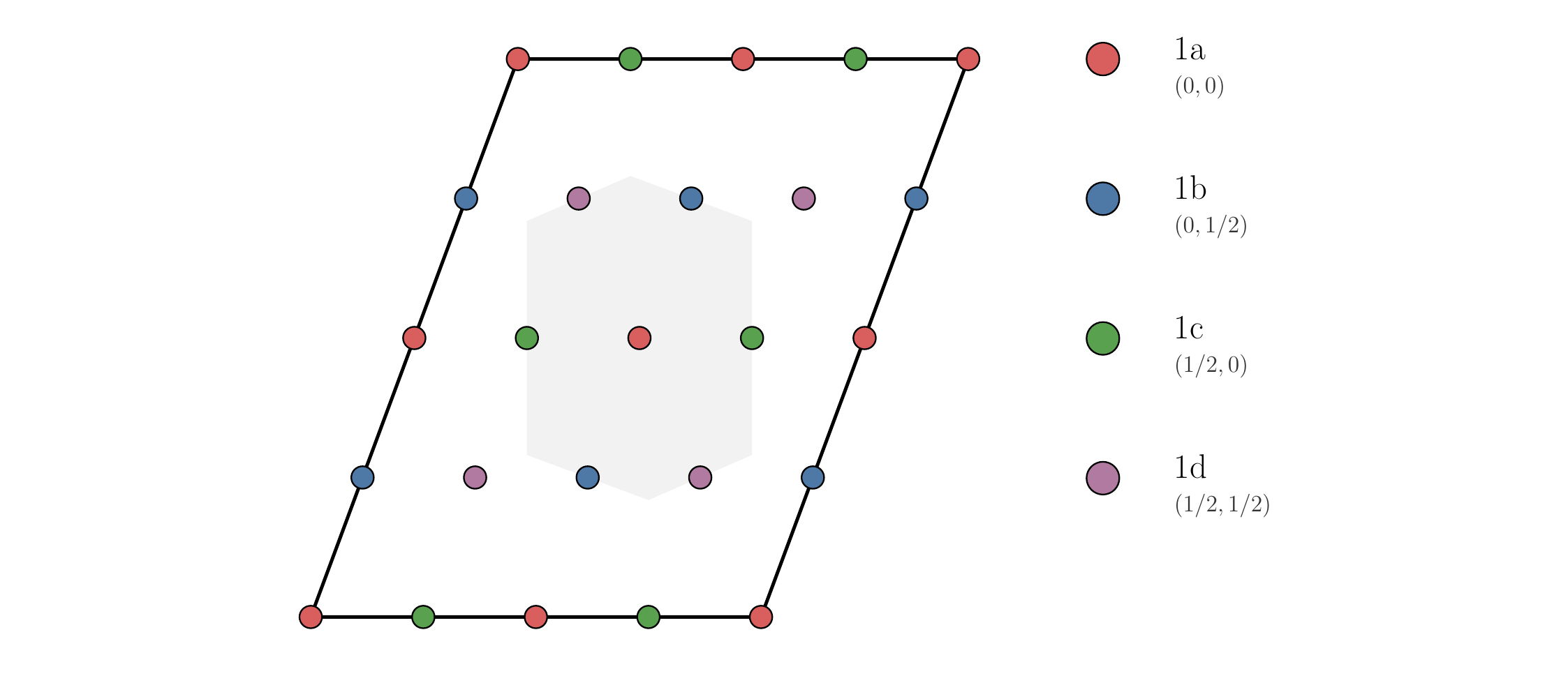}
    \caption{\textbf{Unit cell and Wyckoff positions for $p2$.}
Colours distinguish the four inequivalent $C_2$-centre families $1a$--$1d$, with representative coordinates given in the legend. The black parallelogram bounds the displayed periodic patch, while the pale-grey polygon marks the central primitive Wigner--Seitz cell; repeated boundary markers are lattice translates of the same Wyckoff positions.}
    \label{fig:uc_p2}
\SOneFigureEnd
\SOnePageEnd

\SOnePageStart
\SOneTableStart
\caption{\textbf{$pm$ (No.\ 3) Dark sets.}
Probe positions are Wyckoff families $1a=(0,y)$ and $1b=(\tfrac12,y)$, both on vertical mirrors with site symmetry $m$, and the general family $2c=(x,y),(-x,y)$ with site symmetry $1$. High-symmetry momenta are $\Gamma=(0,0)$, $X=(\pi,0)$, $Y=(0,\pi)$, and $S=(\pi,\pi)$. Entries list the Wyckoff families that are \emph{symmetry-forbidden (dark)} by the local projector for each little-group irrep. At all four momenta the little co-group is $C_s=\{E,\sigma_x\}$, with one-dimensional irreps $A'$ and $A''$ satisfying $\chi(\sigma_x)=+1$ and $-1$, respectively. The general family $2c$ never yields a symmetry-forced node.}
\label{tab:pm_dark_sets}
\vspace{0.1cm}
\begin{ruledtabular}
\begin{tabular}{cll} \\[-3mm]
 & \textbf{Irrep} & \textbf{Dark set} \\[1.5mm] \hline \vspace{-0.2cm} \\
\multirow{2}{*}{\rotatebox[origin=c]{90}{$\Gamma$; $C_s$}}
& $A'$ & $\varnothing$ \\
& $A''$ & $\{1a,\,1b\}$ \\[1.5mm] \hline \vspace{-0.2cm} \\
\multirow{2}{*}{\rotatebox[origin=c]{90}{$X$; $C_s$}}
& $A'$ & $\{1b\}$ \\
& $A''$ & $\{1a\}$ \\[1.5mm] \hline \vspace{-0.2cm} \\
\multirow{2}{*}{\rotatebox[origin=c]{90}{$Y$; $C_s$}}
& $A'$ & $\varnothing$ \\
& $A''$ & $\{1a,\,1b\}$ \\[1.5mm] \hline \vspace{-0.2cm} \\
\multirow{2}{*}{\rotatebox[origin=c]{90}{$S$; $C_s$}}
& $A'$ & $\{1b\}$ \\
& $A''$ & $\{1a\}$ \\[+1.5mm]
\end{tabular}
\end{ruledtabular}
\SOneTableEnd

\SOneFigureStart
    \includegraphics[width=\linewidth]{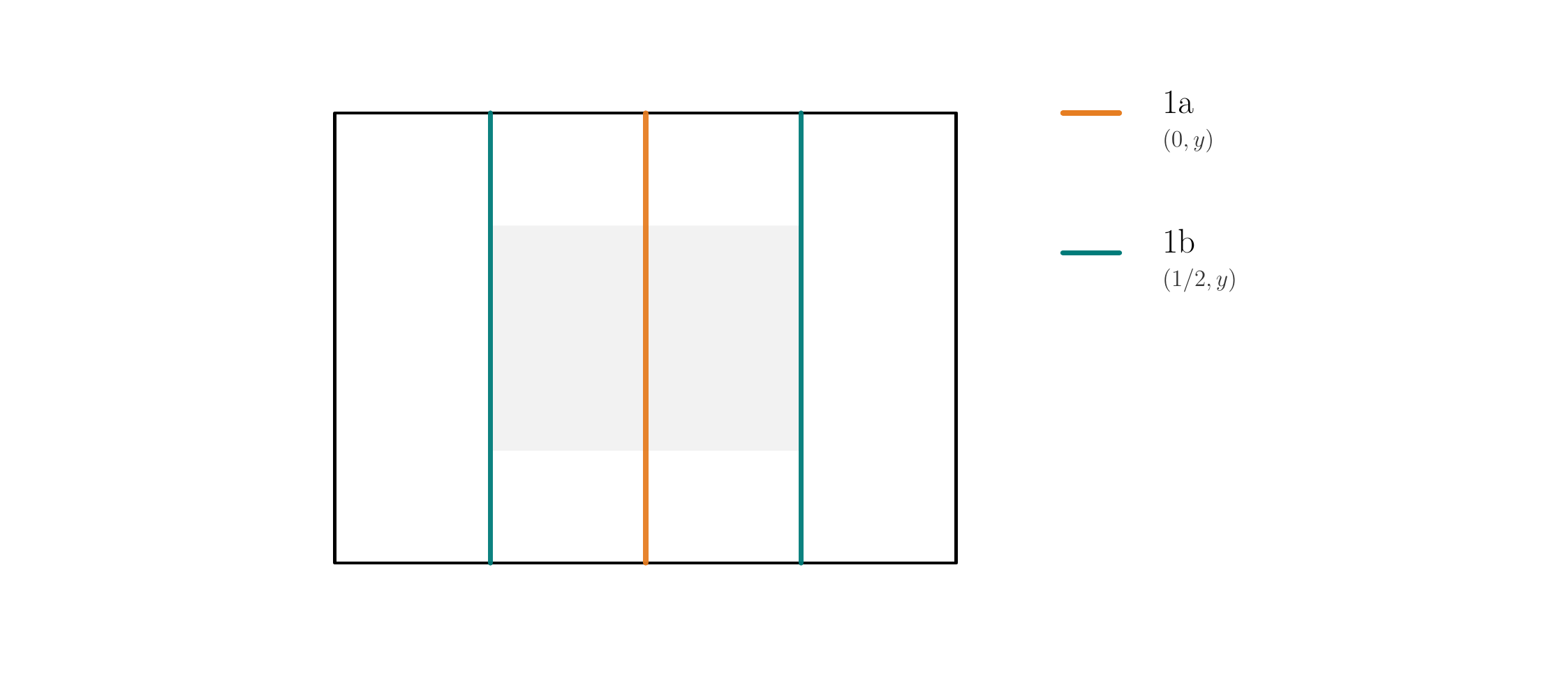}
    \caption{\textbf{Unit cell and Wyckoff positions for $pm$.}
The orange and teal vertical lines are the mirror families $1a$ and $1b$, respectively. Translated affine copies of a given colour belong to the same Wyckoff family and carry the same table label. The pale-grey rectangle marks the central primitive Wigner--Seitz cell.}
    \label{fig:uc_pm}
\SOneFigureEnd
\SOnePageEnd

\SOnePageStart
\SOneTableStart
\caption{\textbf{$cm$ (No.\ 5) Dark sets.}
The mirror Wyckoff family is $2a=\{(0,y),(\tfrac12,y+\tfrac12)\}$ in the conventional centred cell; the two displayed representatives differ by the centring translation and therefore have identical dark status. The general family $4b$ has site symmetry $1$ and never yields a symmetry-forced node. The momentum coordinates $\Gamma=(0,0)$, $X=(\pi,0)$, $Y=(0,\pi)$, and $S=(\pi,\pi)$. The little co-group is $C_s$ at $\Gamma$ and $S$, and $C_1$ at $X$ and $Y$. }
\label{tab:cm_dark_sets}
\vspace{0.1cm}
\begin{ruledtabular}
\begin{tabular}{cll} \\[-3mm]
 & \textbf{Irrep} & \textbf{Dark set} \\[1.5mm] \hline \vspace{-0.2cm} \\
\multirow{2}{*}{\rotatebox[origin=c]{90}{$\Gamma$; $C_s$}}
& $A'$ & $\varnothing$ \\
& $A''$ & $\{2a\}$ \\[1.5mm] \hline \vspace{-0.2cm} \\
\multirow{2}{*}{\rotatebox[origin=c]{90}{$X$; $C_1$}}
& \multirow{2}{*}{$A$} & \multirow{2}{*}{$\varnothing$} \\
& & \\[1.5mm] \hline \vspace{-0.2cm} \\
\multirow{2}{*}{\rotatebox[origin=c]{90}{$Y$; $C_1$}}
& \multirow{2}{*}{$A$} & \multirow{2}{*}{$\varnothing$} \\
& & \\[1.5mm] \hline \vspace{-0.2cm} \\
\multirow{2}{*}{\rotatebox[origin=c]{90}{$S$; $C_s$}}
& $A'$ & $\varnothing$ \\
& $A''$ & $\{2a\}$ \\[+1.5mm]
\end{tabular}
\end{ruledtabular}
\SOneTableEnd

\SOneFigureStart
    \includegraphics[width=\linewidth]{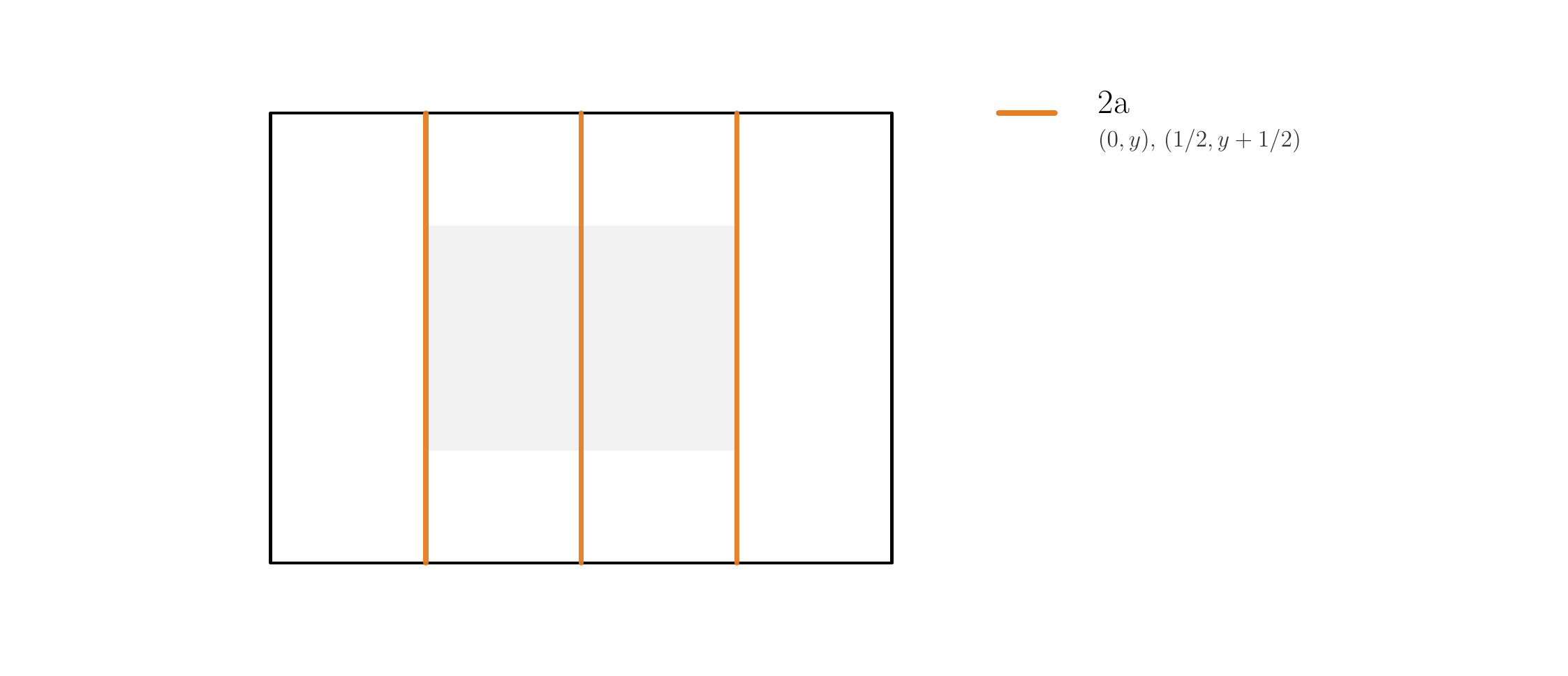}
    \caption{\textbf{Unit cell and Wyckoff positions for $cm$.}
All orange lines are solid members of the single centred mirror orbit $2a$. They are related by centring or lattice translations and have identical dark status. Accordingly, every table entry labelled $2a$ applies to all shown members.}
    \label{fig:uc_cm}
\SOneFigureEnd
\SOnePageEnd

\SOnePageStart
\SOneTableStart
\caption{\textbf{$pmm$ (No.\ 6) Dark sets.}
Probe positions are the Wyckoff families $1a=(0,0)$, $1b=(0,\tfrac12)$, $1c=(\tfrac12,0)$, and $1d=(\tfrac12,\tfrac12)$, all with site symmetry $2mm$; the horizontal-mirror families $2e=(x,0)$ and $2f=(x,\tfrac12)$; and the vertical-mirror families $2g=(0,y)$ and $2h=(\tfrac12,y)$.
High-symmetry momenta are $\Gamma=(0,0)$, $X=(\pi,0)$, $Y=(0,\pi)$, and $M=(\pi,\pi)$; each little co-group is $C_{2v}$.
We choose $B_1$ odd under the horizontal mirror and $B_2$ odd under the vertical mirror.
Entries list the Wyckoff families that are \emph{symmetry-forbidden (dark)} by the local projector.}
\label{tab:pmm_dark_sets}
\vspace{0.1cm}
\begin{ruledtabular}
\begin{tabular}{cll} \\[-3mm]
 & \textbf{Irrep} & \textbf{Dark set} \\[1.5mm] \hline \vspace{-0.2cm} \\
\multirow{4}{*}{\rotatebox[origin=c]{90}{$\Gamma$; $C_{2v}$}}
& $A_1$ & $\varnothing$ \\
& $A_2$ & $\{1a,\,1b,\,1c,\,1d,\,2e,\,2f,\,2g,\,2h\}$ \\
& $B_1$ & $\{1a,\,1b,\,1c,\,1d,\,2e,\,2f\}$ \\
& $B_2$ & $\{1a,\,1b,\,1c,\,1d,\,2g,\,2h\}$ \\[1.5mm] \hline \vspace{-0.2cm} \\
\multirow{4}{*}{\rotatebox[origin=c]{90}{$X$; $C_{2v}$}}
& $A_1$ & $\{1c,\,1d,\,2h\}$ \\
& $A_2$ & $\{1a,\,1b,\,1c,\,1d,\,2e,\,2f,\,2g\}$ \\
& $B_1$ & $\{1a,\,1b,\,1c,\,1d,\,2e,\,2f,\,2h\}$ \\
& $B_2$ & $\{1a,\,1b,\,2g\}$ \\[1.5mm] \hline \vspace{-0.2cm} \\
\multirow{4}{*}{\rotatebox[origin=c]{90}{$Y$; $C_{2v}$}}
& $A_1$ & $\{1b,\,1d,\,2f\}$ \\
& $A_2$ & $\{1a,\,1b,\,1c,\,1d,\,2e,\,2g,\,2h\}$ \\
& $B_1$ & $\{1a,\,1c,\,2e\}$ \\
& $B_2$ & $\{1a,\,1b,\,1c,\,1d,\,2f,\,2g,\,2h\}$ \\[1.5mm] \hline \vspace{-0.2cm} \\
\multirow{4}{*}{\rotatebox[origin=c]{90}{$M$; $C_{2v}$}}
& $A_1$ & $\{1b,\,1c,\,1d,\,2f,\,2h\}$ \\
& $A_2$ & $\{1a,\,1b,\,1c,\,2e,\,2g\}$ \\
& $B_1$ & $\{1a,\,1c,\,1d,\,2e,\,2h\}$ \\
& $B_2$ & $\{1a,\,1b,\,1d,\,2f,\,2g\}$ \\[+1.5mm]
\end{tabular}
\end{ruledtabular}
\SOneTableEnd

\SOneFigureStart
    \includegraphics[width=\linewidth]{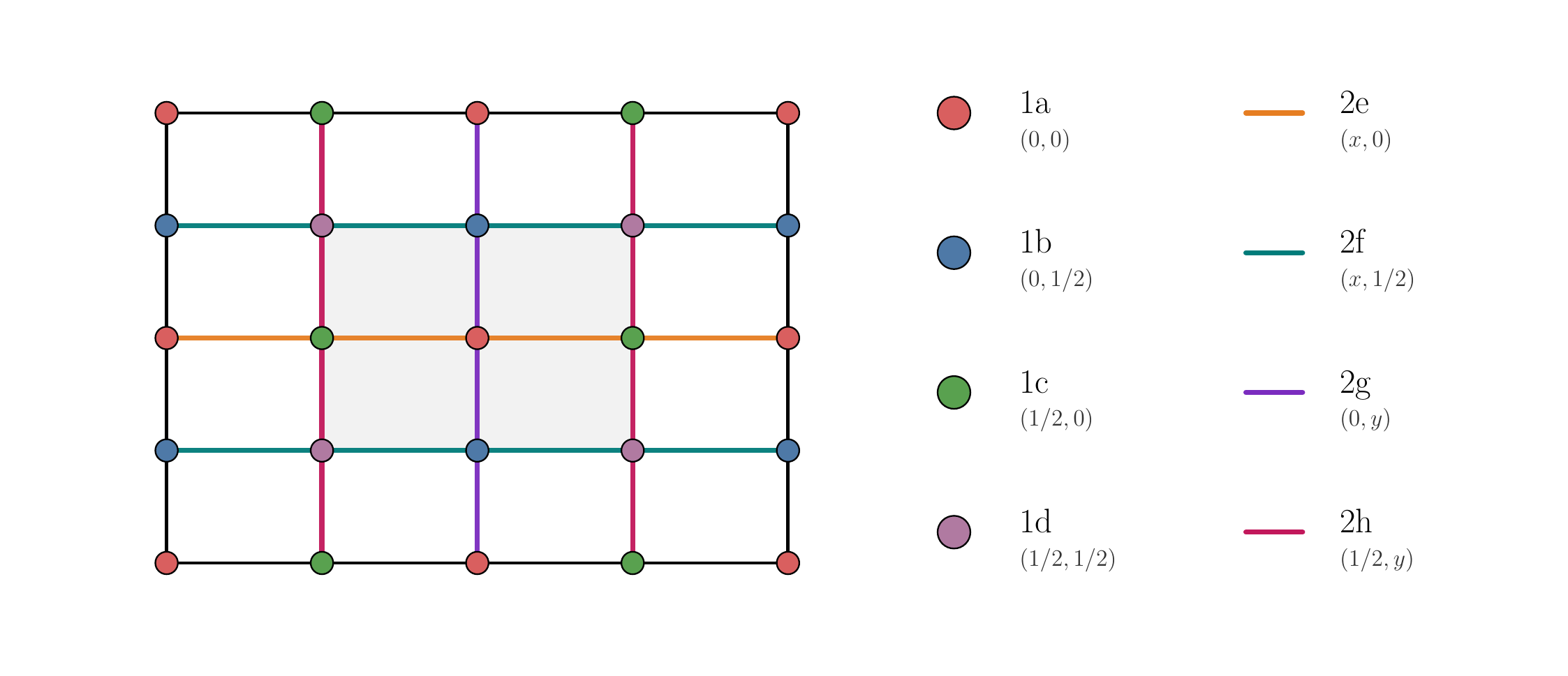}
    \caption{\textbf{Unit cell and Wyckoff positions for $pmm$.}
Marker colours distinguish the four $2mm$ centres $1a$--$1d$. The coloured horizontal and vertical lines identify the mirror families $2e$--$2h$ listed in the legend. Translated copies of a given colour belong to the same family.}
    \label{fig:uc_pmm}
\SOneFigureEnd
\SOnePageEnd

\SOnePageStart
\SOneTableStart
\caption{\textbf{$pmg$ (No.\ 7) Dark sets.}
Probe positions are Wyckoff families $2a$ with $C_2$ centres at $(0,0)$ and $(\tfrac12,0)$, $2b$ with $C_2$ centres at $(0,\tfrac12)$ and $(\tfrac12,\tfrac12)$, and $2c$ on the vertical mirrors with site symmetry $m$.
High-symmetry momenta are $\Gamma=(0,0)$, $X=(\pi,0)$, $Y=(0,\pi)$, and $S=(\pi,\pi)$.
At $X$ and $S$ the unique small irrep is two-dimensional and projective; its local projector has positive rank at every listed probe.}
\label{tab:pmg_dark_sets}
\vspace{0.1cm}
\begin{ruledtabular}
\begin{tabular}{cll} \\[-3mm]
 & \textbf{Irrep} & \textbf{Dark set} \\[1.5mm] \hline \vspace{-0.2cm} \\
\multirow{4}{*}{\rotatebox[origin=c]{90}{$\Gamma$; $C_{2v}$}}
& $A_1$ & $\varnothing$ \\
& $A_2$ & $\{2c\}$ \\
& $B_1$ & $\{2a,\,2b\}$ \\
& $B_2$ & $\{2a,\,2b,\,2c\}$ \\[1.5mm] \hline \vspace{-0.2cm} \\
\multirow{1}{*}{\rotatebox[origin=c]{90}{$X$}}
& $X_5$ (2D projective) & $\varnothing$ \\[1.5mm] \hline \vspace{-0.2cm} \\
\multirow{4}{*}{\rotatebox[origin=c]{90}{$Y$; $C_{2v}$}}
& $A_1$ & $\{2b\}$ \\
& $A_2$ & $\{2b,\,2c\}$ \\
& $B_1$ & $\{2a\}$ \\
& $B_2$ & $\{2a,\,2c\}$ \\[1.5mm] \hline \vspace{-0.2cm} \\
\multirow{1}{*}{\rotatebox[origin=c]{90}{$S$}}
& $S_5$ (2D projective) & $\varnothing$ \\[+1.5mm]
\end{tabular}
\end{ruledtabular}
\SOneTableEnd

\SOneFigureStart
    \includegraphics[width=\linewidth]{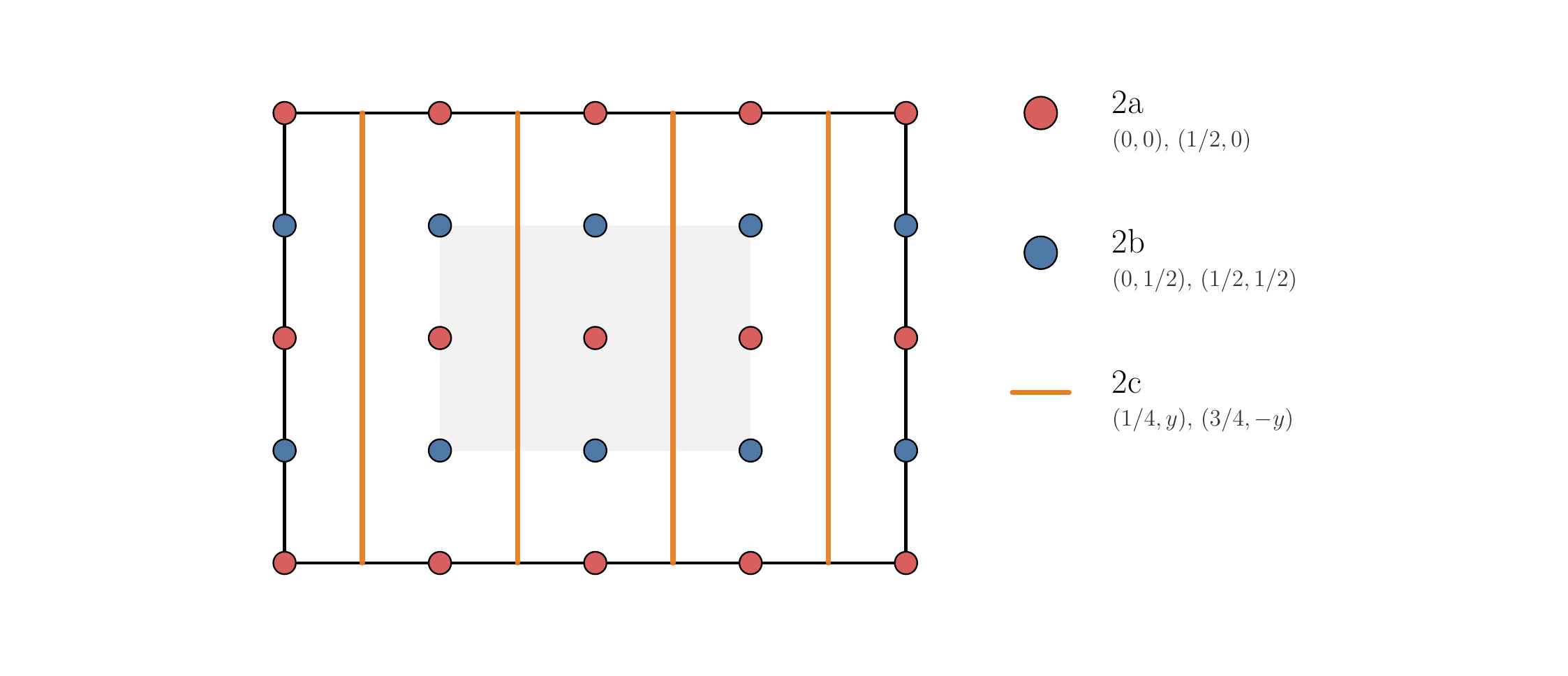}
    \caption{\textbf{Unit cell and Wyckoff positions for $pmg$.}
Red and blue markers show the $C_2$ families $2a$ and $2b$. All orange vertical lines are solid affine members of the mirror orbit $2c$.}
    \label{fig:uc_pmg}
\SOneFigureEnd
\SOnePageEnd

\SOnePageStart
\SOneTableStart
\caption{\textbf{$pgg$ (No.\ 8) Dark sets.}
Probe positions are the $C_2$ Wyckoff families $2a=\{(0,0),(\tfrac12,\tfrac12)\}$ and $2b=\{(\tfrac12,0),(0,\tfrac12)\}$.
The general family $4c$ has site symmetry $1$ and never yields a symmetry-forced node.
At $X$ and $Y$ there is one two-dimensional projective small irrep, denoted $X_5$ and $Y_5$.
At $M$ there are four one-dimensional projective small irreps; in the ordering $(E,C_2,g_x,g_y)$ their character signatures are $\rho_1=(1,-1,-i,i)$, $\rho_2=(1,-1,i,-i)$, $\rho_3=(1,1,-i,-i)$, and $\rho_4=(1,1,i,i)$.}
\label{tab:pgg_dark_sets_complete}
\vspace{0.1cm}
\begin{ruledtabular}
\begin{tabular}{cll} \\[-3mm]
 & \textbf{Irrep} & \textbf{Dark set} \\[1.5mm] \hline \vspace{-0.2cm} \\
\multirow{4}{*}{\rotatebox[origin=c]{90}{$\Gamma$; $C_{2v}$}}
& $A_1$ & $\varnothing$ \\
& $A_2$ & $\varnothing$ \\
& $B_1$ & $\{2a,\,2b\}$ \\
& $B_2$ & $\{2a,\,2b\}$ \\[1.5mm] \hline \vspace{-0.2cm} \\
\multirow{1}{*}{\rotatebox[origin=c]{90}{$X$}}
& $X_5$ (2D projective) & $\varnothing$ \\[1.5mm] \hline \vspace{-0.2cm} \\
\multirow{1}{*}{\rotatebox[origin=c]{90}{$Y$}}
& $Y_5$ (2D projective) & $\varnothing$ \\[1.5mm] \hline \vspace{-0.2cm} \\
\multirow{4}{*}{\rotatebox[origin=c]{90}{$M$}}
& $\rho_1$ & $\{2a\}$ \\
& $\rho_2$ & $\{2a\}$ \\
& $\rho_3$ & $\{2b\}$ \\
& $\rho_4$ & $\{2b\}$ \\[+1.5mm]
\end{tabular}
\end{ruledtabular}
\SOneTableEnd

\SOneFigureStart
    \includegraphics[width=\linewidth]{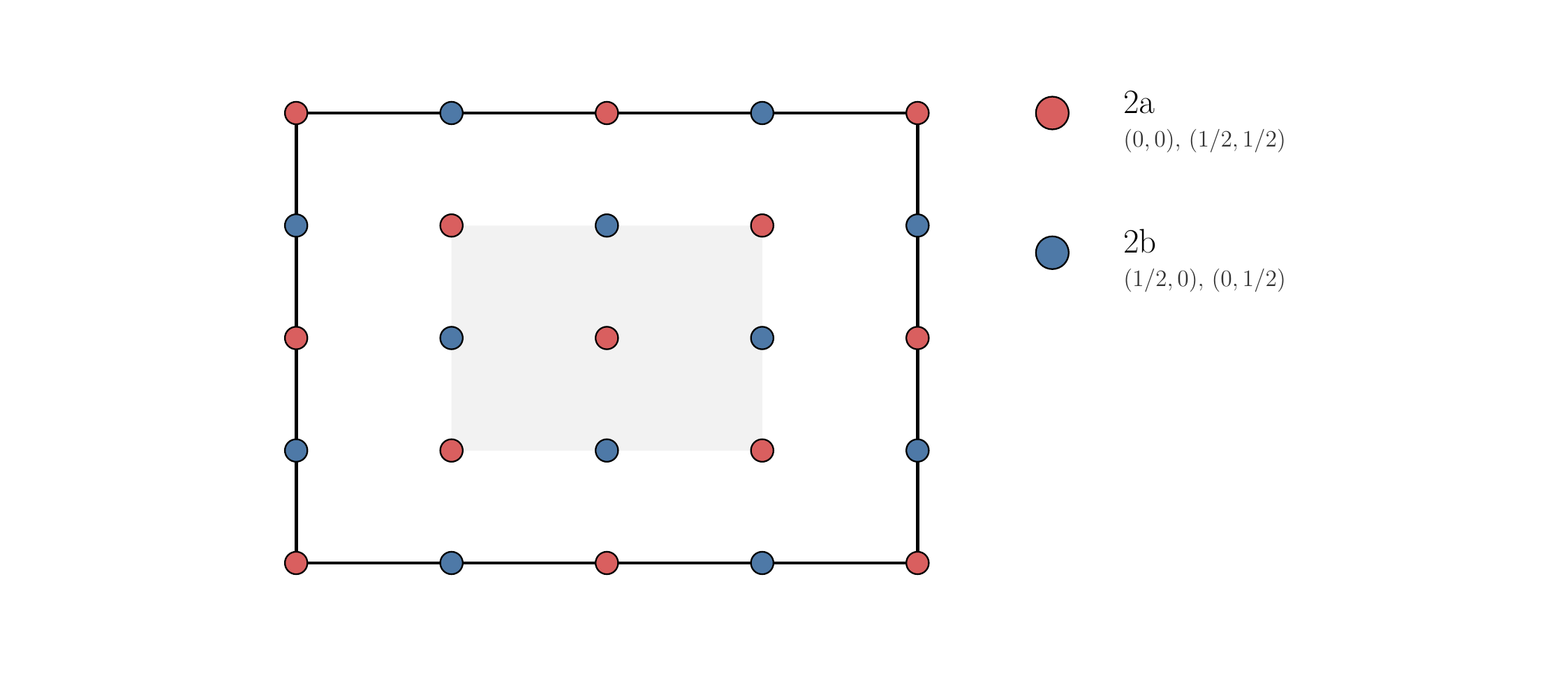}
    \caption{\textbf{Unit cell and Wyckoff positions for $pgg$.}
Red and blue markers distinguish the two $C_2$ Wyckoff families $2a$ and $2b$; repeated markers are lattice translates. The general family $4c$ is omitted because its site symmetry is trivial and it never contributes a symmetry-enforced one-point zero.}
    \label{fig:uc_pgg}
\SOneFigureEnd
\SOnePageEnd

\SOnePageStart
\SOneTableStart
\caption{\textbf{$cmm$ (No.\ 9, $c2mm$) Dark sets.}
The figure uses the standard families $2a=\{(0,0),(\tfrac12,\tfrac12)\}$, $2b=\{(0,\tfrac12),(\tfrac12,0)\}$, the $C_2$ family $4c$, horizontal-mirror family $4d$, and vertical-mirror family $4e$; the general family is $8f$.
At $X,Y$, we resolve $4c$ into the centring-related pairs $4c^{(1)}=\{(\tfrac14,\tfrac14),(\tfrac34,\tfrac34)\}$ and $4c^{(2)}=\{(\tfrac34,\tfrac14),(\tfrac14,\tfrac34)\}$. At $\Gamma,M$, the label $4c$ denotes the complete four-point family.
The momentum coordinates $\Gamma=(0,0)$, $X=(\pi,0)$, $Y=(0,\pi)$, and $M=(\pi,\pi)$ are quoted in the reciprocal basis dual to primitive translations.
The little co-group is $C_{2v}$ at $\Gamma,M$ and $C_2$ at $X,Y$.
We choose $B_1$ odd under the horizontal mirror and $B_2$ odd under the vertical mirror.}
\label{tab:cmm_dark_sets}
\vspace{0.1cm}
\begin{ruledtabular}
\begin{tabular}{cll} \\[-3mm]
 & \textbf{Irrep} & \textbf{Dark set} \\[1.5mm] \hline \vspace{-0.2cm} \\
\multirow{4}{*}{\rotatebox[origin=c]{90}{$\Gamma$; $C_{2v}$}}
& $A_1$ & $\varnothing$ \\
& $A_2$ & $\{2a,\,2b,\,4d,\,4e\}$ \\
& $B_1$ & $\{2a,\,2b,\,4c,\,4d\}$ \\
& $B_2$ & $\{2a,\,2b,\,4c,\,4e\}$ \\[1.5mm] \hline \vspace{-0.2cm} \\
\multirow{4}{*}{\rotatebox[origin=c]{90}{$X$; $C_2$}}
& & \\
& $A$ & $\{2b,\,4c^{(1)}\}$ \\
& $B$ & $\{2a,\,4c^{(2)}\}$ \\
& & \\[1.5mm] \hline \vspace{-0.2cm} \\
\multirow{4}{*}{\rotatebox[origin=c]{90}{$Y$; $C_2$}}
& & \\
& $A$ & $\{2b,\,4c^{(2)}\}$ \\
& $B$ & $\{2a,\,4c^{(1)}\}$ \\
& & \\[1.5mm] \hline \vspace{-0.2cm} \\
\multirow{4}{*}{\rotatebox[origin=c]{90}{$M$; $C_{2v}$}}
& $A_1$ & $\{4c\}$ \\
& $A_2$ & $\{2a,\,2b,\,4c,\,4d,\,4e\}$ \\
& $B_1$ & $\{2a,\,2b,\,4d\}$ \\
& $B_2$ & $\{2a,\,2b,\,4e\}$ \\[+1.5mm]
\end{tabular}
\end{ruledtabular}
\SOneTableEnd

\SOneFigureStart
    \includegraphics[width=\linewidth]{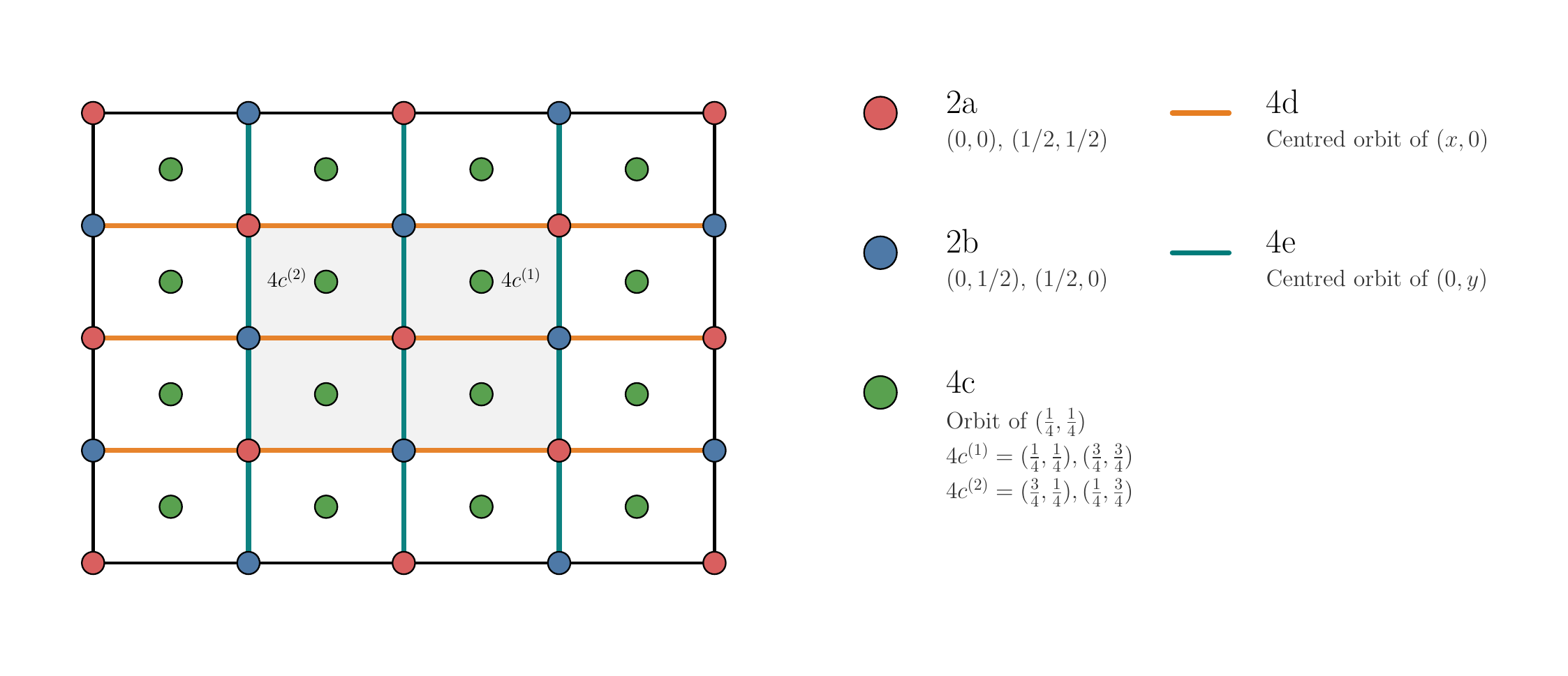}
    \caption{\textbf{Unit cell and Wyckoff positions for $cmm$.}
Colours identify the full families $2a$, $2b$, and $4c$, while orange and teal lines show the horizontal $4d$ and vertical $4e$ mirror families. The nearby labels $4c^{(1)}$ and $4c^{(2)}$ identify the two centring-related pairs required at $X,Y$; at $\Gamma,M$ the table uses the unsplit label $4c$. All displayed $4d$ and $4e$ mirrors form a single member-resolved family.}
    \label{fig:uc_cmm}
\SOneFigureEnd
\SOnePageEnd

\SOnePageStart
\SOneTableStart
\caption{\textbf{$p4$ (No.\ 10) Dark sets.}
Probe positions are $1a=(0,0)$ and $1b=(\tfrac12,\tfrac12)$, with site symmetry $4$, and the two members $2c_x=(\tfrac12,0)$ and $2c_y=(0,\tfrac12)$ of the $2c$ family, with site symmetry $2$.
The general family $4d$ has site symmetry $1$ and never yields a symmetry-forced node.
High-symmetry momenta are $\Gamma=(0,0)$, $X=(\pi,0)$, $Y=(0,\pi)$, and $M=(\pi,\pi)$; the little co-groups are $C_4$ at $\Gamma,M$ and $C_2$ at $X,Y$. At $\Gamma,M$, the label $2c$ denotes both members; the resolved labels $2c_x,2c_y$ are used at $X,Y$.}
\label{tab:p4_dark_sets_pgnotation}
\vspace{0.1cm}
\begin{ruledtabular}
\begin{tabular}{cll} \\[-3mm]
 & \textbf{Irrep} & \textbf{Dark set} \\[1.5mm] \hline \vspace{-0.2cm} \\
\multirow{4}{*}{\rotatebox[origin=c]{90}{$\Gamma$; $C_4$}}
& $A$ & $\varnothing$ \\
& $B$ & $\{1a,\,1b\}$ \\
& $E_+$ & $\{1a,\,1b,\,2c\}$ \\
& $E_-$ & $\{1a,\,1b,\,2c\}$ \\[1.5mm] \hline \vspace{-0.2cm} \\
\multirow{2}{*}{\rotatebox[origin=c]{90}{$X$; $C_2$}}
& $A$ & $\{1b,\,2c_x\}$ \\
& $B$ & $\{1a,\,2c_y\}$ \\[1.5mm] \hline \vspace{-0.2cm} \\
\multirow{2}{*}{\rotatebox[origin=c]{90}{$Y$; $C_2$}}
& $A$ & $\{1b,\,2c_y\}$ \\
& $B$ & $\{1a,\,2c_x\}$ \\[1.5mm] \hline \vspace{-0.2cm} \\
\multirow{4}{*}{\rotatebox[origin=c]{90}{$M$; $C_4$}}
& $A$ & $\{1b,\,2c\}$ \\
& $B$ & $\{1a,\,2c\}$ \\
& $E_+$ & $\{1a,\,1b\}$ \\
& $E_-$ & $\{1a,\,1b\}$ \\[+1.5mm]
\end{tabular}
\end{ruledtabular}
\SOneTableEnd

\SOneFigureStart
    \includegraphics[width=\linewidth]{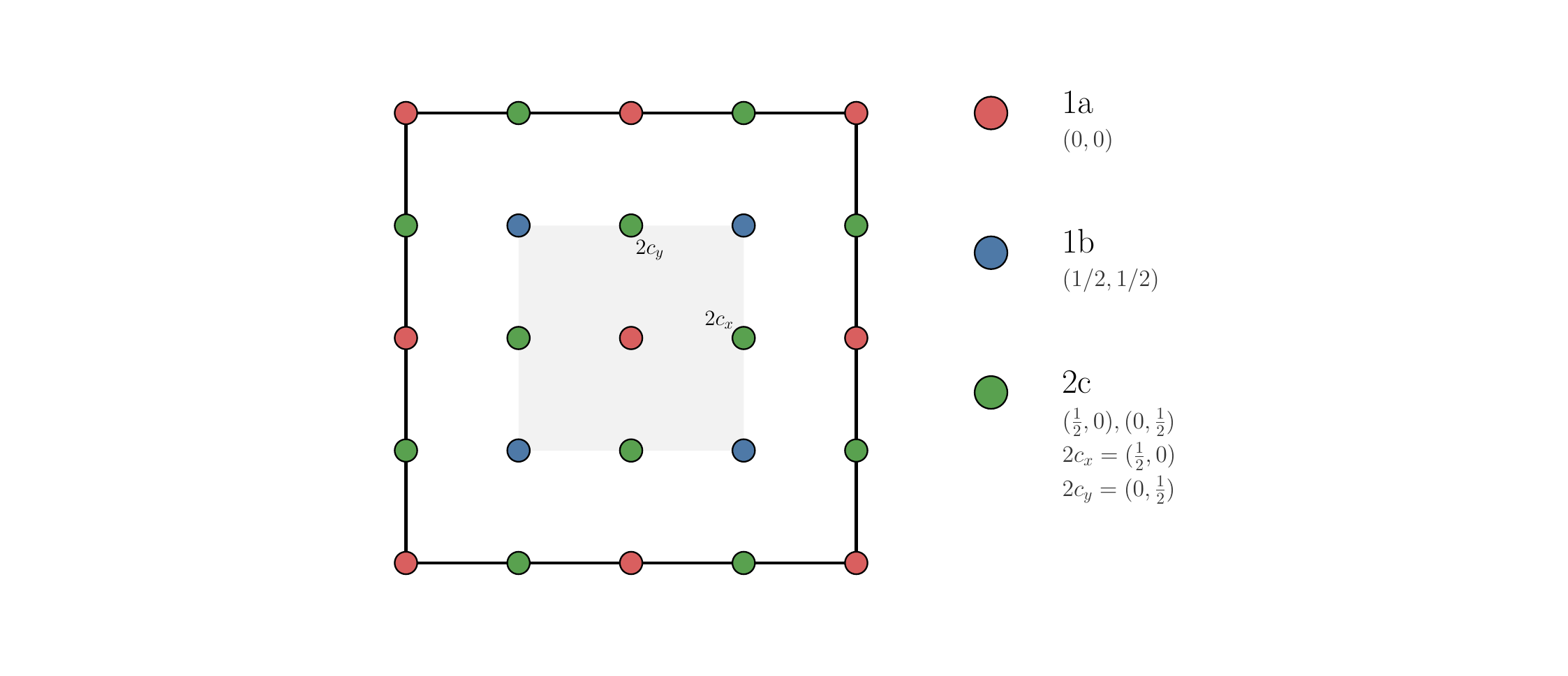}
    \caption{\textbf{Unit cell and Wyckoff positions for $p4$.}
Colours identify $1a$, $1b$, and the complete $2c$ orbit. The nearby labels $2c_x=(\tfrac12,0)$ and $2c_y=(0,\tfrac12)$ mark the two central-cell members used separately at $X,Y$. At $\Gamma,M$, where the two members have the same dark status, the table uses the unsplit label $2c$.}
    \label{fig:uc_p4}
\SOneFigureEnd
\SOnePageEnd

\SOnePageStart
\SOneTableStart
\caption{\textbf{$p4m$ (No.\ 11, $p4mm$) Dark sets.}
Probe families are $1a=(0,0)$ and $1b=(\tfrac12,\tfrac12)$ with site symmetry $4mm$; the two members $2c_x=(\tfrac12,0)$ and $2c_y=(0,\tfrac12)$ of the $2c$ family, with site symmetry $2mm$; axis mirrors $4d$; shifted axis mirrors $4e$; and diagonal mirrors $4f$.
At $X,Y$ we resolve the axis families as $4d_{\mathrm h}:y=0$, $4d_{\mathrm v}:x=0$, $4e_{\mathrm h}:y=\tfrac12$, and $4e_{\mathrm v}:x=\tfrac12$. At $\Gamma,M$, the unsplit labels $2c$, $4d$, and $4e$ denote the complete corresponding families.
The general family $8g$ has site symmetry $1$.
The little co-groups are $C_{4v}$ at $\Gamma,M$ and $C_{2v}$ at $X,Y$.
At $\Gamma,M$ we use the standard convention in which $B_1$ is even under the axis mirrors and odd under the diagonal mirrors; at $X,Y$, $B_1$ is even under the horizontal mirror and $B_2$ is odd under it.}
\label{tab:p4m_dark_sets}
\vspace{0.1cm}
\begin{ruledtabular}
\begin{tabular}{cll} \\[-3mm]
 & \textbf{Irrep} & \textbf{Dark set} \\[1.5mm] \hline \vspace{-0.2cm} \\
\multirow{5}{*}{\rotatebox[origin=c]{90}{$\Gamma$; $C_{4v}$}}
& $A_1$ & $\varnothing$ \\
& $A_2$ & $\{1a,\,1b,\,2c,\,4d,\,4e,\,4f\}$ \\
& $B_1$ & $\{1a,\,1b,\,4f\}$ \\
& $B_2$ & $\{1a,\,1b,\,2c,\,4d,\,4e\}$ \\
& $E$ & $\{1a,\,1b,\,2c\}$ \\[1.5mm] \hline \vspace{-0.2cm} \\
\multirow{4}{*}{\rotatebox[origin=c]{90}{$X$; $C_{2v}$}}
& $A_1$ & $\{1b,\,2c_x,\,4e_{\mathrm v}\}$ \\
& $A_2$ & $\{1a,\,1b,\,2c,\,4d,\,4e_{\mathrm h}\}$ \\
& $B_1$ & $\{1a,\,2c_y,\,4d_{\mathrm v}\}$ \\
& $B_2$ & $\{1a,\,1b,\,2c,\,4d_{\mathrm h},\,4e\}$ \\[1.5mm] \hline \vspace{-0.2cm} \\
\multirow{4}{*}{\rotatebox[origin=c]{90}{$Y$; $C_{2v}$}}
& $A_1$ & $\{1b,\,2c_y,\,4e_{\mathrm h}\}$ \\
& $A_2$ & $\{1a,\,1b,\,2c,\,4d,\,4e_{\mathrm v}\}$ \\
& $B_1$ & $\{1a,\,1b,\,2c,\,4d_{\mathrm v},\,4e\}$ \\
& $B_2$ & $\{1a,\,2c_x,\,4d_{\mathrm h}\}$ \\[1.5mm] \hline \vspace{-0.2cm} \\
\multirow{5}{*}{\rotatebox[origin=c]{90}{$M$; $C_{4v}$}}
& $A_1$ & $\{1b,\,2c,\,4e\}$ \\
& $A_2$ & $\{1a,\,1b,\,2c,\,4d,\,4f\}$ \\
& $B_1$ & $\{1a,\,1b,\,2c,\,4e,\,4f\}$ \\
& $B_2$ & $\{1a,\,2c,\,4d\}$ \\
& $E$ & $\{1a,\,1b\}$ \\[+1.5mm]
\end{tabular}
\end{ruledtabular}
\SOneTableEnd

\SOneFigureStart
    \includegraphics[width=\linewidth]{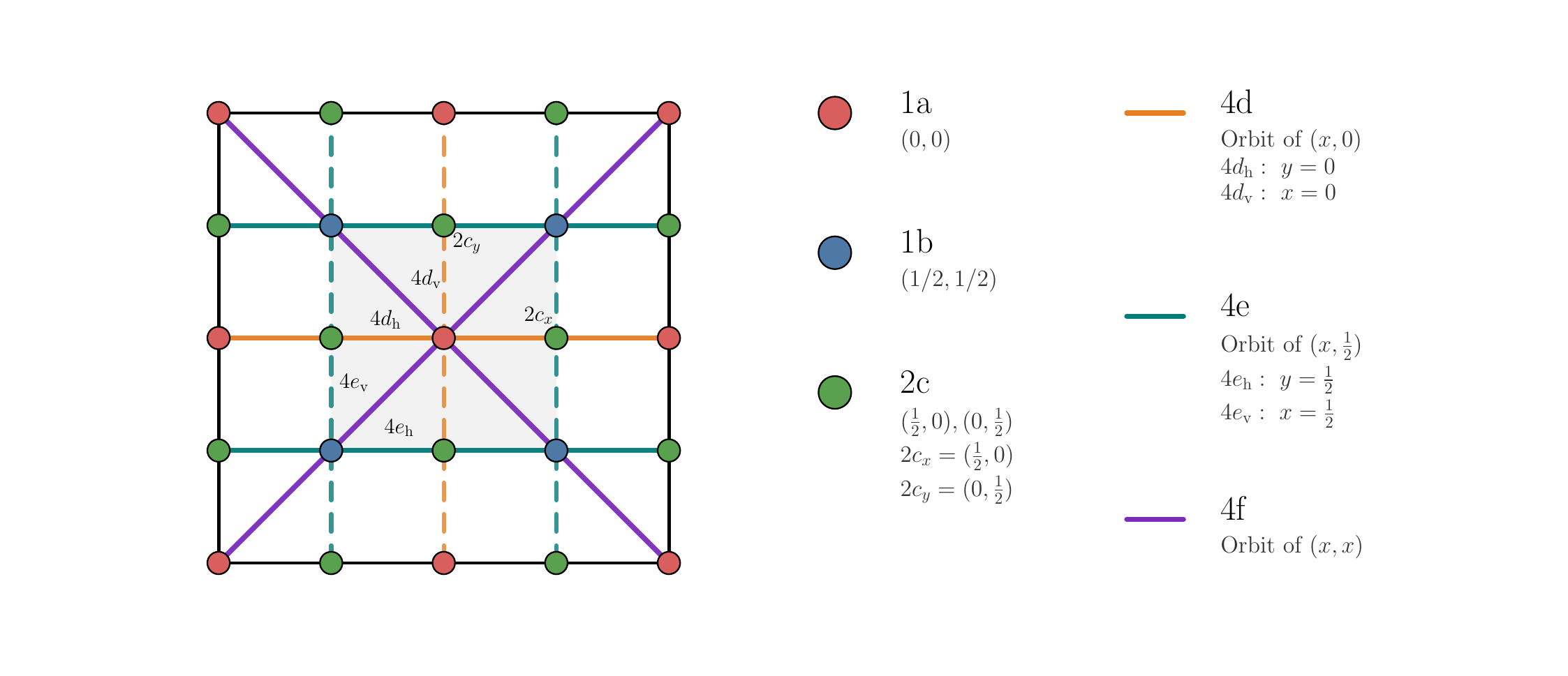}
    \caption{\textbf{Unit cell and Wyckoff positions for $p4m$.}
Marker colours identify $1a$, $1b$, and the full $2c$ orbit; nearby labels distinguish $2c_x$ and $2c_y$. For the axis-mirror families, solid orange horizontal segments are $4d_{\mathrm h}$ and dashed orange vertical segments are $4d_{\mathrm v}$, while solid teal horizontal segments are $4e_{\mathrm h}$ and dashed teal vertical segments are $4e_{\mathrm v}$. The two solid purple diagonals through the origin together form the unsplit family $4f$. The resolved labels are used at $X,Y$, whereas $\Gamma,M$ use the complete-family labels $2c$, $4d$, and $4e$.}
    \label{fig:uc_p4m}
\SOneFigureEnd
\SOnePageEnd

\SOnePageStart
\SOneTableStart
\caption{\textbf{$p4g$ (No.\ 12, $p4gm$) Dark sets.}
Probe families are $2a=\{(0,0),(\tfrac12,\tfrac12)\}$ with site symmetry $4$, $2b=\{(\tfrac12,0),(0,\tfrac12)\}$ with site symmetry $2mm$, and the diagonal-mirror family $4c$; the general family $8d$ has site symmetry $1$.
At $\Gamma$ the small representations are the ordinary $C_{4v}$ irreps.
At $X,Y$ there is one two-dimensional projective small irrep, denoted $X_5,Y_5$.
At $M$ the projective irreps have character signatures in the ordering $(E,C_2,C_4,C_4^{-1},g_x,g_y,m_d,m_{d'})$: $\rho_1=(1,-1,-i,i,-i,i,-1,1)$, $\rho_2=(1,-1,i,-i,i,-i,-1,1)$, $\rho_3=(1,-1,-i,i,i,-i,1,-1)$, $\rho_4=(1,-1,i,-i,-i,i,1,-1)$, and $\rho_5=(2,2,0,0,0,0,0,0)$.}
\label{tab:p4g_dark_sets}
\vspace{0.1cm}
\begin{ruledtabular}
\begin{tabular}{cll} \\[-3mm]
 & \textbf{Irrep} & \textbf{Dark set} \\[1.5mm] \hline \vspace{-0.2cm} \\
\multirow{5}{*}{\rotatebox[origin=c]{90}{$\Gamma$; $C_{4v}$}}
& $A_1$ & $\varnothing$ \\
& $A_2$ & $\{2b,\,4c\}$ \\
& $B_1$ & $\{2a,\,2b,\,4c\}$ \\
& $B_2$ & $\{2a\}$ \\
& $E$ & $\{2a,\,2b\}$ \\[1.5mm] \hline \vspace{-0.2cm} \\
\multirow{1}{*}{\rotatebox[origin=c]{90}{$X$}}
& $X_5$ (2D projective) & $\varnothing$ \\[1.5mm] \hline \vspace{-0.2cm} \\
\multirow{1}{*}{\rotatebox[origin=c]{90}{$Y$}}
& $Y_5$ (2D projective) & $\varnothing$ \\[1.5mm] \hline \vspace{-0.2cm} \\
\multirow{5}{*}{\rotatebox[origin=c]{90}{$M$}}
& $\rho_1$ & $\{2a,\,2b,\,4c\}$ \\
& $\rho_2$ & $\{2a,\,2b,\,4c\}$ \\
& $\rho_3$ & $\{2a\}$ \\
& $\rho_4$ & $\{2a\}$ \\
& $\rho_5$ & $\{2b\}$ \\[+1.5mm]
\end{tabular}
\end{ruledtabular}
\SOneTableEnd

\SOneFigureStart
    \includegraphics[width=\linewidth]{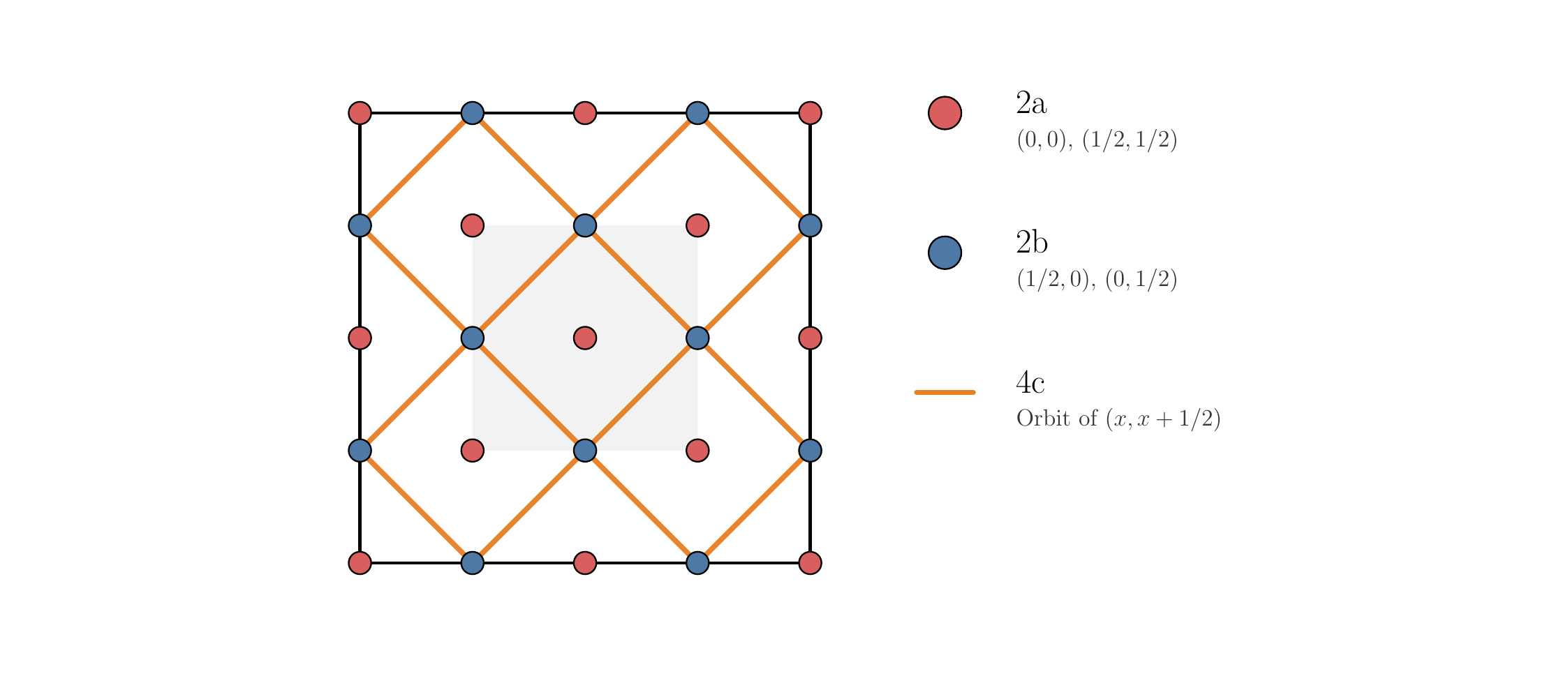}
    \caption{\textbf{Unit cell and Wyckoff positions for $p4g$.}
Red and blue markers identify the $2a$ and $2b$ families. All orange diagonal lines are solid members of the mirror orbit $4c$, related by wallpaper-group symmetries and translations. No member split of $4c$ is required in the table.}
    \label{fig:uc_p4g}
\SOneFigureEnd
\SOnePageEnd

\SOnePageStart
\SOneTableStart
\caption{\textbf{$p3$ (No.\ 13) Dark sets.}
Probe families are $1a=(0,0)$, $1b=(\tfrac13,\tfrac23)$, and $1c=(\tfrac23,\tfrac13)$, all with site symmetry $3$.
The general family $3d$ has site symmetry $1$.
The little co-group is $C_3$ at $\Gamma,K,K'$ and $C_1$ at the edge-centre momenta, which therefore give no nontrivial selection rule.
In the Cartesian axes used in Fig.~\ref{fig:uc_p3}, the primitive reciprocal vectors are $\bm b_1=(0,2\pi)$ and $\bm b_2=(\sqrt{3}\pi,\pi)$, so $K=\tfrac13(\bm b_1+\bm b_2)=(\pi/\sqrt{3},\pi)$ and $K'=-K$.}
\label{tab:p3_dark_sets}
\vspace{0.1cm}
\begin{ruledtabular}
\begin{tabular}{cll} \\[-3mm]
 & \textbf{Irrep} & \textbf{Dark set} \\[1.5mm] \hline \vspace{-0.2cm} \\
\multirow{3}{*}{\rotatebox[origin=c]{90}{$\Gamma$; $C_3$}}
& $A$ & $\varnothing$ \\
& $E_+$ & $\{1a,\,1b,\,1c\}$ \\
& $E_-$ & $\{1a,\,1b,\,1c\}$ \\[1.5mm] \hline \vspace{-0.2cm} \\
\multirow{3}{*}{\rotatebox[origin=c]{90}{$K$; $C_3$}}
& $A$ & $\{1b,\,1c\}$ \\
& $E_+$ & $\{1a,\,1b\}$ \\
& $E_-$ & $\{1a,\,1c\}$ \\[1.5mm] \hline \vspace{-0.2cm} \\
\multirow{3}{*}{\rotatebox[origin=c]{90}{$K'$; $C_3$}}
& $A$ & $\{1b,\,1c\}$ \\
& $E_+$ & $\{1a,\,1c\}$ \\
& $E_-$ & $\{1a,\,1b\}$ \\[+1.5mm]
\end{tabular}
\end{ruledtabular}
\SOneTableEnd

\SOneFigureStart
    \includegraphics[width=\linewidth]{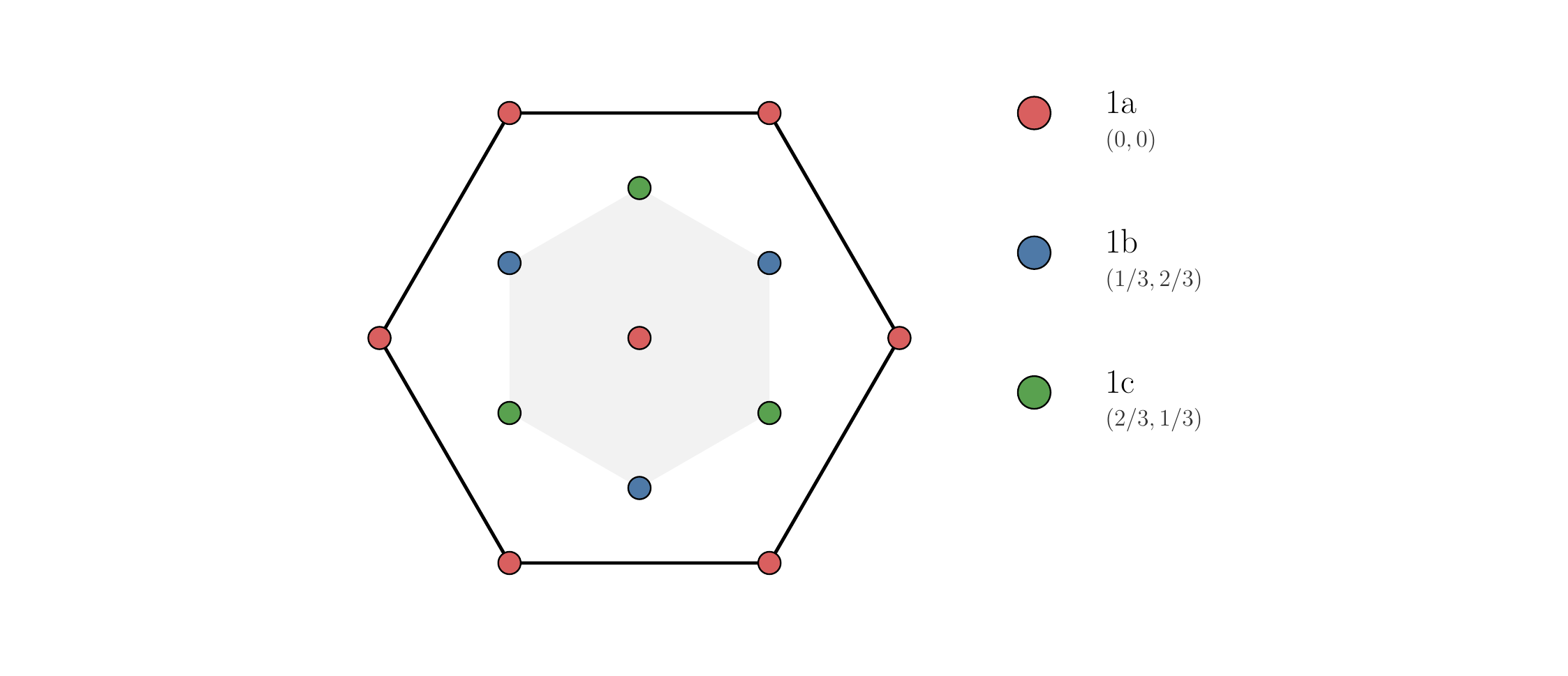}
    \caption{\textbf{Unit cell and Wyckoff positions for $p3$.}
The three colours distinguish the inequivalent $C_3$ centres $1a$, $1b$, and $1c$, with coordinates listed in the legend. In the Cartesian reciprocal convention associated with these axes, $\bm b_1=(0,2\pi)$ and $\bm b_2=(\sqrt{3}\pi,\pi)$, hence $K=(\pi/\sqrt{3},\pi)$ and $K'=-K$. The black hexagon bounds the periodic patch and the pale-grey hexagon is the central primitive Wigner--Seitz cell.}
    \label{fig:uc_p3}
\SOneFigureEnd
\SOnePageEnd

\SOnePageStart
\SOneTableStart
\caption{\textbf{$p3m1$ (No.\ 14) Dark sets.}
Probe families are $1a=(0,0)$, $1b=(\tfrac13,\tfrac23)$, and $1c=(\tfrac23,\tfrac13)$ with site symmetry $3m$, and $3d$ on the three mirror orientations with site symmetry $m$.
The general family $6e$ has site symmetry $1$.
In the Cartesian axes used in Fig.~\ref{fig:uc_p3m1}, take $\bm b_1=(0,2\pi)$ and $\bm b_2=(\sqrt{3}\pi,\pi)$, so $K=(\pi/\sqrt{3},\pi)$, $K'=-K$, and $M_1=\tfrac12\bm b_1=(0,\pi)$.
The little co-groups are $C_{3v}$ at $\Gamma$, $C_3$ at $K,K'$, and $C_s$ at $M_1$.
The vector $M_1$ points along the positive Cartesian $y$ direction, and so $3d_{\parallel}$ denotes the member on the vertical mirror $x=0$, parallel to $M_1$; the other two edge-centre momenta follow by $C_3$ rotation.}
\label{tab:p3m1_dark_sets}
\vspace{0.1cm}
\begin{ruledtabular}
\begin{tabular}{cll} \\[-3mm]
 & \textbf{Irrep} & \textbf{Dark set} \\[1.5mm] \hline \vspace{-0.2cm} \\
\multirow{3}{*}{\rotatebox[origin=c]{90}{$\Gamma$; $C_{3v}$}}
& $A_1$ & $\varnothing$ \\
& $A_2$ & $\{1a,\,1b,\,1c,\,3d\}$ \\
& $E$ & $\{1a,\,1b,\,1c\}$ \\[1.5mm] \hline \vspace{-0.2cm} \\
\multirow{3}{*}{\rotatebox[origin=c]{90}{$K$; $C_3$}}
& $A$ & $\{1b,\,1c\}$ \\
& $E_+$ & $\{1a,\,1b\}$ \\
& $E_-$ & $\{1a,\,1c\}$ \\[1.5mm] \hline \vspace{-0.2cm} \\
\multirow{3}{*}{\rotatebox[origin=c]{90}{$K'$; $C_3$}}
& $A$ & $\{1b,\,1c\}$ \\
& $E_+$ & $\{1a,\,1c\}$ \\
& $E_-$ & $\{1a,\,1b\}$ \\[1.5mm] \hline \vspace{-0.2cm} \\
\multirow{2}{*}{\rotatebox[origin=c]{90}{$M_1$; $C_s$}}
& $A'$ & $\varnothing$ \\
& $A''$ & $\{1a,\,1b,\,1c,\,3d_{\parallel}\}$ \\[+1.5mm]
\end{tabular}
\end{ruledtabular}
\SOneTableEnd

\SOneFigureStart
    \includegraphics[width=\linewidth]{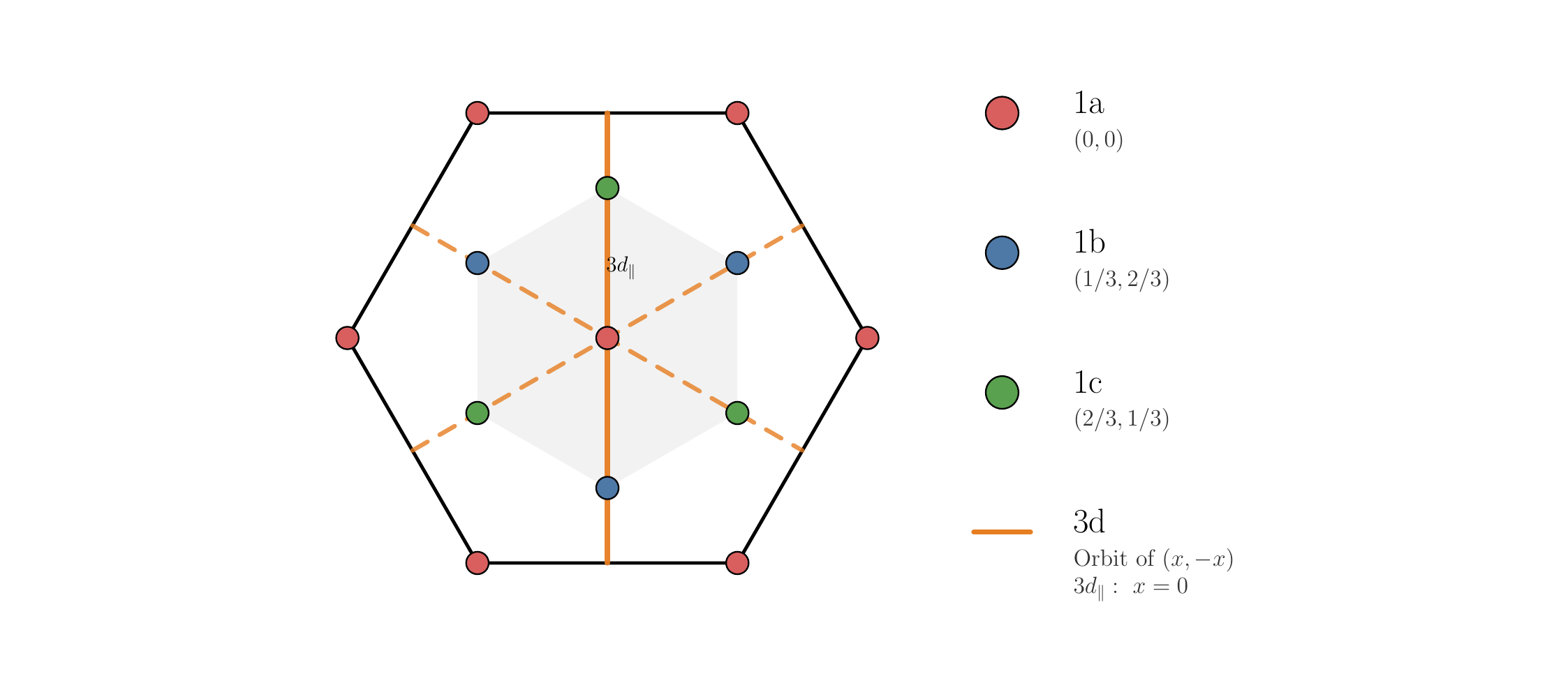}
    \caption{\textbf{Unit cell and Wyckoff positions for $p3m1$.}
Orange lines show the three mirror orientations of the $3d$ orbit. In the Cartesian reciprocal convention $\bm b_1=(0,2\pi)$ and $\bm b_2=(\sqrt{3}\pi,\pi)$, the selected edge-centre momentum is $M_1=\tfrac12\bm b_1=(0,\pi)$ and therefore points vertically. The solid mirror $x=0$ is the member $3d_{\parallel}$, parallel to $M_1$; the two dashed mirrors are the other $C_3$-related members. At $\Gamma$ the table uses the unsplit family $3d$, and the other two edge-centre momenta follow by $C_3$ rotation.}
    \label{fig:uc_p3m1}
\SOneFigureEnd
\SOnePageEnd

\SOnePageStart
\SOneTableStart
\caption{\textbf{$p31m$ (No.\ 15) Dark sets.}
Probe families are $1a=(0,0)$ with site symmetry $3m$, $2b=\{(\tfrac13,\tfrac23),(\tfrac23,\tfrac13)\}$ with site symmetry $3$, and the mirror family $3c$ with site symmetry $m$.
The general family $6d$ has site symmetry $1$.
In the Cartesian axes used in Fig.~\ref{fig:uc_p31m}, take $\bm b_1=(0,2\pi)$ and $\bm b_2=(\sqrt{3}\pi,\pi)$, so $K=(\pi/\sqrt{3},\pi)$, $K'=-K$, and $M_1=\tfrac12\bm b_1=(0,\pi)$.
The little co-groups are $C_{3v}$ at $\Gamma,K,K'$ and $C_s$ at $M_1$.
Because $M_1$ points along the positive Cartesian $y$ direction, $3c_{\perp}$ denotes the member on the horizontal mirror $y=0$, perpendicular to $M_1$; the other two edge-centre momenta follow by $C_3$ rotation.}
\label{tab:p31m_dark_sets}
\vspace{0.1cm}
\begin{ruledtabular}
\begin{tabular}{cll} \\[-3mm]
 & \textbf{Irrep} & \textbf{Dark set} \\[1.5mm] \hline \vspace{-0.2cm} \\
\multirow{3}{*}{\rotatebox[origin=c]{90}{$\Gamma$; $C_{3v}$}}
& $A_1$ & $\varnothing$ \\
& $A_2$ & $\{1a,\,3c\}$ \\
& $E$ & $\{1a,\,2b\}$ \\[1.5mm] \hline \vspace{-0.2cm} \\
\multirow{3}{*}{\rotatebox[origin=c]{90}{$K$; $C_{3v}$}}
& $A_1$ & $\{2b\}$ \\
& $A_2$ & $\{1a,\,2b,\,3c\}$ \\
& $E$ & $\{1a\}$ \\[1.5mm] \hline \vspace{-0.2cm} \\
\multirow{3}{*}{\rotatebox[origin=c]{90}{$K'$; $C_{3v}$}}
& $A_1$ & $\{2b\}$ \\
& $A_2$ & $\{1a,\,2b,\,3c\}$ \\
& $E$ & $\{1a\}$ \\[1.5mm] \hline \vspace{-0.2cm} \\
\multirow{2}{*}{\rotatebox[origin=c]{90}{$M_1$; $C_s$}}
& $A'$ & $\varnothing$ \\
& $A''$ & $\{1a,\,3c_{\perp}\}$ \\[+1.5mm]
\end{tabular}
\end{ruledtabular}
\SOneTableEnd

\SOneFigureStart
    \includegraphics[width=\linewidth]{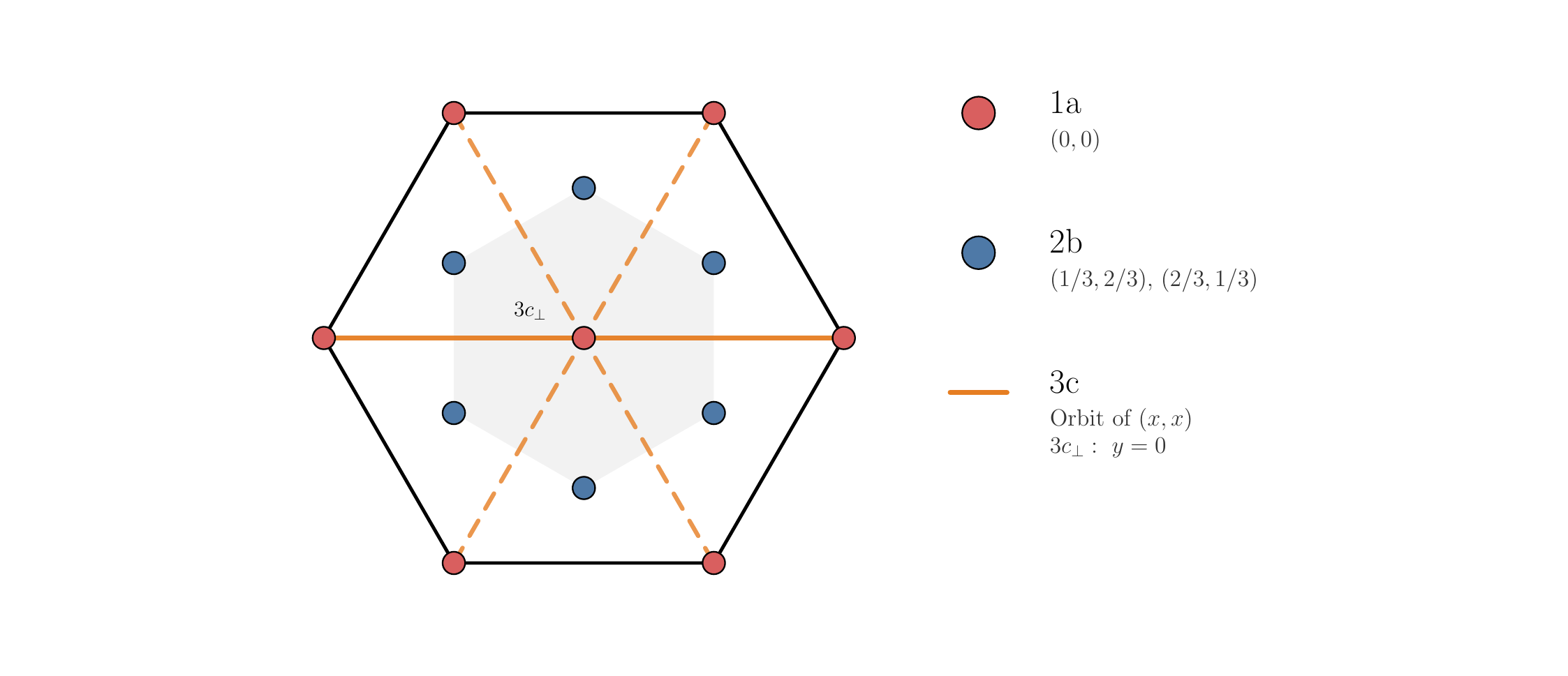}
    \caption{\textbf{Unit cell and Wyckoff positions for $p31m$.}
Orange lines show the three mirror orientations of the $3c$ orbit. In the Cartesian reciprocal convention $\bm b_1=(0,2\pi)$ and $\bm b_2=(\sqrt{3}\pi,\pi)$, the selected edge-centre momentum is $M_1=\tfrac12\bm b_1=(0,\pi)$ and therefore points vertically. The solid mirror $y=0$ is the member $3c_{\perp}$, perpendicular to $M_1$; the two dashed mirrors are the other $C_3$-related members. At $\Gamma,K,K'$ the table uses the unsplit family $3c$.}
    \label{fig:uc_p31m}
\SOneFigureEnd
\SOnePageEnd

\SOnePageStart
\SOneTableStart
\caption{\textbf{$p6$ (No.\ 16) Dark sets.}
Probe families are $1a=(0,0)$ with site symmetry $6$, $2b=\{(\tfrac13,\tfrac23),(\tfrac23,\tfrac13)\}$ with site symmetry $3$, and $3c=\{(\tfrac12,0),(0,\tfrac12),(\tfrac12,\tfrac12)\}$ with site symmetry $2$.
The general family $6d$ has site symmetry $1$.
In the Cartesian axes used in Fig.~\ref{fig:uc_p6}, take $\bm b_1=(0,2\pi)$ and $\bm b_2=(\sqrt{3}\pi,\pi)$, so $K=(\pi/\sqrt{3},\pi)$, $K'=-K$, and $M_1=\tfrac12\bm b_1=(0,\pi)$; $M_1$ points vertically and the other two edge-centre momenta follow by $C_3$ rotation.
At $K,K'$ the two $2b$ members are denoted $2b^{(1)}=(\tfrac13,\tfrac23)$ and $2b^{(2)}=(\tfrac23,\tfrac13)$.
At $M_1$ we write $3c_x=(\tfrac12,0)$, $3c_y=(0,\tfrac12)$, and $3c_d=(\tfrac12,\tfrac12)$.}
\label{tab:p6_dark_sets}
\vspace{0.1cm}
\begin{ruledtabular}
\begin{tabular}{cll} \\[-3mm]
 & \textbf{Irrep} & \textbf{Dark set} \\[1.5mm] \hline \vspace{-0.2cm} \\
\multirow{6}{*}{\rotatebox[origin=c]{90}{$\Gamma$; $C_6$}}
& $A$ & $\varnothing$ \\
& $B$ & $\{1a,\,3c\}$ \\
& $E_1^{+}$ & $\{1a,\,2b,\,3c\}$ \\
& $E_1^{-}$ & $\{1a,\,2b,\,3c\}$ \\
& $E_2^{+}$ & $\{1a,\,2b\}$ \\
& $E_2^{-}$ & $\{1a,\,2b\}$ \\[1.5mm] \hline \vspace{-0.2cm} \\
\multirow{3}{*}{\rotatebox[origin=c]{90}{$K$; $C_3$}}
& $A$ & $\{2b\}$ \\
& $E_{+}$ & $\{1a,\,2b^{(1)}\}$ \\
& $E_{-}$ & $\{1a,\,2b^{(2)}\}$ \\[1.5mm] \hline \vspace{-0.2cm} \\
\multirow{3}{*}{\rotatebox[origin=c]{90}{$K'$; $C_3$}}
& $A$ & $\{2b\}$ \\
& $E_{+}$ & $\{1a,\,2b^{(2)}\}$ \\
& $E_{-}$ & $\{1a,\,2b^{(1)}\}$ \\[1.5mm] \hline \vspace{-0.2cm} \\
\multirow{2}{*}{\rotatebox[origin=c]{90}{$M_1$; $C_2$}}
& $A$ & $\{3c_x,\,3c_d\}$ \\
& $B$ & $\{1a,\,3c_y\}$ \\[+1.5mm]
\end{tabular}
\end{ruledtabular}
\SOneTableEnd

\SOneFigureStart
    \includegraphics[width=\linewidth]{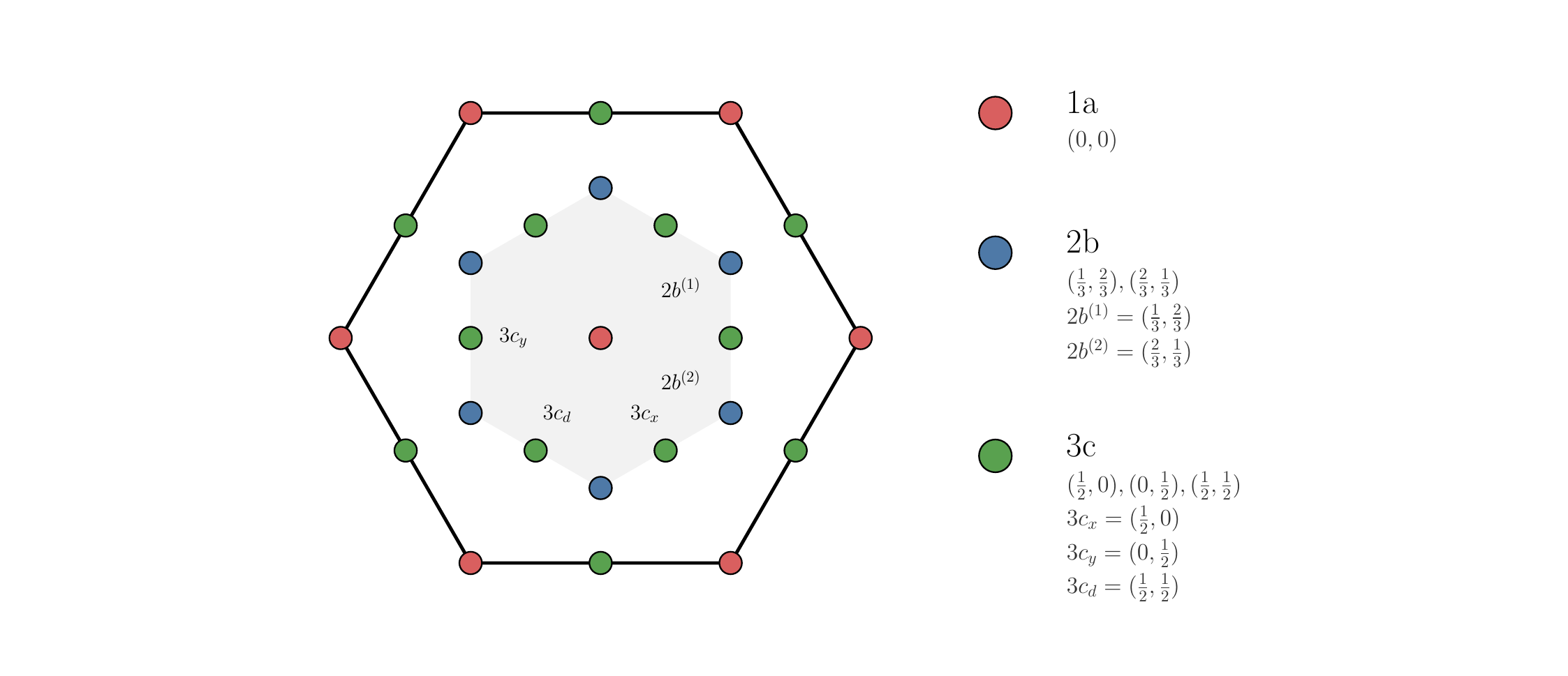}
    \caption{\textbf{Unit cell and Wyckoff positions for $p6$.}
The legend gives the complete $1a$, $2b$, and $3c$ families. In the Cartesian reciprocal convention $\bm b_1=(0,2\pi)$ and $\bm b_2=(\sqrt{3}\pi,\pi)$, one has $K=(\pi/\sqrt{3},\pi)$, $K'=-K$, and $M_1=(0,\pi)$, with $M_1$ directed vertically. Nearby labels identify $2b^{(1)}$ and $2b^{(2)}$, used separately at $K,K'$, and $3c_x$, $3c_y$, and $3c_d$, used at $M_1$. Complete-family labels are retained whenever all members have the same dark status; the other two edge-centre momenta follow by $C_3$ rotation.}
    \label{fig:uc_p6}
\SOneFigureEnd
\SOnePageEnd

\SOnePageStart
\SOneTableStart
\caption{\textbf{$p6m$ (No.\ 17, $p6mm$) Dark sets.}
The standard Wyckoff families are $1a=(0,0)$ with site symmetry $6mm$, $2b=\{(\tfrac13,\tfrac23),(\tfrac23,\tfrac13)\}$ with site symmetry $3m$, $3c=\{(\tfrac12,0),(0,\tfrac12),(\tfrac12,\tfrac12)\}$ with site symmetry $2mm$, and the two mirror families $6d$ and $6e$; the general family is $12f$.
In the Cartesian axes used in Fig.~\ref{fig:uc_p6m}, take $\bm b_1=(0,2\pi)$ and $\bm b_2=(\sqrt{3}\pi,\pi)$, so $K=(\pi/\sqrt{3},\pi)$, $K'=-K$, and $M_1=\tfrac12\bm b_1=(0,\pi)$.
At $M_1$ we write $3c_x=(\tfrac12,0)$, $3c_y=(0,\tfrac12)$, and $3c_d=(\tfrac12,\tfrac12)$.
The vector $M_1$ points vertically, and so the two mirror-orbit subsets fixed by the $M_1$ little group are denoted $6d_{\perp}$ for the horizontal mirror pair and $6e_{\parallel}$ for the vertical mirror pair; in fractional real-space coordinates they may be represented as $6d_{\perp}=\{(0,x),(0,-x)\}$ and $6e_{\parallel}=\{(2x,x),(-2x,-x)\}$. The other two edge-centre momenta follow by $C_3$ rotation.
We choose $B_1$ odd on the $6d$ mirrors and $B_2$ odd on the $6e$ mirrors.}
\label{tab:p6m_dark_sets}
\vspace{0.1cm}
\begin{ruledtabular}
\begin{tabular}{cll} \\[-3mm]
 & \textbf{Irrep} & \textbf{Dark set} \\[1.5mm] \hline \vspace{-0.2cm} \\
\multirow{6}{*}{\rotatebox[origin=c]{90}{$\Gamma$; $C_{6v}$}}
& $A_1$ & $\varnothing$ \\
& $A_2$ & $\{1a,\,2b,\,3c,\,6d,\,6e\}$ \\
& $B_1$ & $\{1a,\,3c,\,6d\}$ \\
& $B_2$ & $\{1a,\,2b,\,3c,\,6e\}$ \\
& $E_1$ & $\{1a,\,2b,\,3c\}$ \\
& $E_2$ & $\{1a,\,2b\}$ \\[1.5mm] \hline \vspace{-0.2cm} \\
\multirow{3}{*}{\rotatebox[origin=c]{90}{$K$; $C_{3v}$}}
& $A_1$ & $\{2b\}$ \\
& $A_2$ & $\{1a,\,2b,\,3c,\,6d\}$ \\
& $E$ & $\{1a\}$ \\[1.5mm] \hline \vspace{-0.2cm} \\
\multirow{3}{*}{\rotatebox[origin=c]{90}{$K'$; $C_{3v}$}}
& $A_1$ & $\{2b\}$ \\
& $A_2$ & $\{1a,\,2b,\,3c,\,6d\}$ \\
& $E$ & $\{1a\}$ \\[1.5mm] \hline \vspace{-0.2cm} \\
\multirow{4}{*}{\rotatebox[origin=c]{90}{$M_1$; $C_{2v}$}}
& $A_1$ & $\{3c_x,\,3c_d\}$ \\
& $A_2$ & $\{1a,\,2b,\,3c,\,6d_{\perp},\,6e_{\parallel}\}$ \\
& $B_1$ & $\{1a,\,3c_y,\,6d_{\perp}\}$ \\
& $B_2$ & $\{1a,\,2b,\,3c_y,\,6e_{\parallel}\}$ \\[+1.5mm]
\end{tabular}
\end{ruledtabular}
\SOneTableEnd

\SOneFigureStart
    \includegraphics[width=\linewidth]{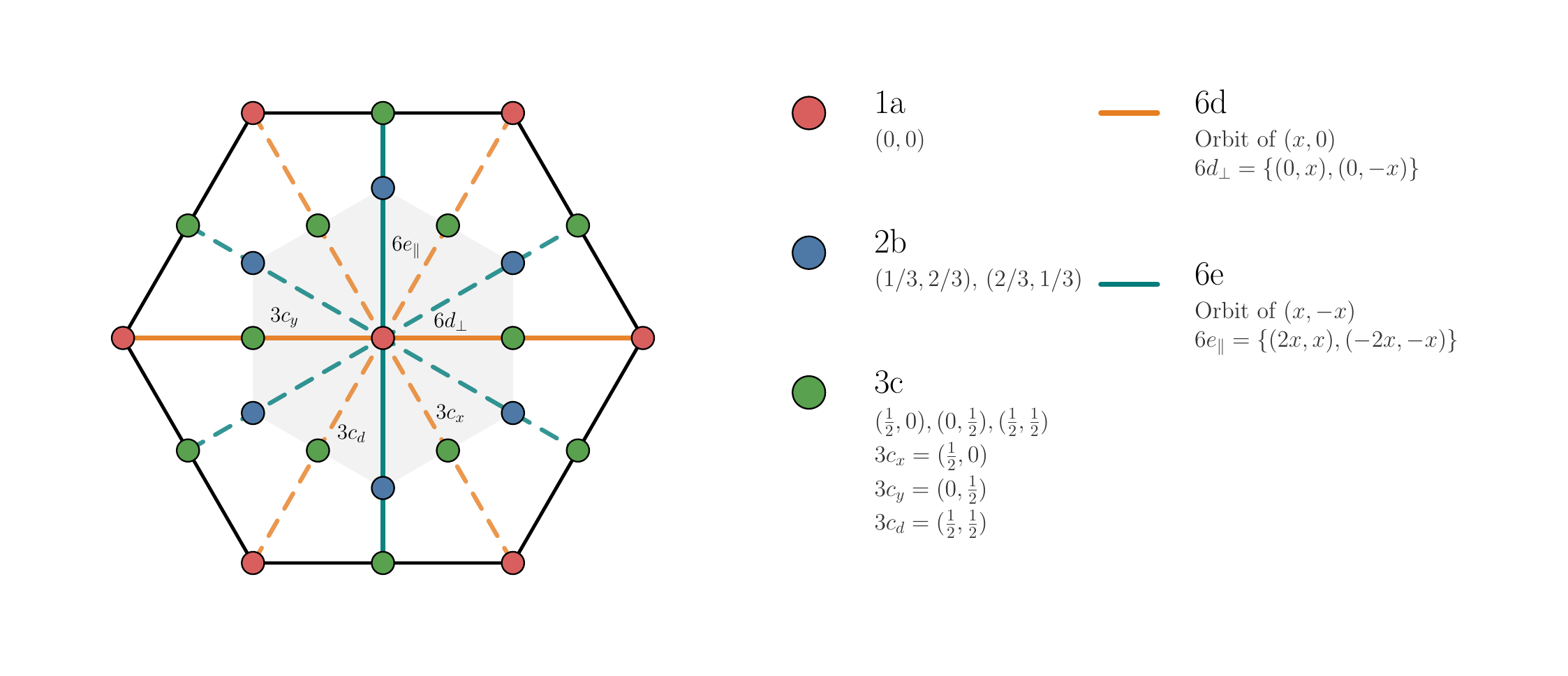}
    \caption{\textbf{Unit cell and Wyckoff positions for $p6m$.}
In the Cartesian reciprocal convention $\bm b_1=(0,2\pi)$ and $\bm b_2=(\sqrt{3}\pi,\pi)$, the selected edge-centre momentum is $M_1=\tfrac12\bm b_1=(0,\pi)$. Marker labels distinguish $3c_x$, $3c_y$, and $3c_d$ for $M_1$; the $2b$ family remains unsplit. The solid horizontal orange mirror is $6d_{\perp}$ and the solid vertical teal mirror is $6e_{\parallel}$. Dashed orange and teal lines are the remaining $C_6$-related members of the $6d$ and $6e$ orbits.}
    \label{fig:uc_p6m}
\SOneFigureEnd
\SOnePageEnd

\section{Gauge invariance of the dark set}
\label{gauge}

 The band representation of a wave function depends on the definition of the symmetries: for instance, whether threefold rotation is defined relative to the origin or relative to another threefold symmetric point in the unit cell. Consequently, one is naturally led to ask the question: are the symmetry-enforced zeros of the wave function gauge invariant? This is particularly important for comparing with scanning tunnelling microscopy which images the charge density --- a gauge-invariant observable.

In this section we show that upon a redefinition of the symmetries, while the irrep of the wave function changes, the physical position of these zeros does not. That is to say, the locations of the symmetry-enforced zeros in a new coordinate system are related to the locations of the zeros in the old coordinate system by the associated coordinate transformation.

\subsection{Proof of gauge invariance}

Using the same notation as in the main text, we let $h=\{R\mid \bm t_h\}$ be a space-group operation, acting on real-space coordinates as $\bm r \mapsto R\bm r+\bm t_h$. Let $\bm k_*$ be a high-symmetry momentum and let $\bm r_*$ be a high-symmetry point in the unit cell. The common stabiliser is $H_{\bm k_*,\bm r_*} = \{h\in G:\ R\bm k_*=\bm k_*+\bm G_h, \,\, R\bm r_*+\bm t_h=\bm r_*\}$. The symmetry action on the Bloch wavefunctions is
\begin{align}
    U_h\Psi_{\bm k_*,a}
    =
    e^{-i\bm k_*\cdot\bm t_h}
    \sum_b [D_{\rho,\bm k_*}(h)]_{ba}\Psi_{\bm k_*,b}.
\end{align}
Since $h\in G_{\bm k_*}$ implies $R\bm k_*=\bm k_*+\bm G_h$, the factors $e^{-i\bm k_*\cdot\bm t_h}$ multiply consistently under the stabiliser group law, the residual reciprocal-lattice phases being equal to unity. More precisely, from the composition rule $\bm t_{hh'}=\bm t_h+R_h\bm t_{h'}$ the assigned phases obey $e^{-i\bm k_*\cdot\bm t_h}e^{-i\bm k_*\cdot\bm t_{h'}}= e^{-i(R_h^{-1}\bm G_h)\cdot\bm t_{h'}}\,e^{-i\bm k_*\cdot\bm t_{hh'}}$.
The allowed local values are obtained by taking the trace of the projector
\begin{align}
    P_{\bm k_*,\bm r_*}
    =
    \tfrac{1}{|H_{\bm k_*,\bm r_*}|}
    \sum_{h\in H_{\bm k_*,\bm r_*}}
    e^{-i\bm k_*\cdot\bm t_h}D_{\rho,\bm k_*}(h).
\end{align}
That is, the selection rule is $m_{\bm k_*,\bm r_*} = \operatorname{Tr}P_{\bm k_*,\bm r_*} \neq 0$.

Now shift the coordinate origin by $\bm\delta$, so that $\bm r'=\bm r-\bm\delta $. The same physical operation $h$ is written in the new coordinates as $h'=\{R\mid \bm t'_h\}$, with $\bm t'_h=\bm t_h+(R-1)\bm\delta$. Indeed,
\begin{align}
    \bm r'
    \mapsto
    R\bm r'+\bm t_h+(R-1)\bm\delta.
\end{align}
If $h$ fixes $\bm r_*$ in the old coordinates, then $h'$ fixes $\bm r'_*= \bm r_*-\bm\delta$ in the new coordinates, since $R\bm r'_*+\bm t'_h = R\bm r_*+\bm t_h-\bm\delta = \bm r_*-\bm\delta = \bm r'_*$. Hence the stabilisers $H_{\bm k_*,\bm r_*}$ and $H'_{\bm k_*,\bm r'_*}$ are the same physical group, written in two coordinate systems. That is, for a site-symmetry element of a symmorphic group the Seitz translation $\bm t_{h'}=(1-R_{h'})\bm r_*$ is a Bravais lattice vector, so the residual phase is unity and $h\mapsto e^{-i\bm k_*\cdot\bm t_h}D_{\rho,\bm k_*}(h)$ is a representation of $H_{\bm k_*,\bm r_*}$.

Now, the matrix representing the full unitary action of the symmetry changes under a Bloch-basis gauge transformation $\Psi'_{\bm k_*}=W_{\bm k_*}^{-1}\Psi_{\bm k_*}$ through
\begin{align}
    e^{-i\bm k_*\cdot\bm t'_h}D'_{\rho,\bm k_*}(h')
    =
    W_{\bm k_*}^{-1}
    e^{-i\bm k_*\cdot\bm t_h}D_{\rho,\bm k_*}(h)
    W_{\bm k_*}.
\end{align}
Equivalently,
\begin{align}
    D'_{\rho,\bm k_*}(h')
    =
    e^{i\bm k_*\cdot(\bm t'_h-\bm t_h)}
    W_{\bm k_*}^{-1}D_{\rho,\bm k_*}(h)W_{\bm k_*}.
\end{align}
The transformed local projector is therefore
\begin{align}
    P'_{\bm k_*,\bm r'_*}
    =
    W_{\bm k_*}^{-1}
    P_{\bm k_*,\bm r_*}
    W_{\bm k_*}.
\end{align}
Hence $m'_{\bm k_*,\bm r'_*}
    =
    \operatorname{Tr}P'_{\bm k_*,\bm r'_*}
    =
    \operatorname{Tr}P_{\bm k_*,\bm r_*}
    =
    m_{\bm k_*,\bm r_*} $. In particular,
\begin{align}
    m_{\bm k_*,\bm r_*}=0
    \quad\Longleftrightarrow\quad
    m'_{\bm k_*,\bm r'_*}=0.
\end{align}
Thus symmetry-enforced zeros are invariant under a passive change of origin. The same argument covers embedding-gauge changes in tight-binding descriptions, as such a change is a unitary transformation $W_{\bm k}$ of the Bloch basis; since the projector changes by conjugation, its rank and trace do not change. Similarly, global reorientation of the rotation axes acts on the operations $\{R\mid\bm t_h\}$ by conjugation with a fixed point-group element and on the Bloch basis by a $\bm k$-independent unitary; the projector again transforms by conjugation, so its trace, and with it the dark set, is unchanged.

\subsection{Worked example: shifting the origin from $1a$ to $1c$ in 1H-TMDs}

Here we consider the example of changing the origin in the 1H-TMD example from the $1a$ position to the $1c$ position, noting that both sites have the exact same triangular symmetry. The selection rules for the origin at $1a$ are discussed in the main text; now we shift the origin by $\bm\delta=\bm r_{\text{1c}}$. For the $C_3$ generator and its square, the old Seitz translations are $\bm t_{\bm r_*,n}$, while the new translations after shifting the origin to $1c$ are (Table~\ref{tab:origin-shift-translations})
\begin{align}
    \bm t'_{\bm r_*,n}
    =
    \bm t_{\bm r_*,n}-\bm t_{\text{1c},n},
    \qquad
    n=1,2.
\end{align}

\begin{table}[t!]
\centering
\caption{{Rotation-centre data for the 1H-TMD triangular lattice in two coordinate systems.} We define the positions $\bm{r}_*$ with the origin at the 1a position. The columns $\bm t_{C_3}$ and $\bm t_{C_3^2}$ are the Seitz translations of $C_3$ and $C_3^2$ when the rotation axis is at $1a$. The columns $\bm t'_{C_3}$ and $\bm t'_{C_3^2}$ give the Seitz translations after shifting the rotation axis to the hollow site $1c$.}
\vspace{0.2cm}
\renewcommand{\arraystretch}{1.4}
\setlength{\tabcolsep}{5pt}
\begin{tabular}{cccccc}
\hline\hline
\textbf{Site}
& \textbf{$\bm r_*$}
& \textbf{$\bm t_{C_3}$}
& \textbf{$\bm t_{C_3^2}$}
& \textbf{$\bm t'_{C_3}$}
& \textbf{$\bm t'_{C_3^2}$} \\
\hline
$1a$
& $0$
& $0$
& $0$
& $-\bm R_1$
& $-\bm R_1-\bm R_2$ \\[1pt]

$1b$
& $\tfrac{1}{3}\bm R_1+\tfrac{2}{3}\bm R_2$
& $\bm R_1+\bm R_2$
& $\bm R_2$
& $\bm R_2$
& $-\bm R_1$ \\[1pt]

$1c$
& $\tfrac{2}{3}\bm R_1+\tfrac{1}{3}\bm R_2$
& $\bm R_1$
& $\bm R_1+\bm R_2$
& $0$
& $0$ \\
\hline\hline
\end{tabular}
\label{tab:origin-shift-translations}
\end{table}

Since the new $C_3$ generator is the old hollow-centred rotation, its character is
\begin{align}
    \chi'_{\bm k}(C_3^n)
    =
    e^{-i\bm k\cdot\bm t_{\text{1c},n}}\chi_{\bm k}(C_3^n),
    \qquad
    n=1,2.
\end{align}
For the valence band at the $K$ point, we use $\chi_K(C_3)=\omega$, $\chi_K(C_3^2)=\omega^2$, $\omega=e^{2\pi i/3}$, and $K\cdot\bm R_1=K\cdot\bm R_2=\tfrac{2\pi}{3}$. Hence, the new characters are $\chi'_K(C_3)
    =
    e^{-iK\cdot\bm R_1}\omega
    =1$ and $\chi'_K(C_3^2)
    =
    e^{-iK\cdot(\bm R_1+\bm R_2)}\omega^2
    =1$.
So, in the $1c$-origin convention, the same valence band has a trivial $C_3$ character at $K$, consistent with the fact that the valence band can be represented by an $s$-like orbital on the hollow site. The new selection-rule index is
\begin{align}
    m'_{\bm k,\bm r_*}
    =
    \tfrac{1}{3}
    \sum_{n=0}^{2}
    e^{-i\bm k\cdot\bm t'_{\bm r_*,n}}\chi'_{\bm k}(C_3^n),
\end{align}
with $\bm t'_{\bm r_*,0}=0$ and $\chi'_{\bm k}(C_3^0)=1$. At $K$, because $\chi'_K(C_3)=\chi'_K(C_3^2)=1$, the shifted-origin columns of Table~\ref{tab:origin-shift-translations} give $m'_{K,\text{1a}}
    = 0$, $m'_{K,\text{1b}}
    =0$, and $m'_{K,\text{1c}}
    =1$, exactly as in the main text. The physical dark and bright sites are unchanged; what changed is the representation of the $C_3$ generator, and that change is exactly cancelled by the change in the Seitz translations entering the selection rule.

\newpage

\newpage
\section{STM simulations}
\label{sm:synthetic-stm}

 In this section, we describe how the simulated STM data are computed and details of the model Hamiltonians used.

\subsection{Calculating STM data from tight-binding models}

The differential tunnelling current imaged in STM is proportional to LDOS$(\text{eV},\bm{r})$, where eV is the bias voltage, and the local density of states (LDOS) is given by the expectation value of the density at fixed energy,
\begin{align}
    \text{LDOS}(E,\bm{r}) = - \tfrac{1}{\pi} \text{Im} \langle c_{\bm{r}}c^\dag_{\bm{r}}\rangle_E = -\tfrac{1}{\pi} \text{Im} G(E, \bm{r})
\end{align}
where $c^\dag_{\bm{r}}$ is the electron creation operator and $G(E,\bm{r})$ is the electron Green's function. To relate the continuum electronic density of states to the wavefunctions of a lattice model, we expand the continuum electron operator in terms of lattice creation operators associated to an orbital $\sigma$ and a lattice site $\bm{R}$, via $ c^\dag_{\bm{r}} = \sum_{\sigma\bm{R}} w_{\sigma\bm{R}}(\bm{r}) c^\dag_{\sigma\bm{R}}$ where $w_{\sigma\bm{R}}(\bm{r})$ are a set of real space orbitals indexed by $\sigma$ and centred at the lattice sites $\bm{R}$, e.g. \cite{Berthod2011TunnelingConductance, Choubey2014Visualization, Kreisel2016QuantitativeSTM, kreisel2021quasi, sobral2023machine, rhodes2024probing, nag2024pomeranchuk, ingham2025group, Holbrook2026}. One obtains
\begin{align}
    \text{LDOS}(E,\bm{r}) &= - \tfrac{1}{\pi} \text{Im} \sum_{\sigma\sigma',\bm{R}\bm{R}'} w^*_{\sigma\bm{R}}(\bm{r}) w_{\sigma'\bm{R}'}(\bm{r}) \langle c_{\sigma\bm{R}} c^\dag_{\sigma'\bm{R}'}\rangle_E \nonumber\\
  &= - \tfrac{1}{\pi} \text{Im} \sum_{\sigma\sigma',\bm{R}\bm{R}'} w^*_{\sigma\bm{R}}(\bm{r}) w_{\sigma'\bm{R}'}(\bm{r}) G(E, \bm{R}-\bm{R}')_{\sigma\sigma'}
\end{align}
where $G(E, \bm{R}-\bm{R}')_{\sigma\sigma'}$ is the Green's function of the lattice model,
\begin{gather}
\label{tb_green}
G(E, \bm{R}-\bm{R}')_{\sigma\sigma'} = \int \frac{d^d\bm{k}}{(2\pi)^d} \left(\frac{1}{E - \mathcal{H}(\bm{k}) + i0} \right)_{\sigma\sigma'} e^{i\bm{k}\cdot(\bm{R}-\bm{R}')}
\end{gather}
where $\mathcal{H}(\bm{k})_{\sigma\sigma'}$ is the tight-binding Hamiltonian, defined as a matrix acting in orbital basis.

\subsection{Numerical implementation}
\label{sm:stm_numerical_implementation}

For the simulated STM data, we evaluate a band-projected version of Eq.~\eqref{tb_green}. Let $\mathcal{B}$ denote the set of tight-binding bands included in the projected LDOS at the chosen probe energy $E$. For a finite momentum mesh $\mathcal{K}$, the lattice Green function used in the simulations is
\begin{align}
G_{\alpha\beta}(E,\bm{\Delta R})
=
\frac{1}{N_{\mathcal{K}}}
\sum_{\bm{k}\in \mathcal{K}}
\sum_{n\in\mathcal{B}}
\frac{
e^{i\bm{k}\cdot\bm{\Delta R}}\,
u_{\alpha n}(\bm{k})u^{*}_{\beta n}(\bm{k})
}{
E+i\eta-\varepsilon_n(\bm{k})
},
\end{align}
where $\alpha,\beta$ label the tight-binding orbitals, $\varepsilon_n(\bm{k})$ is the band energy, $u_{\alpha n}(\bm{k})$ is the corresponding normalised eigenvector, and $\eta$ is the Lorentzian broadening. In figures where the targeted momentum is related to symmetry-equivalent points by point-group rotations, the numerical quadrature is symmetrised by averaging over the corresponding rotated momenta. 

The continuum LDOS is then reconstructed from the lattice Green function by
\begin{align}
\operatorname{LDOS}(E,\bm{r})
=
-\frac{1}{\pi}
\operatorname{Im}
\sum_{\alpha\beta}
\sum_{\bm{R},\bm{R}'}
w^{*}_{\alpha}(\bm{r}-\bm{R}-\bm{\tau}_{\alpha})
w_{\beta}(\bm{r}-\bm{R}'-\bm{\tau}_{\beta})
G_{\alpha\beta}(E,\bm{R}-\bm{R}').
\end{align}

The vectors $\bm{\tau}_{\alpha}$ specify the intracell orbital positions. The real-space sums are evaluated on finite patches large enough that increasing the patch changes the plotted LDOS only below the visual resolution of the figures. All LDOS panels in Figs.~\ref{fig:tmd}, \ref{fig:haldane}, \ref{fig:bhz}, \ref{fig:generalised_haldane}, and \ref{fig:generalised_bhz} are normalised independently according to
\begin{align}
Z_{\mathrm{plot}}(\bm{r})
=
\frac{\operatorname{LDOS}(E,\bm{r})}
{\max_{\bm{r}\in\Omega}\operatorname{LDOS}(E,\bm{r})},
\end{align}
where $\Omega$ is the plotted real-space region.

\subsection{Three-orbital model of TMDs}
\label{sm:tmd}

Our calculations of the LDOS for TMDs employ the minimal three-band model of Ref. \cite{liu2013three}, comprised of $d_{z^2}$, $d_{xy}$, and $d_{x^2-y^2}$ orbitals located on the triangular lattice formed by the tungsten atoms. The model is given by
\begin{align}
\label{threebandH}
H^{\mathrm{NN}}(\boldsymbol{k})=\left[\begin{array}{ccc}
V_0 & V_1 & V_2 \\
V_1^* & V_{11} & V_{12} \\
V_2^* & V_{12}^* & V_{22}
\end{array}\right]
\end{align}
where
\begin{gather}
V_0 =\epsilon_1-\epsilon_0+2 t_0(2 \cos \alpha \cos \beta+\cos 2 \alpha)+2 r_0(2 \cos 3 \alpha \cos \beta+\cos 2 \beta)+2 u_0(2 \cos 2 \alpha \cos 2 \beta+\cos 4 \alpha) \\
\text{Re}\left[V_1\right] =-2 \sqrt{3} t_2 \sin \alpha \sin \beta+2(r_1+r_2) \sin 3 \alpha \sin \beta-2 \sqrt{3} u_2 \sin 2 \alpha \sin 2 \beta, \\
\text{Im}\left[V_1\right] =2 t_1 \sin \alpha(2 \cos \alpha+\cos \beta)+2(r_1-r_2) \sin 3 \alpha \cos \beta+2 u_1 \sin 2 \alpha(2 \cos 2 \alpha+\cos 2 \beta), \\
\text{Re}\left[V_2\right] =2 t_2(\cos 2 \alpha-\cos \alpha \cos \beta)-\tfrac{2}{\sqrt{3}}(r_1+r_2)(\cos 3 \alpha \cos \beta-\cos 2 \beta)+2 u_2(\cos 4 \alpha-\cos 2 \alpha \cos 2 \beta), \\
\text{Im}\left[V_2\right] =2 \sqrt{3} t_1 \cos \alpha \sin \beta+\tfrac{2}{\sqrt{3}}(r_1-r_2) \sin \beta(\cos 3 \alpha+2 \cos \beta)+2 \sqrt{3} u_1 \cos 2 \alpha \sin 2 \beta, \\
V_{11} =\epsilon_2-\epsilon_0+(t_{11}+3 t_{22}) \cos \alpha \cos \beta+2 t_{11} \cos 2 \alpha+4 r_{11} \cos 3 \alpha \cos \beta+2(r_{11}+\sqrt{3} r_{12}) \cos 2 \beta \\
 +(u_{11}+3 u_{22}) \cos 2 \alpha \cos 2 \beta+2 u_{11} \cos 4 \alpha, \\
\text{Re}\left[V_{12}\right] =\sqrt{3}(t_{22}-t_{11}) \sin \alpha \sin \beta+4 r_{12} \sin 3 \alpha \sin \beta+\sqrt{3}(u_{22}-u_{11}) \sin 2 \alpha \sin 2 \beta, \\
\text{Im}\left[V_{12}\right] =4 t_{12} \sin \alpha(\cos \alpha-\cos \beta)+4 u_{12} \sin 2 \alpha(\cos 2 \alpha-\cos 2 \beta), \\
V_{22} =\epsilon_2-\epsilon_0+(3 t_{11}+t_{22}) \cos \alpha \cos \beta+2 t_{22} \cos 2 \alpha+2 r_{11}(2 \cos 3 \alpha \cos \beta+\cos 2 \beta) \\
 +\tfrac{2}{\sqrt{3}} r_{12}(4 \cos 3 \alpha \cos \beta-\cos 2 \beta)+(3 u_{11}+u_{22}) \cos 2 \alpha \cos 2 \beta+2 u_{22} \cos 4 \alpha \\
(\alpha, \beta)=(k_x a/2, \sqrt{3}k_y a/2)
\end{gather}
We perform numerical simulations using parameters for WSe$_2$; when only nearest-neighbour hoppings are retained, the optimal fit to DFT for WSe$_2$ is given by $t_0=-0.146$, $t_1=-0.124$, $t_2=0.507$, $t_{11}=0.117$, $t_{12}=0.127$, $t_{22}=0.015$, $\epsilon_1=0.728$, $\epsilon_2=1.655$ (all units in eV). Our results incorporated next-nearest-neighbour hoppings $r_i$, $u_i$, (values from Ref. \cite{liu2013three}), but the STM simulations are essentially unchanged if only nearest-neighbour terms are used.

For the STM simulations, the continuum orbital envelopes used are
\begin{gather}
w_{z^2}(\bm{r}) = \frac{1}{4}\sqrt{\frac{5}{\pi}}\,g(\bm{r}), \qquad
w_{xy}(\bm{r}) = \frac{1}{2}\sqrt{\frac{15}{2\pi}}\,\frac{2xy}{|\bm{r}|}\,g(\bm{r}), \qquad
w_{x^2-y^2}(\bm{r}) = \frac{1}{2}\sqrt{\frac{15}{2\pi}}\,\frac{x^2-y^2}{|\bm{r}|}\,g(\bm{r}),  \\ 
g(\bm{r}) = \exp(-|\bm{r}|^2/0.2)
\end{gather}
At $\bm{r}=0$, the two in-plane components $w_{xy}$ and $w_{x^2-y^2}$ are set to zero by continuity of the plotted envelope.

The real-space sum is evaluated over the hexagonal patch
\begin{align}
\bm{R}=n\bm{R}_1+m\bm{R}_2,
\qquad
\max(|n|,|m|,|n+m|)\leq 4.
\end{align}
The plotted grid has $401\times 347$ points over the hexagonal region
\begin{align}
-1\leq x\leq 1,
\qquad
-\tfrac{\sqrt{3}}{2}\leq y\leq \tfrac{\sqrt{3}}{2}.
\end{align}
The colour scale is unit-normalised independently for each plot at a given high-symmetry momentum, e.g. $\Gamma$ and $K$.

\newpage
\subsubsection*{Wannier centre of the TMD valence band and the connected conduction doublet}

Let $\omega=e^{2\pi i/3}$. At the three $C_3$ centres $1a$, $1c$, and $1b$, the $C_3$ phases differ at $K$ and $K'$. Note that we use the high-symmetry-position labels of the layer group $p\bar{6}m2$ rather than of the 3D space group $P\bar{6}m2$ (No.~187), as we work in two dimensions; in the space-group notation, $(\tfrac{1}{3}, \tfrac{2}{3}, 0)$ is the $1e$ rather than the $1b$ position. A $d_{z^2}$ orbital transforms in the $A'$ irreducible representation of $D_{3h}$; such an orbital situated at the three $C_3$ centres induces the $C_3$ characters
\begin{align}
A'@1a &: \Gamma:1,\quad K:1,\quad K':1, \\
A'@1c &: \Gamma:1,\quad K:\omega,\quad K':\omega^2, \\
A'@1b &: \Gamma:1,\quad K:\omega^2,\quad K':\omega.
\end{align}
Likewise, an $E'$ doublet (the $d_{xy}/d_{x^2-y^2}$ orbitals) centred at these same positions gives
\begin{align}
E'@1a &: \Gamma:\omega^\pm,\quad K:\omega\oplus \omega^2,\quad K':\omega^2\oplus \omega, \\
E'@1c &: \Gamma:\omega^\pm,\quad K:1\oplus \omega^2,\quad K':1\oplus \omega, \\
E'@1b &: \Gamma:\omega^\pm,\quad K:1\oplus \omega,\quad K':1\oplus \omega^2.
\end{align}
For the isolated valence band of the three-band model one finds the $C_3$ characters
\begin{align}
\Gamma &: 1, &
K &: \omega, &
K' &: \omega^2.
\end{align}
This matches $A'@1c$. Thus, the valence-band Wannier centre is at the hollow-site position $1c$; since the atomic positions of the metals are at $1a$ and the chalcogens are at $1b$, this implies the valence band is described by an obstructed atomic limit. For the connected upper two-band subspace one instead finds
\begin{align}
\Gamma &: \omega^\pm, &
K &: 1\oplus \omega^2, &
K' &: 1\oplus \omega.
\end{align}
Among the multiplicity-one $C_3$-centred candidates, this matches $E'@1c$. Therefore, the connected conduction doublet is also centred at $1c$.

\subsubsection*{Unobstructed counterfactual}
\label{sm:tmd_unobstructed_counterfactual}

To highlight how the local density of states would appear for WSe$_2$ in the case of a hypothetical unobstructed band, we also consider a deliberately simplified counterfactual model. In the physical three-band model of WSe$_2$, the isolated valence band has representation data consistent with an $A'$ orbital centred at the hollow $1c$ position, while the connected upper two-band subspace has the symmetry data of an $E'$ doublet centred at $1c$. The counterfactual model instead realises an $E'$ doublet of $D_{3h}$, built from the same $(d_{xy},d_{x^2-y^2})$ orbital sector, while using the same real-space orbital envelopes and STM reconstruction as in the physical TMD calculation. The inter-orbital hybridisations in Eq.~\eqref{threebandH} are removed, the $d_{z^2}$ band is pushed away, and the doublet dispersion is chosen so that the $K$ valley is spectrally clean. Thus the model is a simplified unobstructed atomic doublet rather than the obstructed valence-band representation of WSe$_2$.

\begin{figure*}[t!]
    \centering
    \includegraphics[width=0.86\linewidth]{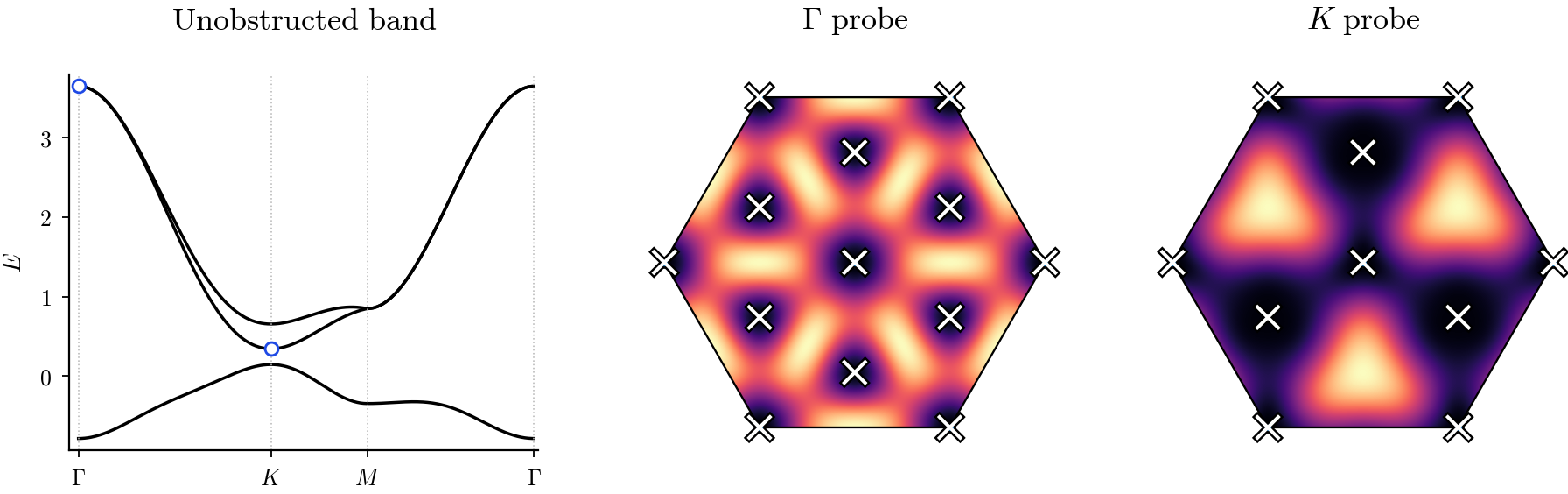}
    \caption{\textbf{Unobstructed hypothetical TMD model.} Bandstructure (left), as well as the LDOS at $\Gamma$ (middle) and $K$ (right) for a hypothetical model of TMDs in which the hoppings were purely intra orbital type, so that the bands are atomic in nature. The LDOS at $K$ resembles that of the actual TMD model, since the bright/dark spots are determined by the symmetry properties --- which remain the same. The difference manifests near the $\Gamma$ point, where the differing symmetry representation results in nodes at the 1a, 1b, and 1c positions, in contrast with the nodeless LDOS in Fig. \ref{fig:tmd}.}
    \label{fig:unobstructed_tmd}
\end{figure*}

Let $V_0(\bm{k})$ denote the $d_{z^2}$ diagonal entry of Eq.~\eqref{threebandH}. We define the counterfactual Hamiltonian in the same orbital basis $(d_{z^2},d_{xy},d_{x^2-y^2})$ by setting all inter-orbital matrix elements to zero and writing
\begin{align}
H_{\mathrm{cf}}(\bm{k})
=
\begin{pmatrix}
H_{z^2}^{\mathrm{cf}}(\bm{k}) & 0 \\
0 & H_{E}^{\mathrm{cf}}(\bm{k})
\end{pmatrix}.
\end{align}
The $d_{z^2}$ block is shifted down in energy,
\begin{align}
H_{z^2}^{\mathrm{cf}}(\bm{k})
=
(\epsilon_1-1.0)
+
\tfrac{1}{2}\left(V_0(\bm{k})-\epsilon_1\right),
\end{align}
where $\epsilon_1=0.728\,\mathrm{eV}$ is the onsite energy of the $d_{z^2}$ orbital in the three-band model. This places the $d_{z^2}$ band below the $d_{xy}/d_{x^2-y^2}$ sector, so that the latter can be imaged without spectral overlap.

The in-plane doublet is replaced by a simple triangular-lattice doublet dispersion,
\begin{align}
H_E^{\mathrm{cf}}(\bm{k})
=
\left(E_E+T_E f_1(\bm{k})\right)\mathbbm{1}_2
+
\lambda s_1(\bm{k})\tau_y,
\end{align}
where $\tau_y$ is a Pauli matrix acting in the real doublet basis $(d_{xy},d_{x^2-y^2})$. We take $E_E=1.55$ eV, $T_E=1.55$ eV, and $\lambda=0.18$ eV. The two scalar form factors are
\begin{align}
f_1(\bm{k})
&=
2\left[
\cos(\bm{k}\cdot\bm{R}_1)
+
\cos(\bm{k}\cdot\bm{R}_2)
+
\cos(\bm{k}\cdot(\bm{R}_1+\bm{R}_2))
\right],
\\
s_1(\bm{k})
&=
\tfrac{1}{3}\left[
\sin(\bm{k}\cdot\bm{R}_1)
+
\sin(\bm{k}\cdot\bm{R}_2)
-
\sin(\bm{k}\cdot(\bm{R}_1+\bm{R}_2))
\right].
\end{align}
The first form factor gives a maximum at $\Gamma$ and a minimum at $K/K'$, while the second is a $C_3$-invariant odd function of momentum. Consequently $s_1(-\bm{k})=-s_1(\bm{k})$, and since $\tau_y^*=-\tau_y$, the Hamiltonian obeys time-reversal symmetry,
\begin{align}
H_E^{\mathrm{cf}}(-\bm{k})=
H_E^{\mathrm{cf}}(\bm{k})^*
\end{align}
This term therefore splits the doublet at $K$ without breaking time reversal. By contrast, the $E$ doublet at $\Gamma$ remains degenerate because $s_1(\Gamma)=0$ and the little group at $\Gamma$ has a two-dimensional $E$ irrep.

The $\Gamma$ probe images the degenerate $E$ doublet at $E=3.650\,\mathrm{eV}$. The $K$ probe images only the lower split state of the doublet, at $E=0.344\,\mathrm{eV}$. The figure is generated from the same band-resolved Green-function LDOS procedure used in the other STM simulations. At $\Gamma$, the projected subspace contains both members of the degenerate $E$ doublet. At $K$, the projected subspace contains only the lower split $K$ state. 

Using the dark set formalism, the crossed positions in the $\Gamma$-probe panel are the symmetry-enforced zeros of the imaged $E$ doublet: within the plotted hexagonal cell they occur at the metal sites together with the centres of both the upward- and downward-pointing triangular plaquettes. For the lower split $K$ state, the crossed positions are the metal sites together with the centres of the downward-pointing triangular plaquettes. These are the high-symmetry positions at which the corresponding Bloch wavefunctions vanish identically by symmetry.

\subsection{Haldane model}
\label{sm:haldane_numerics}

For the Haldane simulations, we use the triangular Bravais lattice
\begin{align}
\bm{R}_1=(1,0),
\qquad
\bm{R}_2=\left(-\tfrac12,\tfrac{\sqrt{3}}{2}\right),
\end{align}
with reciprocal vectors defined by
$\bm{G}_i\cdot\bm{R}_j=2\pi\delta_{ij}$. The hexagon centre is the
$1a$ position, while the two honeycomb sublattices are embedded at
\begin{align}
\bm{\tau}_A
&=\tfrac13(\bm{R}_1+2\bm{R}_2)
\quad (1b),
&
\bm{\tau}_B
&=\tfrac13(2\bm{R}_1+\bm{R}_2)
\quad (1c).
\end{align}
Here $1b$ and $1c$ refer to the $p3$ labelling appropriate to the
generic model with both $M$ and $\phi$ nonzero. When the symmetry is
enhanced to $p6$, these two positions form the single Wyckoff orbit
$2b$.

We work in the sublattice basis
$\Psi_{\bm{k}}=(c_{A\bm{k}},c_{B\bm{k}})^{\mathsf T}$, with the
embedded Bloch convention
\begin{align}
c_{\mu\bm{k}}
=
N^{-1/2}
\sum_{\bm{R}}
e^{-i\bm{k}\cdot(\bm{R}+\bm{\tau}_\mu)}
c_{\mu\bm{R}},
\qquad
\mu=A,B.
\end{align}
The three nearest-neighbour vectors point from sublattice $A$ to
sublattice $B$,
\begin{align}
\{\bm{\alpha}_n\}_{n=1}^{3}
=
\{
\tfrac13(\bm{R}_1-\bm{R}_2),
-\tfrac13(2\bm{R}_1+\bm{R}_2),
\tfrac13(\bm{R}_1+2\bm{R}_2)
\},
\end{align}
and the oriented next-nearest-neighbour vectors are
\begin{align}
\{\bm{\beta}_n\}_{n=1}^{3}
=
\{\bm{R}_1,\bm{R}_2,-\bm{R}_1-\bm{R}_2\}.
\end{align}
The hoppings along $+\bm{\beta}_n$ carry phase $+\phi$ on sublattice
$A$ and phase $-\phi$ on sublattice $B$; the reverse hoppings carry
the complex-conjugate phases. Defining
\begin{align}
f_{\bm{k}}
=
\sum_{n=1}^{3}e^{i\bm{k}\cdot\bm{\alpha}_n}, \,\,\,\,\,\,\,\,\,\,\,\,\,\,\,\,\,\,
g_{\bm{k}}
=
\sum_{n=1}^{3}e^{i\bm{k}\cdot\bm{\beta}_n},
\end{align}
the Hamiltonian is
\begin{align}
H_{\mathrm{H}}(\bm{k})
=
\begin{pmatrix}
d_0(\bm{k})+d_z(\bm{k}) & -t_1f_{\bm{k}}\\
-t_1f_{\bm{k}}^{*} & d_0(\bm{k})-d_z(\bm{k})
\end{pmatrix},
\end{align}
where
\begin{align}
d_0(\bm{k})
=
2t_2\cos\phi\,\Re g_{\bm{k}},
\quad
\bm{d}(\bm{k})
=
\left(
-t_1\Re f_{\bm{k}},
\,t_1\Im f_{\bm{k}},
\,d_z(\bm{k})
\right),
\quad
d_z(\bm{k})
&=
M-2t_2\sin\phi\,\Im g_{\bm{k}}.
\end{align}

Noting that $\bm{K}=\tfrac13(\bm{G}_1+\bm{G}_2)$, $\bm{K}'=-\bm{K}$, then $f_{\bm{K}}=f_{\bm{K}'}=0$,
$g_{\bm{K}}=3e^{2\pi i/3}$, and
$g_{\bm{K}'}=3e^{-2\pi i/3}$. The Dirac masses $d_z(\pm\bm{K})$ are therefore
\begin{align}
m_K = M-3\sqrt{3}t_2\sin\phi,
\quad
m_{K'} = M+3\sqrt{3}t_2\sin\phi.
\end{align}
The Chern number is nonzero when $m_K$ and $m_{K'}$ have opposite
signs. For the occupied lower band,
\begin{align}
C
=
-\tfrac{1}{4\pi}
\int_{\mathrm{BZ}}d^2k\,
\hat{\bm{d}}\cdot
(
\partial_{k_x}\hat{\bm{d}}
\times
\partial_{k_y}\hat{\bm{d}}
).
\end{align}
For the parameter set for the topological regime quoted below this gives $C=-1$, in agreement with the inference from the sign of the mass terms.

The continuum orbital attached to each honeycomb site is an $s$-like
Gaussian,
\begin{align}
w_s(\bm{r})
=
\tfrac{1}{\sqrt{2\pi}\sigma_H}
\exp(-|\bm{r}|^2/(2\sigma_H^2)),
\qquad
\sigma_H=0.65.
\end{align}
The real-space patch used in the Haldane LDOS contraction contains
all cells satisfying
\begin{align}
\max(|n|,|m|,|n+m|)\leq 5,
\qquad
\bm{R}=n\bm{R}_1+m\bm{R}_2,
\end{align}
and the Green function is computed for displacements satisfying
\begin{align}
\max(|n|,|m|,|n+m|)\leq 7.
\end{align}

The LDOS is evaluated using the upper band with Lorentzian broadening $\eta=0.025$ and a $48\times48$ uniform Brillouin-zone mesh. The plotted real-space grid is $81\times81$ over $-3\leq x,y\leq 3$ before restricting to the displayed hexagonal region. The $K$-probe panel is evaluated at the valley energy; since $K$ and $K'$ are degenerate for both main-text parameter sets, the energy-resolved LDOS contains both valley contributions. In Fig.~\ref{fig:haldane}, we used the parameters $t_1 = 1.0$ and $t_2 = 0.15$ for both scenarios, setting $\phi = 0$, $M = 0.575$, $m_K = 0.575$, and $m_{K'} = 0.575$ for the trivial case ($C=0$), and $\phi = \pi/3$, $M = 0$, $m_K = -0.675$, and $m_{K'} = 0.675$ for the Chern case ($C=-1$ for the lower band and $C=1$ for the upper band).

\newpage
\subsection{BHZ model}
\label{sm:bhz_numerics}

For the BHZ simulations, we use a square Bravais lattice $\bm{R}_1=(1,0),
\bm{R}_2=(0,1)$ with one site per unit cell. For clarity, we explicitly state the high-symmetry momenta
\begin{align}
\Gamma=(0,0),
\qquad
X=(\pi,0),
\qquad
Y=(0,\pi),
\qquad
M=(\pi,\pi).
\end{align}
The tight-binding basis is
\begin{align}
(s_\uparrow,p_{+,\uparrow},s_\downarrow,p_{-,\downarrow}).
\end{align}
where $p_\pm = p_x\pm ip_y$ and in which the spin-conserving BHZ Hamiltonian is
\begin{align}
H_{\mathrm{BHZ}}(\bm{k})
=
\begin{pmatrix}
h_{\uparrow}(\bm{k}) & 0\\
0 & h_{\downarrow}(\bm{k})
\end{pmatrix}
+
D(\cos k_x+\cos k_y)\mathbbm{1}_4,
\end{align}
with
\begin{align}
h_{\uparrow}(\bm{k})
&=
\begin{pmatrix}
m_{\bm{k}} & A\sin k_x-iA\sin k_y\\
A\sin k_x+iA\sin k_y & -m_{\bm{k}}
\end{pmatrix},
\end{align}
where $m_{\bm{k}}=-(m+\cos k_x+\cos k_y)$ and for which $h_{\downarrow}(\bm{k})
=
h_{\uparrow}^{*}(-\bm{k})$. The continuum orbitals are taken to be
\begin{align}
w_s(\bm{r})
&=
\frac{1}{\sqrt{\pi}\sigma_B}
\exp(-|\bm{r}|^2/(2\sigma_B^2)),
\\
w_{p_+}(\bm{r})
&=
\frac{\sqrt{2}}{\sqrt{\pi}\sigma_B^2}
(x+iy)
\exp(-|\bm{r}|^2/(2\sigma_B^2)),
\\
w_{p_-}(\bm{r})
&=
\frac{\sqrt{2}}{\sqrt{\pi}\sigma_B^2}
(x-iy)
\exp(-|\bm{r}|^2/(2\sigma_B^2)),
\end{align}
with $\sigma_B=0.40$. The real-space patch used in the BHZ LDOS contraction contains the square-lattice cells
\begin{align}
\bm{R}=n\bm{R}_1+m\bm{R}_2,
\qquad
-4\leq n,m\leq 4.
\end{align}

The main-text BHZ figure uses the parameters $A = 1.0$ and $D = 0.2$ for both scenarios, setting $m = 2.5$ and $\nu = 0$ for the trivial case, and $m = 1.5$ and $\nu = 1$ for the QSH case. The probe energies are
\begin{align}
E_{\Gamma}^{\mathrm{triv}}=-4.1,
\qquad
E_{M}^{\mathrm{triv}}=-0.9,
\qquad
E_{\Gamma}^{\mathrm{QSH}}=-3.1,
\qquad
E_{M}^{\mathrm{QSH}}=-0.9.
\end{align}
The finite-energy LDOS uses $\eta=0.025$, a $48\times48$ uniform Brillouin-zone mesh, $C_4$-rotation averaging, and an $81\times81$ real-space grid over $-1\leq x,y\leq 1$. Each panel is independently normalised to its maximum inside the plotted square.

\newpage

\section{Dark set imaging with further symmetry breaking}
\label{sm:generalised}

\begin{figure*}[h!]
    \centering
    \includegraphics[width=0.82\linewidth]{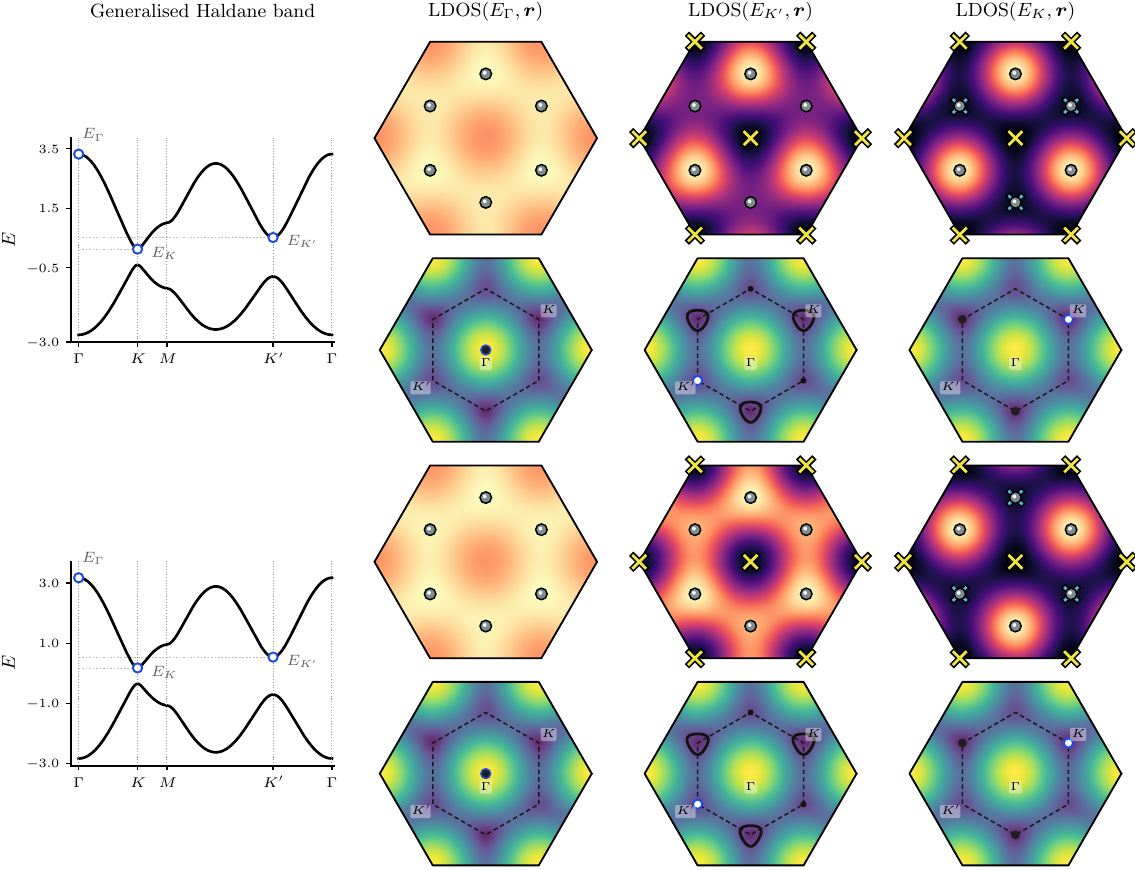}
    \caption{\textbf{Generalised Haldane model.} Top: $C=0$ case, Bottom: $|C|=1$ case. Left: cut bandstructure in each case with high-symmetry points marked with blue circles. Subsequent columns left-right: LDOS computed at the energies associated to high-symmetry momenta, with the associated Fermi contour underneath. The boundary of the first Brillouin zone is denoted by a dashed black line. The LDOS involves integrating over all states with a fixed energy; the associated Fermi contour is represented by a solid grey line. At the $K$ point, the contour is a set of small dots at the three equivalent $K$ points. At $K'$, the contour comprises small dots at the $K'$ corner and pockets encircling $K$. At the $\Gamma$ point the contour is a small dot at the centre of the Brillouin zone.}
    \label{fig:generalised_haldane}
\end{figure*}

In the generalised Haldane model in which both a chiral hopping and Semenoff mass are present, $K$ and $K'$ are no longer degenerate as they are in Fig. \ref{fig:haldane}. In Fig. \ref{fig:generalised_haldane} we show imaging of the charge density at the high-symmetry points in the generalised Haldane model. In the trivial case, as one lowers the bias energy to first image $K$ then $K'$, only one sublattice stays bright. By contrast, when the bias energy intersects $K'$ in the topological case, the other sublattice suddenly turns bright --- providing a sharp signature of the Chern number. In experiment, the valley-resolved LDOS can also be explored through strain-induced pseudomagnetic fields, which have been used to resolve the sublattice structure of pseudo-Landau levels in buckled graphene superlattices \cite{mao2020evidence}.

In this simulation, we used the parameters $t_1 = 1.0$ and probed the lowest band for both scenarios, setting $t_2 = 0.06$, $\phi = 0.675$, $M = 0.46$, $C = 0$, $m_K = 0.265$, and $m_{K'} = 0.655$ for the trivial case, and $t_2 = 0.09$, $\phi = 1.25$, $M = 0.18$, $C = -1$, $m_K = -0.264$, and $m_{K'} = 0.624$ for the Chern case.

\begin{figure*}[h!]
    \centering
    \includegraphics[width=0.9\linewidth]{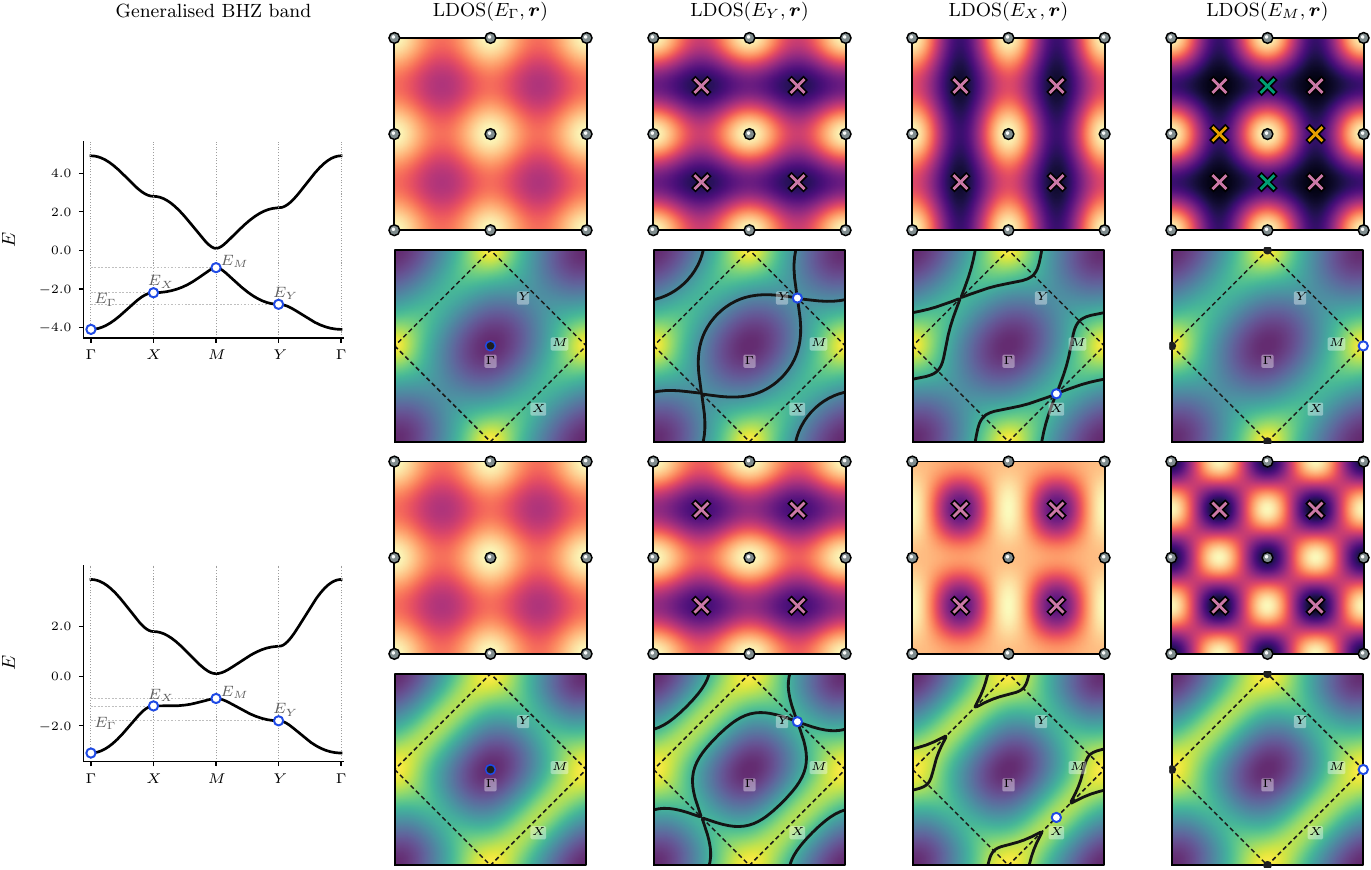}
    \caption{\textbf{Generalised BHZ model.} Top: $\nu=0$ case, Bottom: $\nu=1$ case. Left: cut bandstructure in each case with high-symmetry points marked with blue circles. Plotting convention is the same as in Fig. \ref{fig:generalised_haldane}. Subsequent columns from left-right: LDOS and associated energy contours. At the $M$ point, the contour is a set of small dots at the four equivalent $M$ points. At $X$/$Y$, the contours touch $X$/$Y$ but also extend along an arc in the interior of the Brillouin zone. Since the Fermi contour is not solely composed of the high-symmetry momentum, the LDOS does not strictly vanish at the indicated nodes, but since the density of states is dominated by the $X$/$Y$ points in each case, images are dominated by the bright/dark features at $X$/$Y$. }
    \label{fig:generalised_bhz}
\end{figure*}

Similarly, we consider a generalised BHZ model in which $C_4$ symmetry is broken to lift the degeneracy of $X$ and $Y$ present in Fig. \ref{fig:bhz}. We add the $C_4$-breaking scalar perturbation
\begin{align}
H_{\mathrm{gen}}(\bm{k})
=
H_{\mathrm{BHZ}}(\bm{k})
+
\delta(\cos k_x-\cos k_y)\mathbbm{1}_4.
\end{align}
This perturbation shifts the energies at $X$ and $Y$ without changing the high-symmetry eigenvectors, because it is proportional to the identity in orbital space. Fig.~\ref{fig:generalised_bhz} uses the scalar perturbation $\delta=-0.15$; the remaining parameters are the same as in the ordinary BHZ simulation described in the previous section.

In these examples, several of the high-symmetry points being imaged are not energetically isolated, yet crucially the qualitative bright/dark structure is dominated by the selection rules near the relevant high-symmetry momenta.

\newpage
\section{Momentum-resolved wavefunctions}
\label{sm:wavefunction}

    The STM simulations we have presented represent the LDOS, which involves averaging the wavefunction density across a constant energy contour in the Brillouin zone. In cases where this contour includes points away from a high-symmetry momentum, the dark set does not produce exact zeros of the LDOS. A more general and explicit visualisation of the dark set can be arrived at through a momentum-resolved plot of the charge density. In this section, we show $|\Psi_{\bm{k}_*}(\bm{r})|^2$ at all the high-symmetry momenta $\bm{k}_*$ for the models we have considered, allowing a momentum-resolved illustration of the dark sets. Specifically, at fixed high-symmetry crystal momentum $\bm{k}_\star$, we plot the probability density $\rho_{n\bm{k}_\star}(\bm r)$ obtained by reconstructing the Bloch eigenstate from its local-orbital amplitudes. We have
    \begin{align}
    \Psi_{n\bm{k}_\star}(\bm r)
    =\sum_{\bm R,\alpha}e^{i\bm{k}_\star\cdot\bm R}
     u_{n\alpha}(\bm{k}_\star)
     w_{\alpha}(\bm r-\bm R-\bm\tau_\alpha),\quad
    \rho_{n\bm{k}_\star}(\bm r)
    =|\Psi_{n\bm{k}_\star}(\bm r)|^2.
    \end{align}
    where $\bm R$ are the Bravais-lattice vectors, $\bm\tau_\alpha$ are the orbital positions, $u_{n\alpha}$ are the wavefunctions found by direct diagonalisation of the Bloch Hamiltonian. Crosses mark symmetry-enforced zeros.

\begin{figure*}[h!]
    \centering
    \includegraphics[width=\linewidth]{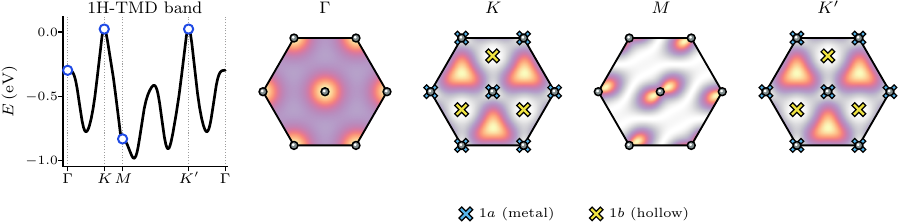}
    \caption{\textbf{1H-TMD Bloch wavefunctions at high-symmetry momenta.} Amplitude $|\Psi_{\bm{k}_*}(\bm{r})|^2$ of the 1H-TMD wavefunctions at the high-symmetry momenta $\bm{k}_*=\Gamma, K, M, K',$ with the associated energies indicated on the bandstructure.}
    \label{fig:hsp-tmds}
\end{figure*}

\begin{figure*}[h!]
    \centering
    \includegraphics[width=\linewidth]{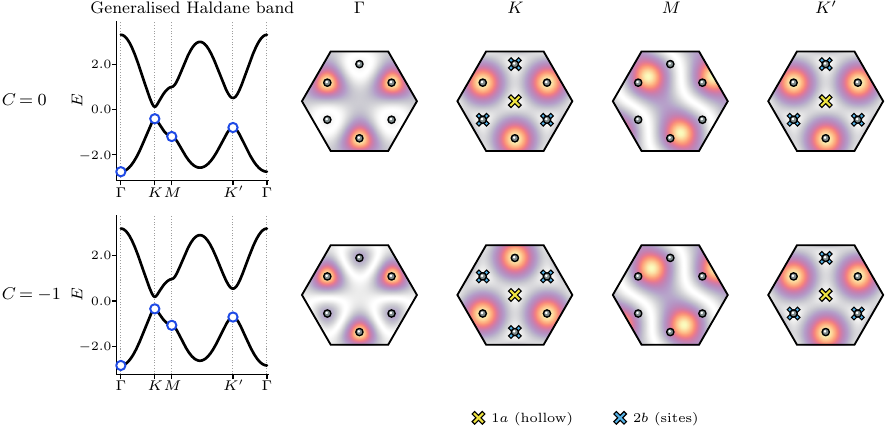}
    \caption{\textbf{Generalised Haldane Bloch wavefunctions at high-symmetry momenta.} Amplitude $|\Psi_{\bm{k}_*}(\bm{r})|^2$ of the generalised Haldane wavefunctions at the high-symmetry momenta $\bm{k}_*=\Gamma, K, M, K',$ with the associated energies indicated on the bandstructure.}
    \label{fig:hsp-haldane}
\end{figure*}

\begin{figure*}[h!]
    \centering
    \includegraphics[width=\linewidth]{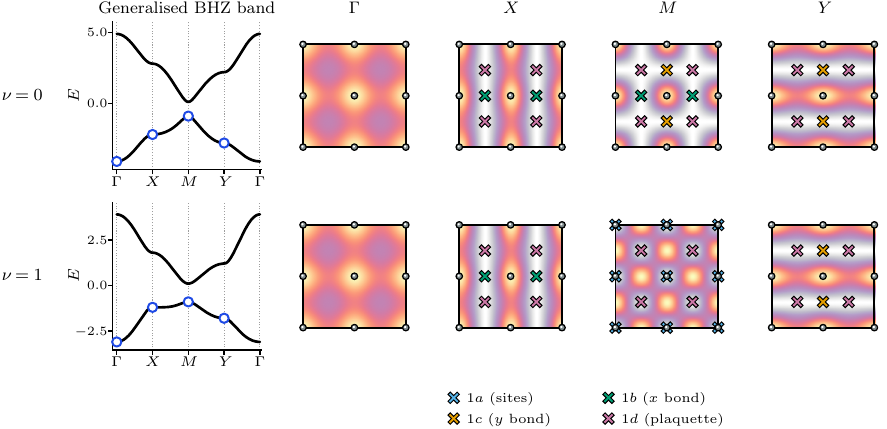}
    \caption{\textbf{Generalised BHZ Bloch wavefunctions at high-symmetry momenta.} Amplitude $|\Psi_{\bm{k}_*}(\bm{r})|^2$ of the generalised BHZ wavefunctions at the high-symmetry momenta $\bm{k}_*=\Gamma, X, M, Y,$ with the associated energies indicated on the bandstructure.}
    \label{fig:hsp-bhz}
\end{figure*}
\begin{figure*}[h!]
    \centering
    \includegraphics[width=\linewidth]{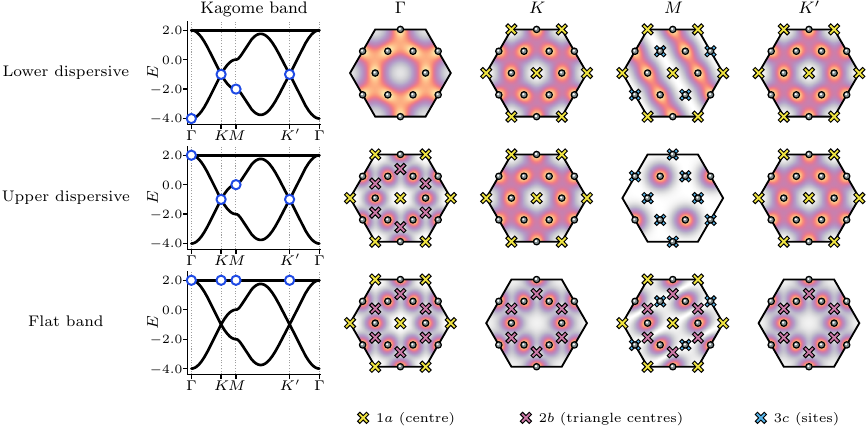}
    \caption{\textbf{Kagome Bloch wavefunctions at high-symmetry momenta.} Amplitude $|\Psi_{\bm{k}_*}(\bm{r})|^2$ of the kagome wavefunctions at the high-symmetry momenta $\bm{k}_*=\Gamma, K, M, K',$ with the associated energies indicated on the bandstructure.}
    \label{fig:hsp-kagome}
\end{figure*}

\FloatBarrier
\section{Derivation of the Kane-Mele invariant as a zero count}
\label{sm:kane-mele}

In this section, we derive that the Kane-Mele invariant can be rewritten as a count of the number of dark set zeros at an inversion centre, for systems with inversion symmetry. The proof assumes the Fu-Kane indicator, which relates the Kane-Mele invariant to the inversion eigenvalues; in forthcoming work we present a proof which does not assume, but rather derives the Fu-Kane indicator using zero counting. Fix one inversion centre $\bm{r}$.  For an occupied
Kramers pair $a$ at a TRIM $\bm{\Lambda}$, let
$\zeta_{I,a}(\bm{\Lambda})=\pm1$ be the inversion
eigenvalue.  We define
\begin{align}
d_a(\bm{\Lambda};\bm{r})
=
\tfrac{1}{2}
\left(
1-e^{-2i\bm{\Lambda}\cdot\bm{r}}\zeta_{I,a}(\bm{\Lambda})
\right).
\end{align}
The variable $d_a(\bm{\Lambda};\bm{r})$ equals one
precisely when inversion about $\bm{r}$ forces the
Kramers pair to vanish at $\bm{r}$, and equals zero
otherwise.  Summing over occupied Kramers pairs and TRIM,
\begin{align}
\mathcal{N}(\bm{r})
=
\sum_{\bm{\Lambda}\in{\rm TRIM}}
\sum_{a=1}^{N}
d_a(\bm{\Lambda};\bm{r})
\pmod2.
\end{align}
The function $\mathcal{N}(\bm{r})$ counts, at the fixed inversion centre
$\bm{r}$, the number of inversion-forced dark occupied
Kramers pairs accumulated over the four TRIM.

Since
\begin{align}
(-1)^{d_a(\bm{\Lambda};\bm{r})}
=
e^{-2i\bm{\Lambda}\cdot\bm{r}}\zeta_{I,a}(\bm{\Lambda}),
\end{align}
we obtain
\begin{align}
(-1)^{\mathcal{N}(\bm{r})}
&=
\left[
\prod_{\bm{\Lambda}\in{\rm TRIM}}
e^{-2i\bm{\Lambda}\cdot\bm{r}}
\right]^N
\prod_{\bm{\Lambda},a}
\zeta_{I,a}(\bm{\Lambda}).
\end{align}
The four two-dimensional TRIM obey
\begin{align}
\sum_{\bm{\Lambda}\in{\rm TRIM}}\bm{\Lambda}
=
\bm{b}_1+\bm{b}_2,
\end{align}
where $\bm{b}_1,\bm{b}_2$ are the reciprocal lattice
vectors.  Since $\bm{r}$ is an inversion centre,
$2\bm{r}$ is a Bravais-lattice vector.  Therefore
\begin{align}
\prod_{\bm{\Lambda}\in{\rm TRIM}}
e^{-2i\bm{\Lambda}\cdot\bm{r}}
=
e^{-2i(\bm{b}_1+\bm{b}_2)\cdot\bm{r}}=1.
\end{align}
Using the Fu--Kane inversion product,
\begin{align}
(-1)^\nu
=
\prod_{\bm{\Lambda},a}
\zeta_{I,a}(\bm{\Lambda}),
\end{align}
we conclude that
\begin{align}
\boxed{
\mathcal{N}(\bm{r})=\nu\pmod2.
}
\end{align}
At any fixed inversion centre, the parity of the number
of inversion-forced dark occupied Kramers pairs over all
four TRIM is the Kane--Mele invariant.

A position with a larger site stabiliser may have
additional zeros enforced by rotations or mirrors.  Those
additional zeros are not included in $\mathcal{N}(\bm{r})$ unless
they are already forced by the inversion subgroup; for this reason, the diagnostic is sharpest when probing an inversion centre with no other symmetry.

\subsection{Reduction in the presence of $C_4$ symmetry}

In the standard $C_4$-symmetric BHZ geometry, the site and plaquette centres are also $C_4$ centres at ${\Gamma}$ and ${M}$.  A zero at one of these positions can therefore be forced by the inclusion of $C_4$ characters in the projector sum. That is, a location can be bright when accounting solely for the constraints implied by inversion symmetry, but forced-dark when $C_4$ is accounted for; such zeros are removed if $C_4$ symmetry is broken. The upshot is that the origin and plaquette centres are ambiguous as a probe of the inversion characters, as their darkness is also determined by $C_4$.

The bond centres ${B}_x$ and ${B}_y$ avoid this ambiguity: at ${\Gamma}$ and ${M}$ they probe the inversion condition without an additional $C_4$ constraint. On the other hand, exact $C_4$ symmetry relates ${X}$ and ${Y}$, so an energy-resolved STM image does not isolate one from the other while preserving the symmetry. Yet, since the occupied inversion-eigenvalue multisets at ${X}$ and ${Y}$ are then identical, their contribution to the fixed-bond zero parity is a known constant --- this suggests that it should be possible to rewrite the invariant in a way that makes it sufficient to image only $\Gamma$ and $M$. The reduction below demonstrates this, rewriting the inversion zero count using only bond-centre data at ${\Gamma}$ and ${M}$.

Now assume that inversion is accompanied by $C_4$
symmetry on a square lattice.  Choose primitive reciprocal
vectors $\bm{b}_1,\bm{b}_2$, and write
\begin{align}
{\Gamma}&=\bm{0},
&
{X}&=\tfrac12\bm{b}_1,
&
{Y}&=\tfrac12\bm{b}_2,
&
{M}&=\tfrac12(\bm{b}_1+\bm{b}_2).
\end{align}
Let the two bond-centred inversion centres be
\begin{align}
{B}_x=\tfrac12\bm{a}_1,
\qquad
{B}_y=\tfrac12\bm{a}_2.
\end{align}
At ${B}_x$, the inversion phases at
$({\Gamma},{X},{Y},{M})$ are
$(+1,-1,+1,-1)$.  Therefore
\begin{align}
d_a({\Gamma};{B}_x)
&=
\tfrac12
\left(
1-\zeta_{I,a}({\Gamma})
\right),
\\
d_a({X};{B}_x)
&=
\tfrac12
\left(
1+\zeta_{I,a}({X})
\right),
\\
d_a({Y};{B}_x)
&=
\tfrac12
\left(
1-\zeta_{I,a}({Y})
\right),
\\
d_a({M};{B}_x)
&=
\tfrac12
\left(
1+\zeta_{I,a}({M})
\right).
\end{align}
At ${B}_y$, the roles of ${X}$ and ${Y}$ are
exchanged.

Since $C_4$ maps ${X}$ to ${Y}$, the occupied
inversion eigenvalue multisets at ${X}$ and ${Y}$
are identical.  Hence, at either fixed bond centre,
\begin{align}
\sum_{a=1}^{N}
d_a({X};{B}_{x,y})
+
\sum_{a=1}^{N}
d_a({Y};{B}_{x,y})
=
N.
\end{align}
The ${X}$ and ${Y}$ contribution to the fixed-bond
zero parity is therefore a known constant.

Define
\begin{align}
z_\Gamma^{\rm bond}
&=
\sum_{a=1}^{N}
d_a({\Gamma};{B}_x)
=
\sum_{a=1}^{N}
d_a({\Gamma};{B}_y),
\\
z_M^{\rm bond}
&=
\sum_{a=1}^{N}
d_a({M};{B}_x)
=
\sum_{a=1}^{N}
d_a({M};{B}_y).
\end{align}
Then the fixed-centre formula reduces to
\begin{align}
\nu
\equiv
N+z_\Gamma^{\rm bond}+z_M^{\rm bond}
\pmod2.
\end{align}
Thus, in a $C_4$-symmetric system, the $X,Y$ data can be
eliminated from the bond-centred zero count.  The
remaining information is the bond-centre contrast at
$\Gamma$ and $M$, together with the known constant $N$.

Equivalently, let $N_\Gamma^{\rm bond}$ and
$N_M^{\rm bond}$ denote the total number of dark
occupied Kramers pairs over the two $C_4$-related bond
centres at $\Gamma$ and $M$.  Since the two bonds are
symmetry-related at both $\Gamma$ and $M$,
\begin{align}
N_\Gamma^{\rm bond}
&=
2z_\Gamma^{\rm bond},
&
N_M^{\rm bond}
&=
2z_M^{\rm bond}.
\end{align}
The same formula may be written as
\begin{align}
\nu
\equiv
N+
\tfrac12
\left(
N_\Gamma^{\rm bond}+N_M^{\rm bond}
\right)
\pmod2.
\end{align}

For a single occupied Kramers pair, $N=1$.  If
$z_\Gamma,z_M\in\{0,1\}$ denote whether the bond pair is
dark at $\Gamma$ and at $M$, then
\begin{align}
\boxed{
\nu
\equiv
1+z_\Gamma+z_M
\pmod2.
}
\end{align}
Thus the topological phase has the same bond-centre
contrast at $\Gamma$ and $M$, while the trivial phase has
opposite bond-centre contrast.

\newpage
\section{Sublattice interference effect on the s-orbital kagome lattice}
\label{sm:kagome}

\subsection{Geometry and tight-binding model}

We take the triangular Bravais lattice, with real space lattice vectors $\bm{a}_1 = (1,0)$ and $\bm{a}_2 = (\tfrac{1}{2},\tfrac{\sqrt{3}}{2})$, and with reciprocal lattice vectors $\bm{b}_1 = (2\pi,-\tfrac{2\pi}{\sqrt{3}})$ and $\bm{b}_2 = (0,\tfrac{4\pi}{\sqrt{3}})$. The three kagome sites in one unit cell are located at the $3c$ positions,
\begin{align}
\bm{r}_A &= \tfrac{1}{2}\bm{a}_1 = (\tfrac{1}{2},0), &
\bm{r}_B &= \tfrac{1}{2}\bm{a}_2 = (\tfrac{1}{4},\tfrac{\sqrt{3}}{4}), &
\bm{r}_C &= \tfrac{1}{2}(\bm{a}_1+\bm{a}_2) = (\tfrac{3}{4},\tfrac{\sqrt{3}}{4}).
\end{align}
For the nearest-neighbour kagome tight-binding model, it is convenient to define
\begin{align}
\bm{d}_{AB} &= \tfrac{1}{2}(\bm{a}_1-\bm{a}_2) = (\tfrac{1}{4},-\tfrac{\sqrt{3}}{4}), \\
\bm{d}_{BC} &= \tfrac{1}{2}\bm{a}_1 = (\tfrac{1}{2},0), \\
\bm{d}_{CA} &= \tfrac{1}{2}\bm{a}_2 = (\tfrac{1}{4},\tfrac{\sqrt{3}}{4}).
\end{align}
In the $(A,B,C)$ basis the Bloch Hamiltonian is
\begin{align}
H_0(\bm{k}) =
-2t
\begin{pmatrix}
0 & \cos(\bm{k}\cdot \bm{d}_{AB}) & \cos(\bm{k}\cdot \bm{d}_{CA}) \\
\cos(\bm{k}\cdot \bm{d}_{AB}) & 0 & \cos(\bm{k}\cdot \bm{d}_{BC}) \\
\cos(\bm{k}\cdot \bm{d}_{CA}) & \cos(\bm{k}\cdot \bm{d}_{BC}) & 0
\end{pmatrix}.
\end{align}
The three inequivalent $M$ points are
\begin{align}
\bm{M}_1 &= \tfrac{1}{2}\bm{b}_2 = (0,\tfrac{2\pi}{\sqrt{3}}), \\
\bm{M}_2 &= \tfrac{1}{2}(\bm{b}_1+\bm{b}_2) = (\pi,\tfrac{\pi}{\sqrt{3}}), \\
\bm{M}_3 &= \tfrac{1}{2}\bm{b}_1 = (\pi,-\tfrac{\pi}{\sqrt{3}}).
\end{align}
We choose the labelling so that $\bm{M}_1$ singles out sublattice $A$, while $\bm{M}_2$ and $\bm{M}_3$ are obtained by $C_3$ rotation.

\subsection{Induced representations at $M_1$}
\label{sm:kag-induction}

We now consider the properties of the wavefunction at $\bm{k}=\bm{M}_1=(0,\tfrac{2\pi}{\sqrt{3}})$. The Bloch functions are (see Eq. \eqref{bloch_induced})
\begin{align}
|\alpha\rangle \equiv |\alpha,\bm{M}_1\rangle
= \sum_{\bm{R}} e^{i \bm{M}_1 \cdot (\bm{R}+\bm{r}_\alpha)} |\bm{R},\alpha\rangle,
\qquad
\alpha=A,B,C,
\end{align}
where $\alpha$ indexes the three sublattices and $|\bm{R},\alpha\rangle$ correspond to localised $s$-like orbitals at the 3c positions. The little co-group of the $M$-point is $C_{2v}=\{E,C_2,\sigma_1,\sigma_2\}$. We now follow the standard site-symmetry-induction construction of a band representation from a localised orbital on the $3c$ Wyckoff orbit \cite{CanoBradlyn2021,Cano2018,Zak1981}; we choose $\sigma_1$ to be the mirror that fixes the $A$ site and swaps $B \leftrightarrow C$, and $\sigma_2=C_2\sigma_1$. Under $C_2$,
\begin{align}
A &\mapsto A-\bm{a}_1, &
B &\mapsto B-\bm{a}_2, &
C &\mapsto C-\bm{a}_1-\bm{a}_2.
\end{align}
At $\bm{M}_1$ the corresponding Bloch phases are
\begin{align}
e^{-i\bm{M}_1\cdot \bm{a}_1} &= 1, \\
e^{-i\bm{M}_1\cdot \bm{a}_2} &= -1, \\
e^{-i\bm{M}_1\cdot (\bm{a}_1+\bm{a}_2)} &= -1,
\end{align}
and so the representation of $C_2$ symmetry in the sublattice basis is
\begin{align}
D(C_2)=
\begin{pmatrix}
1 & 0 & 0 \\
0 & -1 & 0 \\
0 & 0 & -1
\end{pmatrix}.
\end{align}
For $\sigma_1$, we have
\begin{align}
A &\mapsto A, &
B &\mapsto C-\bm{a}_2, &
C &\mapsto B+\bm{a}_1-\bm{a}_2.
\end{align}
Accounting for the Bloch phases appropriately, one finds
\begin{align}
D(\sigma_1)=
\begin{pmatrix}
1 & 0 & 0 \\
0 & 0 & -1 \\
0 & -1 & 0
\end{pmatrix}.
\end{align}
For $\sigma_2$, we have
\begin{align}
A &\mapsto A-\bm{a}_1, &
B &\mapsto C-\bm{a}_1, &
C &\mapsto B-\bm{a}_1,
\end{align}
and the Bloch phase is now $+1$, giving
\begin{align}
D(\sigma_2)=
\begin{pmatrix}
1 & 0 & 0 \\
0 & 0 & 1 \\
0 & 1 & 0
\end{pmatrix}.
\end{align}
The characters of the three-dimensional band representation induced from an $s$ orbital on the $3c$ orbit are given by the traces of these matrices,
\begin{align}
\chi^{(s@3c)}_{\bm{M}_1}(E,C_2,\sigma_1,\sigma_2)=(3,-1,1,1).
\end{align}
Using the $C_{2v}$ character table
\begin{align}
\chi_{A_1} &= (1,1,1,1), &
\chi_{A_2} &= (1,1,-1,-1), \\
\chi_{B_1} &= (1,-1,1,-1), &
\chi_{B_2} &= (1,-1,-1,1),
\end{align}
we can use the orthogonality of characters \cite{Serre1977} to decompose these characters into irreps; one immediately finds
\begin{align}
\chi^{(s@3c)}_{\bm{M}_1}=A_1\oplus B_1\oplus B_2.
\end{align}
Thus the $s@3c$ elementary band representation contributes exactly three one-dimensional irreducible representations at each $M$ point
\begin{align}
(s@3c)\uparrow G\Big|_{\bm{M}_1}=A_1\oplus B_1\oplus B_2.
\end{align}
These are the symmetry representations which appear in the kagome $s$ orbital tight-binding model near the $M$-point.

\subsection{Dark set on the three kagome sublattices}

We now compute the dark set index
\begin{align}
m_\rho(\bm{M}_1,\bm{r})
=
\tfrac{1}{|H_{\bm{M}_1,\bm{r}}|}
\sum_{h\in H_{\bm{M}_1,\bm{r}}}
e^{-i\bm{M}_1\cdot (1-R_h)\bm{r}}\,
\chi_\rho(h).
\end{align}
We first consider sublattice $A$; for $\bm{r}=\bm{r}_A$, all four little-group elements fix $A$ modulo lattice translation, so
\begin{align}
H_{\bm{M}_1,A}=C_{2v}.
\end{align}
The translation phase is trivial for all four elements, and so
\begin{align}
m_\rho(\bm{M}_1,A)
=
\tfrac{1}{4}
\Big[
\chi_\rho(E)+\chi_\rho(C_2)+\chi_\rho(\sigma_1)+\chi_\rho(\sigma_2)
\Big].
\end{align}
Consulting the character table for $C_{2v}$ allows us to evaluate this for different $\rho$; this gives the value of the index as
\begin{align}
m_{A_1}(\bm{M}_1,\bm{r}_A) &= 1, \\
m_{A_2}(\bm{M}_1,\bm{r}_A) &= 0, \\
m_{B_1}(\bm{M}_1,\bm{r}_A) &= 0, \\
m_{B_2}(\bm{M}_1,\bm{r}_A) &= 0.
\end{align}
Hence of the four irreps of $C_{2v}$, at $\bm{M}_1$ all are dark on sublattice $A$ except the $A_1$ irrep.

At the $B$ sublattice $\bm{r}=\bm{r}_B$, only $E$ and $C_2$ fix the point modulo a lattice vector, so
\begin{align}
H_{\bm{M}_1,B}=\{E,C_2\}\cong C_2.
\end{align}
The $C_2$ action sends $B\mapsto B-\bm{a}_2$, i.e. $(1-C_2)\bm{r}_B=\bm{a}_2$, yielding the Bloch phase
\begin{align}
e^{-i\bm{M}_1\cdot \bm{a}_2}=-1.
\end{align}
Hence
\begin{align}
m_\rho(\bm{M}_1,B)
=
\tfrac{1}{2}
\Big[
\chi_\rho(E)-\chi_\rho(C_2)
\Big].
\end{align}
Therefore, using the character table for $C_{2v}$ again
\begin{align}
m_{A_1}(\bm{M}_1,B) &= 0, \\
m_{A_2}(\bm{M}_1,B) &= 0, \\
m_{B_1}(\bm{M}_1,B) &= 1, \\
m_{B_2}(\bm{M}_1,B) &= 1.
\end{align}
So the $A_1$ state is dark on $B$, while both $B_1$ and $B_2$ are allowed there. Exactly the same logic holds for $\bm{r}=\bm{r}_C$, because $C_2$ sends $C\mapsto C-\bm{a}_1-\bm{a}_2$ and
\begin{align}
e^{-i\bm{M}_1\cdot (\bm{a}_1+\bm{a}_2)}=-1.
\end{align}
Thus
\begin{align}
m_\rho(\bm{M}_1,C)
=
\tfrac{1}{2}
\Big[
\chi_\rho(E)-\chi_\rho(C_2)
\Big],
\end{align}
and therefore
\begin{align}
m_{A_1}(\bm{M}_1,C) &= 0, \\
m_{A_2}(\bm{M}_1,C) &= 0, \\
m_{B_1}(\bm{M}_1,C) &= 1, \\
m_{B_2}(\bm{M}_1,C) &= 1.
\end{align}

\subsection{Site-resolved dark sets at the $M$ points}

Collecting the three site tests at $\bm{M}_1$, we obtain
\begin{align}
D_{A_1}(\bm{M}_1) &= \{B,C\}, \\
D_{B_1}(\bm{M}_1) &= \{A\}, \\
D_{B_2}(\bm{M}_1) &= \{A\}.
\end{align}
Thus the $A_1$ state is supported only on the singled-out sublattice $A$, whereas the $B_1$ and $B_2$ states avoid $A$ and live on the complementary pair $(B,C)$.

By $C_3$ symmetry, the other two $M$ points give the cyclic permutations
\begin{align}
D_{A_1}(\bm{M}_2) &= \{C,A\}, &
D_{B_1}(\bm{M}_2) &= D_{B_2}(\bm{M}_2)=\{B\}, \\
D_{A_1}(\bm{M}_3) &= \{A,B\}, &
D_{B_1}(\bm{M}_3) &= D_{B_2}(\bm{M}_3)=\{C\}.
\end{align}
Hence, for each $M$ patch, one irreducible representation is bright on exactly one kagome sublattice, while the other two are bright on exactly two sublattices.

\subsection{Identification with the nearest-neighbour kagome bands}

At $\bm{M}_1$ one has
\begin{align}
\bm{M}_1\cdot \bm{d}_{AB} &= -\tfrac{\pi}{2}, \\
\bm{M}_1\cdot \bm{d}_{CA} &= \tfrac{\pi}{2}, \\
\bm{M}_1\cdot \bm{d}_{BC} &= 0,
\end{align}
so the nearest-neighbour Hamiltonian becomes
\begin{align}
H_0(\bm{M}_1)=
-2t
\begin{pmatrix}
0 & 0 & 0 \\
0 & 0 & 1 \\
0 & 1 & 0
\end{pmatrix}.
\end{align}
Its three eigenstates are
\begin{align}
|p_1\rangle &= |A\rangle, &
E_{p_1} &= 0, \\
|m_1\rangle &= \tfrac{1}{\sqrt{2}}(|B\rangle+|C\rangle), &
E_{m_1} &= -2t, \\
|f_1\rangle &= \tfrac{1}{\sqrt{2}}(|B\rangle-|C\rangle), &
E_{f_1} &= +2t.
\end{align}
Up to the overall hopping-sign convention, these are the middle ($p$-type VHS), lower ($m$-type VHS), and upper (flat-band) states. Comparing with the symmetry matrices above, one finds the symmetry eigenvalues:
\begin{align}
|p_1\rangle &: (1,1,1,1)=A_1, \\
|m_1\rangle &: (1,-1,-1,1)=B_2, \\
|f_1\rangle &: (1,-1,1,-1)=B_1.
\end{align}
Thus, at $\bm{M}_1$,
\begin{align}
D_{p_1} &= \{B,C\}, \\
D_{m_1} &= \{A\}, \\
D_{f_1} &= \{A\}.
\end{align}
In other words, the $p$-type Van Hove state is supported on a single kagome sublattice, while the $m$-type Van Hove state avoids that sublattice and lives on the complementary pair. 

One interesting aspect of this result is its robustness --- unlike the flat band of the kagome lattice, which becomes dispersive upon the addition of hoppings other than the simplest nearest-neighbour term, the sublattice interference effect is purely a statement about the band representations at the $M$-point, and is therefore an \textit{exact} statement in the presence of symmetries. Adding complicated additional hoppings or interaction-induced band renormalisation can change the energetic ordering of the irreps from those in the nearest-neighbour model, but their dark sets remain the same.

On the other hand, when $C_2$ or mirror $\sigma_{1,2}$ are broken, the sublattice texture can change --- for instance, it has been previously noted that a breathing distortion of the kagome lattice, which breaks $C_2$ and retains mirror, can mix the $m$- and $p$-type VHS while leaving the sublattice structure of the flat band intact \cite{nag2024pomeranchuk, beck2026kekule}.

\subsection{Distinguishing the $m$-type band from the flat band}

On the three kagome sites alone, the $B_1$ and $B_2$ irreps look identical: both are dark on $A$ at $\bm{M}_1$. To distinguish them one may use an additional high-symmetry probe, namely the midpoint of the $BC$ bond,
\begin{align}
\bm{P}_A = \tfrac{1}{2}(\bm{r}_B+\bm{r}_C)
= (\tfrac{1}{2},\tfrac{\sqrt{3}}{4}).
\end{align}
For this point,
\begin{align}
H_{\bm{M}_1,\bm{P}_A}=\{E,\sigma_2\},
\end{align}
and the twist is trivial, so
\begin{align}
m_\rho(\bm{M}_1,\bm{P}_A)=\tfrac{1}{2}
\Big[
1+\chi_\rho(\sigma_2)
\Big].
\end{align}
Hence
\begin{align}
m_{A_1}(\bm{M}_1,\bm{P}_A) &= 1, \\
m_{A_2}(\bm{M}_1,\bm{P}_A) &= 0, \\
m_{B_1}(\bm{M}_1,\bm{P}_A) &= 0, \\
m_{B_2}(\bm{M}_1,\bm{P}_A) &= 1.
\end{align}
So the lower $m$-type state $|m_1\rangle$ is symmetry-allowed on the $BC$ bond midpoint, while the flat-band state $|f_1\rangle$ is symmetry-forced-dark there.

\newpage

\section{Hartree renormalisation of the flat bands in magic-angle bilayer graphene}
\label{sm:tbg}

This section describes the continuum Hamiltonian, the flat-band symmetry representations and dark sets, the one-shot Hartree calculation used, and the atomistic real-space embedding used for the wavefunction panels in Fig. \ref{fig:tbg}. 

\subsection{Continuum model}
\label{sm:tbg_continuum}

We use the single-valley continuum Dirac description of a twisted graphene bilayer \cite{LopesDosSantos2007Twist,Mele2010Commensuration,BistritzerMacDonald2011,LopesDosSantos2012Continuum}. We write the Hamiltonian for valley $\tau=+1$ throughout; the opposite valley is generated by time reversal, which also complex-conjugates all crystalline eigenvalues. The reciprocal primitive vectors of an untwisted graphene layer are denoted $\bm{G}_1,\bm{G}_2$, and the moiré reciprocal lattice vectors are
\begin{align}
\bm{g}_i=(R_{\theta/2}-R_{-\theta/2})\bm{G}_i,\qquad i=1,2,
\label{eq:tbg_moire_reciprocal}
\end{align}
with $\bm{a}_i\cdot\bm{g}_j=2\pi\delta_{ij}$. We define,
\begin{align}
\bm{q}_1=-\tfrac{1}{3}(\bm{g}_1+\bm{g}_2),\qquad \bm{q}_2=\bm{q}_1+\bm{g}_1,\qquad \bm{q}_3=\bm{q}_1+\bm{g}_2,
\label{eq:tbg_q_vectors_revised}
\end{align}
and the high symmetry momenta
\begin{align}
K_M:\bm{k}=0,\qquad \Gamma_M:\bm{k}=\bm{q}_1,\qquad M_M:\bm{k}=\bm{q}_1+\tfrac{1}{2}\bm{g}_1,\qquad K'_M:\bm{k}=-\bm{q}_1.
\label{eq:tbg_hsp_gauge_revised}
\end{align}

The plane-wave basis is $|\bm{g},\ell,s\rangle$, where $\bm{g}=m\bm{g}_1+n\bm{g}_2$, $\ell=1,2$ labels the layer, and $s=A,B$ labels the microscopic graphene sublattice. Using $\bm{\sigma}=(\sigma_x,\sigma_y)$, we define the intralayer Hamiltonian blocks
\begin{align}
H_{11}(\bm{k}+\bm{g})=-\hbar v_F\bm{\sigma}\cdot R_{-\theta/2}(\bm{k}+\bm{g}),\qquad H_{22}(\bm{k}+\bm{g})=-\hbar v_F\bm{\sigma}\cdot R_{\theta/2}(\bm{k}+\bm{g}+\bm{q}_1),
\label{eq:tbg_dirac_blocks_revised}
\end{align}
where subscripts $1,2$ index the layers, and
\begin{align}
H_{12}(\bm{g},\bm{g}')=T_1\delta_{\bm{g}',\bm{g}}+T_2\delta_{\bm{g}',\bm{g}+\bm{g}_1}+T_3\delta_{\bm{g}',\bm{g}+\bm{g}_2},\qquad H_{21}=H_{12}^{\dagger}.
\label{eq:tbg_interlayer_revised}
\end{align}
The tunnelling matrices are then
\begin{align}
T_1=\begin{pmatrix}w_{AA}&w_{AB}\\ w_{AB}&w_{AA}\end{pmatrix},\qquad T_2=\begin{pmatrix}w_{AA}&w_{AB}\omega^{-1}\\ w_{AB}\omega&w_{AA}\end{pmatrix},\qquad T_3=\begin{pmatrix}w_{AA}&w_{AB}\omega\\ w_{AB}\omega^{-1}&w_{AA}\end{pmatrix}.
\label{eq:tbg_tunnel_revised}
\end{align}
The numerical parameters used for the magic-angle bandstructure are
\begin{align}
\theta=1.05^{\circ},\qquad w_{AA}=80\,\mathrm{meV},\qquad w_{AB}=110\,\mathrm{meV},\qquad v_F=8.8\times10^5\,\mathrm{m}\,\mathrm{s}^{-1},\qquad a_0=0.246\,\mathrm{nm}.
\label{eq:tbg_parameters_revised}
\end{align}
For the Hartree calculation, reciprocal vectors are retained within the radial cutoff $R_{\mathrm{fac}}=3.75$ about the physical Dirac momentum of each layer.  This gives $55$ layer-1 and $48$ layer-2 reciprocal vectors, hence a one-valley Hamiltonian dimension of $206$; the residual $C_3$ covariance error at $K_M$ is $2.5\times10^{-13}$.  The atomistic density panels use the separate high-symmetry-point-centred basis described in Sec.~\ref{sm:tbg_density}.

\subsection{One-shot Hartree bands}
\label{sm:tbg_hartree}

Electrostatics is known to strongly reshape the active bands away from charge neutrality \cite{GuineaWalet2018,RademakerAbaninMellado2019,CeaWaletGuinea2019,Goodwin2020Hartree,CeaPantaleonWaletGuinea2022}.  A complete mean-field calculation requires a choice of reference density matrix and, in general, a self-consistency loop.  We use a charge-neutrality reference density in the sense reviewed in Ref.~\cite{Kwan2025MeanFieldGuide}, but retain only the Hartree term and diagonalise once rather than perform a fully self-consistent calculation.  Thus we compute
\begin{align}
H^{\mathrm{1shot}}(\nu)=H^0+H_H[\delta P_\nu],
\qquad
\delta P_\nu=P_\nu-P_{\nu=0},
\label{eq:tbg_one_shot_definition}
\end{align}
with no exchange term and no update of $\delta P_\nu$ after diagonalisation.  This construction isolates the electrostatic form-factor mechanism responsible for the momentum-dependent reshaping, rather than attempting a self-consistent Hartree or Hartree--Fock prediction.

\label{sm:tbg_filling}
The interacting panel in Fig.~\ref{fig:tbg} is evaluated at electron filling $\nu=2$ with a spin-valley-symmetric occupation.  Each of the four spin-valley flavours contributes one half electron per moiré unit cell to the central conduction band.  On a uniform $N\times N$ mesh of the moiré Brillouin zone, the noninteracting central-conduction-band energies are sorted independently in the two valleys and the lowest half of the mesh points are occupied.  Equivalently,
\begin{align}
f_{\tau\bm{k}}=\Theta(\mu_\tau-\epsilon^0_{\tau c\bm{k}}),
\qquad
\tfrac{1}{N^2}\sum_{\bm{k}}f_{\tau\bm{k}}=\tfrac{1}{2},
\label{eq:tbg_nu2_occupation}
\end{align}
where $c$ labels the central conduction band.  Let $\eta=\pm$ label the two time-reversed valleys generated from the $\tau=+1$ Hamiltonian of Sec.~\ref{sm:tbg_continuum}.  The source density matrix is
\begin{align}
\delta P_{\nu=2}=\tfrac{2}{N^2}\sum_{\eta=\pm}\sum_{\bm{k}}f_{\eta\bm{k}}|u^0_{\eta c\bm{k}}\rangle\langle u^0_{\eta c\bm{k}}|,
\label{eq:tbg_deltaP_nu2}
\end{align}
where $|u^0_{\eta c\bm{k}}\rangle$ is the normalised noninteracting eigenvector and the prefactor $2$ is the spin degeneracy.  This contains exactly two added electrons per moiré unit cell.  At charge neutrality $P_\nu=P_{\nu=0}$, so the Hartree correction vanishes by construction.

For a moiré reciprocal vector $\delta\bm{g}$, the plane-wave density form factor and excess-density harmonic are
\begin{align}
F_{\eta n\bm{k}}(\delta\bm{g})
&=\sum_{\ell,s}\sum_{\bm{g}}u^*_{\eta n\bm{k}}(\bm{g},\ell,s)u_{\eta n\bm{k}}(\bm{g}+\delta\bm{g},\ell,s),
\label{eq:tbg_form_factor_revised}
\\
\delta n(\delta\bm{g})
&=\tfrac{2}{N^2}\sum_{\eta=\pm}\sum_{\bm{k}}f_{\eta\bm{k}}F_{\eta c\bm{k}}(\delta\bm{g}),
\label{eq:tbg_density_harmonic_nu2}
\end{align}
where terms outside the finite plane-wave basis are omitted.  We use $N=72$ in the results presented in the main text, and no finite-temperature smearing.  The uniform component $\delta\bm{g}=0$ is removed because it is cancelled by the neutralising background or absorbed into the chemical potential.

\label{sm:tbg_hartree_stars}
The dual-gate screened Coulomb kernel is
\begin{align}
V(g)=\tfrac{2\pi e^2}{\epsilon_r g}\tanh(gd_{\mathrm{sc}}),
\qquad g=|\delta\bm g|,
\label{eq:tbg_dual_gate_kernel}
\end{align}
where $e^2$ denotes $e^2/(4\pi\epsilon_0)$ in units of $\mathrm{meV}\,\mathrm{nm}$.  We use $\epsilon_r=10$ and $d_{\mathrm{sc}}=30\,\mathrm{nm}$.  The resulting Hartree matrix is diagonal in layer and microscopic sublattice and is taken to be the same in both graphene layers,
\begin{align}
[H_H]_{\bm{g}\ell s,\bm{g}'\ell's'}
=
\delta_{\ell\ell'}\delta_{ss'}
\tfrac{1}{\mathcal A_M}\sum_{\delta\bm{g}\ne0}V(|\delta\bm{g}|)\,\delta n(\delta\bm{g})\,\delta_{\bm{g}',\bm{g}+\delta\bm{g}},
\label{eq:tbg_hartree_matrix_revised}
\end{align}
where $\mathcal A_M$ is the moiré unit-cell area.  The microscopic interlayer separation is neglected in the interaction kernel, which is appropriate on the moiré length scale.  After constructing Eq.~\eqref{eq:tbg_hartree_matrix_revised}, we impose Hermiticity and average the matrix over the three $C_3$ rotations to remove residual quadrature and cutoff asymmetries.

The main text calculation retains the first two stars $\mathcal S_1\cup\mathcal S_2$ of the triangular moiré reciprocal lattice, with $g_s/|\bm g_1|=1$ and $\sqrt{3}$ and six vectors in each star. The first two stars give $96.80\%$ of the three-star value of the relative $K_M-\Gamma_M$ shift; adding the third star, $g_3/|\bm g_1|=2$, changes either central band by at most $0.85\,\mathrm{meV}$ along the complete $K_M\!\to\!\Gamma_M\!\to\!M_M\!\to\!K_M$ path.  Replacing the $72\times72$ source mesh by a $60\times60$ mesh changes the quoted high-symmetry shifts by less than $2\times10^{-4}\,\mathrm{meV}$. The qualitative reshaping is also robust to the screening parameters --- at fixed $d_{\mathrm{sc}}=30\,\mathrm{nm}$, varying $\epsilon_r$ from $8$ to $12$ changes $\Delta_{K\Gamma}$ from $29.10$ to $19.64\,\mathrm{meV}$ and $\Delta_{M\Gamma}$ from $29.37$ to $19.80\,\mathrm{meV}$, close to the expected $1/\epsilon_r$ scaling.  At fixed $\epsilon_r=10$, varying $d_{\mathrm{sc}}$ from $5$ to $30\,\mathrm{nm}$ changes $\Delta_{K\Gamma}$ only from $23.280$ to $23.466\,\mathrm{meV}$; the finite-wavevector interaction is already nearly saturated at the shortest screening length considered.  These variations alter the overall Hartree scale but not the pronounced minimum of the shift near $\Gamma_M$.

\begin{figure*}[t]
\centering
\includegraphics[width=\textwidth]{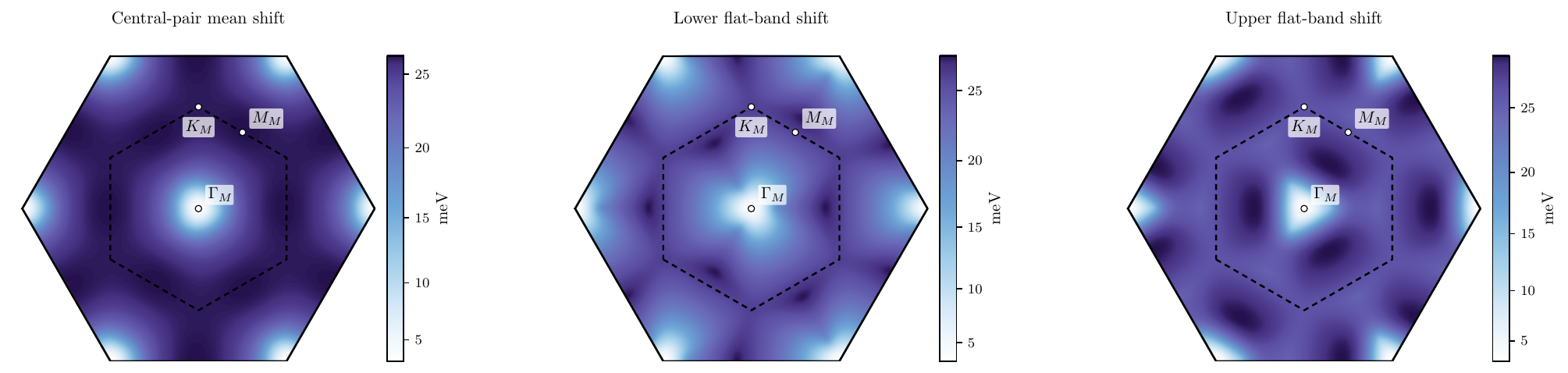}
\caption{One-shot band shifts throughout the hexagonal moiré Brillouin zone.  Left: mean shift $\Delta\bar E$ of the central pair.  Middle and right: shifts of the lower and upper central bands after diagonalising $H^0+H_H$.  The central-pair mean has a pronounced minimum at $\Gamma_M$, while the separate branches display the threefold redistribution of the common outer-Brillouin-zone shift.}
\label{fig:tbg_exact_shift_bz}
\end{figure*}

\label{sm:tbg_bz_maps}
For the Brillouin-zone maps, $\bm{\kappa}$ denotes Cartesian momentum measured from $\Gamma_M$.  We diagonalise the full one-shot Hamiltonian throughout the hexagonal moiré Brillouin zone and define
\begin{align}
\Delta E_\pm(\bm{\kappa})
&=E^{\mathrm{1shot}}_\pm(\bm{\kappa})-E^0_\pm(\bm{\kappa}),
&
\Delta\bar E(\bm{\kappa})
&=\tfrac{1}{2}\{\Delta E_+(\bm{\kappa})+\Delta E_-(\bm{\kappa})\}.
\label{eq:tbg_exact_shift_maps}
\end{align}
The raw central-pair mean shifts are $2.516\,\mathrm{meV}$ at $\Gamma_M$, $25.982\,\mathrm{meV}$ at $K_M$, and $26.189\,\mathrm{meV}$ at $M_M$.  Thus $\Delta_{K\Gamma}\equiv\Delta\bar E(K_M)-\Delta\bar E(\Gamma_M)=23.466\,\mathrm{meV}$ and $\Delta_{M\Gamma}\equiv\Delta\bar E(M_M)-\Delta\bar E(\Gamma_M)=23.673\,\mathrm{meV}$.  The separate lower- and upper-band shifts need not be maximal at $K_M$: the symmetry-preserving Hartree potential produces a common shift of the Dirac doublet there, whereas away from the corner the branches can mix and redistribute the shift, giving the threefold structure in Fig.~\ref{fig:tbg_exact_shift_bz}.

For the band path used in the main text, the noninteracting zero is the mean energy of the central pair at $K_M$.  The Hartree spectrum is shifted by a constant so that the central-pair mean at $\Gamma_M$ aligns with the noninteracting value; this removes no momentum-dependent information.

\begin{figure*}[t]
\centering
\includegraphics[width=0.82\textwidth]{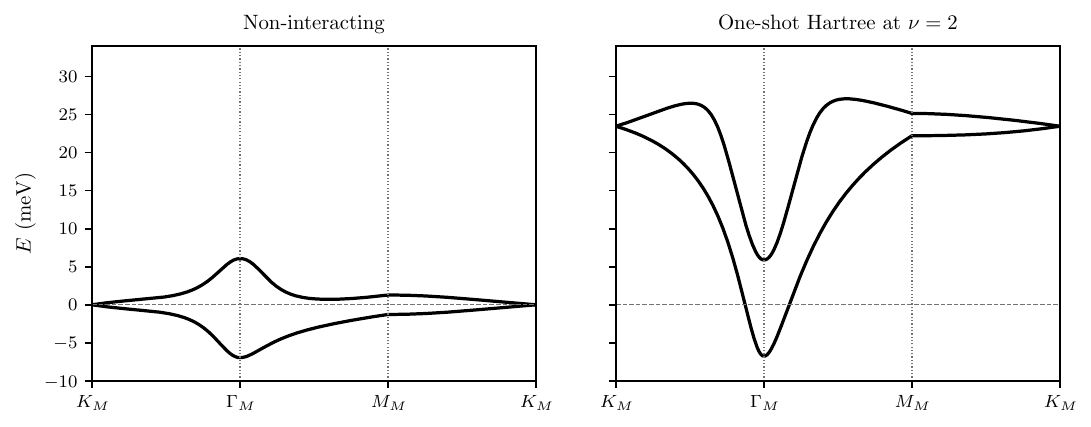}
\caption{Noninteracting bands and the charge-neutrality-subtracted one-shot Hartree bands at $\nu=2$.  The source density is evaluated on a $72\times72$ mesh and the first two moiré reciprocal stars are retained.  The Hartree panel is aligned at $\Gamma_M$, highlighting a relative displacement of $23.47\,\mathrm{meV}$ at $K_M$ and $23.67\,\mathrm{meV}$ at $M_M$.}
\label{fig:tbg_production_bands}
\end{figure*}

\subsection{Flat-band symmetry representations}
\label{sm:tbg_symmetry_representations}

The one-valley and two-valley symmetry groups must be distinguished.  A single valley has magnetic space group $P6'2'2$, generated by $C_{3z}$, $C_{2x}$, $C_{2z}\mathcal T$, and moiré translations.  Restoring the time-reversed valley gives the ordinary space group $P622$, whose point group is $D_6$.  The flat-band representations were identified in the early symmetry analyses of Refs.~\cite{ZouPoVishwanathSenthil2018,KangVafek2018,PoZouVishwanathSenthil2018,PoZouSenthilVishwanath2019,SongWangShiLiFangBernevig2019}; the same representation content is central to the heavy-fermion descriptions of Refs.~\cite{SongBernevig2022,ShiDai2022,CalugaruBorovkovLauColemanSongBernevig2023}.

For one valley, the central pair transforms as
\begin{align}
\Gamma_M:&\ \Gamma_1\oplus\Gamma_2,
&K_M:&\ K_2K_3,
&M_M:&\ M_1\oplus M_2.
\label{eq:tbg_single_valley_irreps}
\end{align}
The time-reversed corner $K'_M$ carries the conjugate co-representation and is left implicit below.  The two $\Gamma_M$ states have $C_{3z}$ eigenvalue $1$ and opposite $C_{2x}$ parities; the $K_2K_3$ co-representation has $C_{3z}$ eigenvalues $\omega$ and $\omega^2$; and the two $M_M$ states have opposite $C_{2x}$ parities.  After the opposite valley is added, the four flat bands transform as
\begin{align}
\Gamma_M:&\ \Gamma_1\oplus\Gamma_2\oplus\Gamma_3\oplus\Gamma_4,
&K_M:&\ 2K_3,
&M_M:&\ M_1\oplus M_2\oplus M_3\oplus M_4.
\label{eq:tbg_two_valley_irreps}
\end{align}
At a selected $K_M$ corner, the little co-group is the six-element proper-rotation group $D_3$, generated in our convention by $C_{3z}$ and $C_{2y}$.  Although $D_3$ is abstractly isomorphic to $C_{3v}$, its order-two elements here are three-dimensional rotations, as we emphasise below.  The crystallographic irrep $K_3$ is also denoted $E$, with $\chi_E(C_{3z})=-1$ and $\chi_E(C_{2y})=0$.  The operations $C_{2x}$ and $C_{2z}$ exchange $K_M$ and $K'_M$.  Table~\ref{tab:tbg_irreps_full} gives the remaining generator characters at $\Gamma_M$ and $M_M$.

Using the conventional crystallographic $C_{3z}$ generator, the
central flat-band subspace in valley $+$ has the
$C_{3z}$-eigenvalue multisets
\begin{align}
\Lambda_{\Gamma_M}=\{1,1\},\qquad
\Lambda_{K_M}=\Lambda_{K'_M}=\{\omega,\omega^2\}.
\label{eq:tbg_c3_eigenvalues_revised}
\end{align}
Time reversal complex-conjugates the eigenvalues and exchanges the laboratory-frame $K_M$ and $K'_M$ points. Since the corner eigenvalues form the conjugation-invariant set $\{\omega,\omega^2\}$, the unordered eigenvalue set is the same at the two corners.

\begin{table*}[t]
\centering
\small
\renewcommand{\arraystretch}{1.45}
\setlength{\tabcolsep}{8pt}
\begin{tabular*}{\textwidth}{@{\extracolsep{\fill}}c c c c c}
\hline\hline
Momentum & Irrep & $\chi(C_{3z})$ & $\chi(C_{2z})$ & $\chi(C_{2x})$ \\
\hline
$\Gamma_M$ & $\Gamma_1\;(A_1)$ & $1$ & $+1$ & $+1$ \\
$\Gamma_M$ & $\Gamma_2\;(A_2)$ & $1$ & $+1$ & $-1$ \\
$\Gamma_M$ & $\Gamma_3\;(B_2)$ & $1$ & $-1$ & $-1$ \\
$\Gamma_M$ & $\Gamma_4\;(B_1)$ & $1$ & $-1$ & $+1$ \\
\hline
$K_M$ & $K_3\;(E)$ & $-1$ & -- & -- \\
\hline
$M_M$ & $M_1\;(A)$ & -- & $+1$ & $+1$ \\
$M_M$ & $M_2\;(B_1)$ & -- & $+1$ & $-1$ \\
$M_M$ & $M_3\;(B_2)$ & -- & $-1$ & $-1$ \\
$M_M$ & $M_4\;(B_3)$ & -- & $-1$ & $+1$ \\
\hline\hline
\end{tabular*}
\caption{\textbf{Two-valley flat-band representation data for magic-angle graphene.} The first label in the Irrep column is the crystallographic small-representation label; the label in parentheses is the Mulliken notation for the associated little co-group.  The little co-groups are $D_6$ at $\Gamma_M$, $D_3\ni C_{3z},C_{2y}$ at $K_M$, and $D_2 \ni C_{2y}$ at $M_M$.  In the axis convention of the $P622$ character table, $C_{2x}$ belongs to the $C_2'$ class and $C_{2y}$ to the $C_2''$ class.  The dashes in the $K_M$ row indicate that $C_{2z}$ and $C_{2x}$ exchange the two corners.  The flat band manifold contains one copy of each listed $\Gamma_M$ and $M_M$ irrep and two copies of $K_3$.}
\label{tab:tbg_irreps_full}
\end{table*}

\subsubsection*{Three-dimensional meaning of the order-two characters}
\label{sm:tbg_order_two_3d}

The order-two characters are physically important, but their nodal consequences live partly outside the two-dimensional, layer-projected density plotted in Fig.~\ref{fig:tbg}.  The distinction follows from the three-dimensional action of the point group.  The abstract groups $D_6$ and $C_{6v}$ are isomorphic, and restricting the coordinates to the plane $z=0$ makes an in-plane twofold rotation look like a mirror.  Their fixed sets in the physical bilayer are nevertheless different.  In $D_6$,
\begin{align}
C_{2x}:(x,y,z)\mapsto(x,-y,-z),
\qquad
\mathcal L_x=\{(x,0,0):x\in\mathbb R\},
\label{eq:tbg_c2x_fixed_axis}
\end{align}
so the fixed positions lie on an axis in the midplane between the layers.  An actual $C_{6v}$ mirror would instead act as $(x,y,z)\mapsto(x,-y,z)$ and fix the whole vertical plane $y=0$, including its intersections with both graphene sheets.

Let $g=\{C_2|\bm t_g\}$ leave $\bm k_*$ invariant and let $p_g=\pm1$ be the physical eigenvalue of a one-dimensional small representation.  At a point of the affine fixed set of $g$, the usual local selection rule reads
\begin{align}
\Phi_{\bm k_*}(\bm r_*)
=e^{-i\bm k_*\cdot\bm t_g}p_g\Phi_{\bm k_*}(\bm r_*).
\label{eq:tbg_3d_order_two_projector}
\end{align}
Thus the full three-dimensional amplitude vanishes whenever the local parity
\begin{align}
\pi_g(\bm k_*,\bm r_*)=e^{-i\bm k_*\cdot\bm t_g}p_g
\label{eq:tbg_local_order_two_parity}
\end{align}
is $-1$.  For the origin-centred $C_{2x}$ axis at $\Gamma_M$, this reduces to
\begin{align}
p_x=-1\quad\Longrightarrow\quad \Phi_{\bm k_*}(x,0,0)=0.
\label{eq:tbg_c2x_midplane_node}
\end{align}
The same statement holds on translated in-plane twofold axes, with the Bloch factor in Eq.~\eqref{eq:tbg_local_order_two_parity} deciding which parity is locally dark.

Another way to see this is to let $\varphi_t$ be an orbital in the top layer and $\varphi_b$ its $C_{2x}$ partner in the bottom layer, with
\begin{align}
\widehat C_{2x}\varphi_t=s_x\varphi_b,
\qquad s_x=\pm1.
\end{align}
The sign $s_x$ includes both the microscopic transformation of the carbon $p_z$ orbital and any basis convention.  The parity eigenstates are
\begin{align}
\Phi_{p_x}=\tfrac{1}{\sqrt2}(\varphi_t+p_xs_x\varphi_b).
\end{align}
On $\mathcal L_x$, invariance of the coordinate implies $\varphi_t=s_x\varphi_b$, and therefore
\begin{align}
\Phi_{p_x}(\bm r\in\mathcal L_x)
=\tfrac{1+p_x}{\sqrt2}\varphi_t(\bm r).
\label{eq:tbg_c2x_orbital_cancellation}
\end{align}
The odd state is forced to vanish exactly at the midpoint between the two $C_{2x}$-related orbitals, and more generally along the full fixed axis.  This result is independent of the orbital sign $s_x$.

The layer-projected density used in Fig.~\ref{fig:tbg} does not sample this fixed axis.  It evaluates the amplitudes on the two sheets with $z=\pm d/2$ at two-dimensional coordinates $\bm{\rho}$, and adds their densities incoherently,
\begin{align}
\rho_{\mathrm{lay}}(\bm\rho)
=|\Phi(\bm\rho,d/2)|^2+|\Phi(\bm\rho,-d/2)|^2.
\label{eq:tbg_layer_projected_density_order_two}
\end{align}
On the layer planes, $C_{2x}$ relates the two terms and fixes their relative phase; it does not fix either point individually.  The sign $p_x$ is lost when the layer densities are added, while the destructive interference in Eq.~\eqref{eq:tbg_c2x_orbital_cancellation} occurs only in the interlayer region where the two orbital tails overlap coherently.  Hence different $C_{2x}$ eigenvalues cannot be distinguished from the layer-projected density.  In the convention of Table~\ref{tab:tbg_irreps_full}, the $C_{2x}$-odd one-dimensional irreps are $\Gamma_2\;(A_2)$, $\Gamma_3\;(B_2)$, $M_2\;(B_1)$, and $M_3\;(B_2)$ --- each has an interlayer nodal axis whenever the local Bloch phase is trivial.  The $K_3\;(E)$ doublet contains one even and one odd eigenvector of $C_{2y}$, so its two-dimensional projected density remains bright on the corresponding axis even though its odd member is individually dark there.

There is an analogous, but valley-sensitive, statement for $C_{2z}$,
\begin{align}
C_{2z}:(x,y,z)\mapsto(-x,-y,z),
\qquad
\mathcal L_z=\{(0,0,z):z\in\mathbb R\}.
\label{eq:tbg_c2z_fixed_axis}
\end{align}
At $\Gamma_M$ and $M_M$, a full two-valley state with local $C_{2z}$ parity $-1$ vanishes on the vertical rotation axis, including its intersections with the two graphene sheets.  In the continuum theory, however, $C_{2z}$ exchanges valleys and sublattices.  If $\Phi_+$ is a valley $+$ amplitude and $\widehat C_{2z}\Phi_+=s_z\Phi_-$, then the coherent parity eigenstates are
\begin{align}
\Phi_{p_z}=\tfrac{1}{\sqrt2}(\Phi_++p_zs_z\Phi_-),
\end{align}
and on $\mathcal L_z$ one obtains
\begin{align}
\Phi_{p_z}(\bm r\in\mathcal L_z)
=\tfrac{1+p_z}{\sqrt2}\Phi_+(\bm r).
\label{eq:tbg_c2z_valley_cancellation}
\end{align}
Thus a $C_{2z}$-odd valley-coherent state has a vertical nodal line.  A single-valley wavefunction is not a $C_{2z}$ eigenstate, while the valley-incoherent density
\begin{align}
\rho_{\mathrm{val}}(\bm r)=|\Phi_+(\bm r)|^2+|\Phi_-(\bm r)|^2
\label{eq:tbg_valley_incoherent_density}
\end{align}
contains no interference term and is blind to $p_z$.  At $K_M$, $C_{2z}$ exchanges $K_M$ and $K'_M$, so a $C_{2z}$ parity can only be assigned to a coherent superposition over the two-corner star, not to a state at one corner.  The $C_{2z}$-odd one-dimensional irreps at the time-reversal-invariant momenta are $\Gamma_3\;(B_2)$, $\Gamma_4\;(B_1)$, $M_3\;(B_2)$, and $M_4\;(B_3)$.

The order-two characters therefore produce \textit{three-dimensional dark sets} ---  the $C_{2x}$ characters control interlayer-parity nodes on midplane axes, while the $C_{2z}$ characters control vertical nodes of valley-coherent states.  Neither effect is represented in the plots used here, which retain one valley and project the density onto the two graphene layers before adding the layers incoherently.  The $C_3$ analysis below instead concerns zeros already present in the layer-resolved amplitudes at the maximal stacking centres, and is directly visible in those plots.

\FloatBarrier

\subsection{Calculation of the flat-band dark sets}
\label{sm:tbg_dark_sets}

We apply the main-text stabiliser projector to one embedded layer-sublattice component at a time.  In valley $+$,
\begin{align}
\psi_{1s}(\bm r)=\sum_{\bm g}u_{\bm g,1s}(\bm k)e^{i(\bm k+\bm g)\cdot\bm r},\qquad \psi_{2s}(\bm r)=\sum_{\bm g}u_{\bm g,2s}(\bm k)e^{i(\bm k+\bm g+\bm q_1)\cdot\bm r}.
\label{eq:tbg_channel_embedding_revised}
\end{align}
Let $\xi_A=1$, $\xi_B=\omega$, and $\bm t_3(\bm r_*)=(1-C_3)\bm r_*$.  In the conventional $C_3$ gauge of Eq.~\eqref{eq:tbg_c3_eigenvalues_revised}, the local phase is
\begin{align}
\zeta_{\ell s}(\bm k_*,\bm r_*)=\omega\xi_s\exp\{i(\bm k_*+\delta_{\ell,2}\bm q_1)\cdot\bm t_3(\bm r_*)\}.
\label{eq:tbg_local_chiral_phase_revised}
\end{align}
For a flat-band subspace $\mathcal S$, write $\Lambda_{\mathcal S}=\{\lambda_a\}_{a=1}^{\dim\mathcal S}$ for its set of $C_{3z}$ eigenvalues, counted with multiplicity; thus $\lambda_a$ is the rotation eigenvalue of the $a$th symmetry-resolved state in $\mathcal S$.  The index gives
\begin{align}
m_{\mathcal S,\ell s}(\bm k_*,\bm r_*)=\sum_{\lambda_a\in\Lambda_{\mathcal S}}\tfrac{1}{3}\{1+\lambda_a\zeta_{\ell s}^{-1}+(\lambda_a\zeta_{\ell s}^{-1})^2\}.
\label{eq:tbg_component_projector_revised}
\end{align}
The component is bright precisely when one $\lambda_a = \zeta_{\ell s}$.  The motivation for this slight reformulation of the equation in the main text is that this form is basis-independent at the Dirac points.

The maximal high-symmetry positions of the moiré cell are
\begin{align}
\bm r_{AA}=0,\qquad \bm r_{AB}=\tfrac{1}{3}(\bm a_1+2\bm a_2),\qquad \bm r_{BA}=\tfrac{1}{3}(2\bm a_1+\bm a_2),
\end{align}
and the three saddle positions $\bm s_1=\bm a_1/2$, $\bm s_2=\bm a_2/2$, $\bm s_3=(\bm a_1+\bm a_2)/2$.  The conventional local $C_3$ phases are listed in Table~\ref{tab:tbg_local_phases_revised}.

\begin{table*}[t]
\centering
\scriptsize
\renewcommand{\arraystretch}{1.48}
\setlength{\tabcolsep}{3.5pt}
\begin{tabular*}{\textwidth}{@{\extracolsep{\fill}}c|ccc|ccc|ccc}
\hline\hline
Channel & $\Gamma_M,AA$ & $\Gamma_M,AB$ & $\Gamma_M,BA$ & $K_M,AA$ & $K_M,AB$ & $K_M,BA$ & $K'_M,AA$ & $K'_M,AB$ & $K'_M,BA$ \\
\hline
$1A$ & $\omega$ & $\omega^2$ & $1$ & $\omega$ & $\omega$ & $\omega$ & $\omega$ & $1$ & $\omega^2$ \\
$1B$ & $\omega^2$ & $1$ & $\omega$ & $\omega^2$ & $\omega^2$ & $\omega^2$ & $\omega^2$ & $\omega$ & $1$ \\
$2A$ & $\omega$ & $1$ & $\omega^2$ & $\omega$ & $\omega^2$ & $1$ & $\omega$ & $\omega$ & $\omega$ \\
$2B$ & $\omega^2$ & $\omega$ & $1$ & $\omega^2$ & $1$ & $\omega$ & $\omega^2$ & $\omega^2$ & $\omega^2$ \\
\hline\hline
\end{tabular*}
\caption{Local $C_3$ phases of the four embedded layer-sublattice components in valley $+$.  A component is bright when its entry matches one of the flat-band eigenvalues in Eq.~\eqref{eq:tbg_c3_eigenvalues_revised}.}
\label{tab:tbg_local_phases_revised}
\end{table*}

Combining this table with Eq.~\eqref{eq:tbg_c3_eigenvalues_revised} gives the channel-resolved result in Table~\ref{tab:tbg_dark_channels_revised}.  This table samples amplitudes on the graphene layers and therefore records the $C_3$-enforced layer-sublattice zeros relevant to Fig.~\ref{fig:tbg}.  The order-two characters add the three-dimensional nodal axes derived in Sec.~\ref{sm:tbg_order_two_3d}: a $C_{2x}$-odd state is dark on the appropriate interlayer fixed axis, and a $C_{2z}$-odd coherent two-valley state is dark on the corresponding vertical axis.  Those nodes are absent from the present layer-projected, single-valley visualisation.  At $K_M$, the $K_3\;(E)$ subspace contains both $C_{2y}$ parities, so its subspace density has no order-two-enforced zero.  The AB and BA centres are exchanged by the relevant order-two operations, which relate their amplitudes rather than imposing an additional one-point zero on a fixed layer.

\begin{table*}[t]
\centering
\small
\renewcommand{\arraystretch}{1.48}
\setlength{\tabcolsep}{7pt}
\begin{tabular*}{\textwidth}{@{\extracolsep{\fill}}c|cccc}
\hline\hline
Position & $\Gamma_M$ & $K_M$ & $K'_M$ & $M_M$ \\
\hline
$AA$ & $1A,1B,2A,2B$ & None & None & None \\
$AB$ & $1A,2B$ & $2B$ & $1A$ & None \\
$BA$ & $1B,2A$ & $2A$ & $1B$ & None \\
$\bm s_1,\bm s_2,\bm s_3$ & None & None & None & None \\
\hline\hline
\end{tabular*}
\caption{Component-resolved dark sets of the central two flat bands in one valley.  The entries list the dark layer-sublattice components.}
\label{tab:tbg_dark_channels_revised}
\end{table*}

The layer-projected density is the sum over all four layer-sublattice components.  It vanishes at the positions $D_{\mathrm{lay}}$ when every component is dark, and therefore
\begin{align}
D_{\mathrm{lay}}(\Gamma_M)=\{AA\},\qquad D_{\mathrm{lay}}(K_M)=D_{\mathrm{lay}}(K'_M)=D_{\mathrm{lay}}(M_M)=\varnothing.
\label{eq:tbg_total_dark_set_revised}
\end{align}
While the wavefunction has components with exact zeros at AB and BA, other components remain bright, and so  Eq.~\eqref{eq:tbg_total_dark_set_revised} is the complete symmetry-enforced dark set for the layer-projected, valley-diagonal density at the maximal in-plane positions used in Fig.~\ref{fig:tbg}. 

\subsection{Wavefunction maps}
\label{sm:tbg_density}

The exact $C_3$ dark-set statement concerns the pointwise embedded continuum amplitude in a fixed valley; the time-reversed valley has the conjugate amplitude and therefore the same density.  The real-space panels in Fig.~\ref{fig:tbg} show the atomistically embedded density of the selected flat-band state at one high-symmetry momentum and energy.

We follow the microscopic reconstruction used in Refs. \cite{calugaru2022spectroscopy,hong2022detecting}; the continuum eigenvector is first converted into a complex amplitude on every microscopic carbon site by restoring the rapidly varying valley momentum and the plane-wave phases.  These complex site amplitudes are then used as coefficients of overlapping carbon-centred orbitals, and the orbitals are summed before the modulus square is taken.  This reconstructs the wavefunction itself within each layer and retains interference between neighbouring carbon sites and between the two sublattices.  For the plots used here, the $z$ coordinate is suppressed and the two layer densities are added incoherently, so the interlayer $C_{2x}$ cancellation and the valley-coherent $C_{2z}$ cancellation of Sec.~\ref{sm:tbg_order_two_3d} are not retained.

Let $\bm{A}_1,\bm{A}_2$ be graphene direct vectors dual to $\bm{G}_1,\bm{G}_2$, and define the sublattice displacement
\begin{align}
\bm{\delta}_{AB}=\tfrac{1}{3}\bm{A}_1-\tfrac{1}{3}\bm{A}_2.
\end{align}
The microscopic carbon positions in layer $\ell$ are
\begin{align}
\bm{R}^{\ell A}_{mn}
&=R_{\alpha_\ell}(m\bm{A}_1+n\bm{A}_2),
\\
\bm{R}^{\ell B}_{mn}
&=R_{\alpha_\ell}(m\bm{A}_1+n\bm{A}_2+\bm{\delta}_{AB}),
\label{eq:tbg_carbon_positions_revised}
\end{align}
where $m,n\in\mathbb Z$, $\alpha_1=\theta/2$, and $\alpha_2=-\theta/2$.  The layer-dependent valley momenta are
\begin{align}
\bm{K}_1=R_{\theta/2}\bm{K}_0,
\,\,\,\,\,\,\,\,\,\,\,\,\,\,\,\bm{K}_2=R_{-\theta/2}\bm{K}_0,\,\,\,\,\,\,\,\,\,\,\,\,\,\,\,
\bm{K}_0=-\tfrac{1}{3}(\bm{G}_1+\bm{G}_2).
\end{align}
For a continuum eigenvector $u_{n\bm{k}}(\bm{g},\ell,s)$, with band index $n$, moiré momentum $\bm{k}$, layer $\ell$, and sublattice $s$, define the microscopic momentum carried by the corresponding plane-wave component as
\begin{align}
\bm{p}^{\ell}_{\bm{k}\bm{g}}
=\bm{K}_\ell+\bm{k}+\bm{g}+\delta_{\ell,2}\bm{q}_1.
\end{align}
The complex coefficient assigned to a carbon site $\bm R$ is then
\begin{align}
\mathcal A^{\ell s}_{n\bm{k}}(\bm{R})
=
\sum_{\bm{g}}
 u_{n\bm{k}}(\bm{g},\ell,s)
 e^{i\bm{p}^{\ell}_{\bm{k}\bm{g}}\cdot\bm{R}}.
\label{eq:tbg_atomistic_amplitude_revised}
\end{align}
This is the continuum envelope evaluated on the microscopic lattice with the valley-scale Bloch phase restored.

We represent each carbon-centred orbital by the two-dimensional Gaussian
\begin{align}
\phi_\sigma(\bm{r})
=
\exp[-|\bm{r}|^2/(2\sigma^2)],
\end{align}
with the choice of $\sigma=0.060\,\mathrm{nm}$. The reconstructed complex wavefunction in layer $\ell$ is
\begin{align}
\Phi^\ell_{n\bm{k}}(\bm{r})
=
\sum_{p,q\in\mathbb Z}
\sum_{s=A,B}
\mathcal A^{\ell s}_{n\bm{k}}(\bm{R}^{\ell s}_{pq})
\phi_\sigma(\bm{r}-\bm{R}^{\ell s}_{pq}),
\label{eq:tbg_splatted_layer_field_revised}
\end{align}
where $p,q$ label graphene unit cells.  The sum is coherent over all carbon sites and both sublattices in a given layer. For a selected state or degenerate subspace $\mathcal S$, the plotted density is
\begin{align}
\rho^{\mathrm{atom}}_{\mathcal S}(\bm{r})
=
\sum_{n\in\mathcal S}
\sum_{\ell=1}^{2}
|\Phi^\ell_{n\bm{k}}(\bm{r})|^2.
\label{eq:tbg_atomistic_density_revised}
\end{align}
Eq.~\eqref{eq:tbg_atomistic_density_revised} adds different bands and the two layers incoherently, while retaining all carbon-site interference within each layer through Eq.~\eqref{eq:tbg_splatted_layer_field_revised}.

The wavefunction panels in Fig.~\ref{fig:tbg} are evaluated at the selected high-symmetry momentum and band energy.  Their eigenvectors are calculated in a high-symmetry-point-centred radial basis with $R_{\mathrm{fac}}=2.5$, for which the dimensions are $80$ at $K_M$ and $84$ at $\Gamma_M$.  The basis is closed under $C_3$ at the relevant momentum to residuals below $2\times10^{-13}$.  The atomistic density is evaluated on a $330\times330$ grid over the moiré Wigner--Seitz cell, averaged over the three $C_3$ rotations, and normalised independently in each panel,
\begin{align}
Z_{\mathrm{plot}}(\bm{r})
=
\tfrac{\rho^{\mathrm{atom}}_{\mathcal S}(\bm{r})}
{\max_{\bm{r}\in\Omega_M}\rho^{\mathrm{atom}}_{\mathcal S}(\bm{r})},
\end{align}
where $\Omega_M$ is the plotted moiré Wigner--Seitz cell.  The exact continuum envelope for the $\Gamma_M$ wavefunction vanishes to numerical precision at the $AA$ position. 

\newpage

\section{A more general proof}
\label{sm:general-proof}

In SM Sec.~\ref{supp-wallpaper}, we verified the central theorem in two dimensions by explicitly computing the dark set for every irrep at every high-symmetry momentum of the 17 wallpaper groups.  Here we give a less constructive proof for symmorphic wallpaper groups. The idea is that each real-space probe defines a linear functional of the little-group character.  We prove that, for every symmorphic wallpaper little co-group, a sufficiently rich finite set probe positions makes the resulting map injective on the character space. The labelled zero pattern of these indices distinguishes the corresponding physical irrep.  For nonsymmorphic groups, fixed-point-free symmetries are absent from every one-point stabiliser; the final subsection explains this obstruction, observable in the tables of Sec.~\ref{supp-wallpaper}.

\subsection{Linear-algebra formulation of the uniqueness problem}

A class function on $G$ is a complex-valued function on the group that is constant on conjugacy classes.  Characters are the most important examples: if $\rho$ is a representation of $G$, then its character
\begin{align}
\chi_{\rho}(g)=\operatorname{Tr}\rho(g)
\end{align}
depends only on the conjugacy class of $g$.  Since a finite group has finitely many conjugacy classes, the space $\operatorname{Class}(G)$ of class functions is finite-dimensional.

For a probe position $\bm r_i$, define its stabiliser $H_i$ and local phase
\begin{align}
\vartheta_i(h)
=
 e^{-i\bm k_\ast\cdot(1-R_h)\bm r_i},
\qquad h\in H_i.
\end{align}
For an irrep $\rho$ of $G$, the local index is
\begin{align}
m_i(\rho)
=
\tfrac{1}{|H_i|}
\sum_{h\in H_i}
\vartheta_i(h)\chi_{\rho}(h).
\end{align}
This expression is linear in $\chi_{\rho}$, so the uniqueness problem is naturally a problem in linear algebra on $\operatorname{Class}(G)$.

\subsection{Restriction, induction, and Frobenius reciprocity}

If $\rho$ is a representation of $G$, then
\begin{align}
\operatorname{Res}^{G}_{H_i}\rho
\end{align}
denotes the same representation viewed only on $H_i$.  If $\sigma$ is a representation of $H_i$, then
\begin{align}
\operatorname{Ind}^{G}_{H_i}\sigma
\end{align}
is the representation of $G$ induced from $\sigma$.

With the standard inner product on class functions,
\begin{align}
\langle f,g\rangle_G
=
\tfrac{1}{|G|}
\sum_{h\in G}f(h)^\ast g(h),
\end{align}
the index of Eq. \eqref{indicator} can be written as the inner product
\begin{align}
m_i(\rho)
=
\left\langle
\vartheta_i^\ast,
\operatorname{Res}^{G}_{H_i}\chi_{\rho}
\right\rangle_{H_i}.
\end{align}
As shown in Sec.~\ref{gauge}, $\vartheta_i$ is a one-dimensional representation of the stabiliser.  Frobenius reciprocity therefore gives
\begin{align}
\left\langle
\sigma,
\operatorname{Res}^{G}_{H}\chi
\right\rangle_H
=
\left\langle
\operatorname{Ind}^{G}_{H}\sigma,
\chi
\right\rangle_G,
\end{align}
and hence
\begin{align}
m_i(\rho)
=
\left\langle
\operatorname{Ind}^{G}_{H_i}\vartheta_i^\ast,
\chi_{\rho}
\right\rangle_G.
\end{align}
Define the probe-induced class function
\begin{align}
\Psi_i
\equiv
\chi_{\operatorname{Ind}^{G}_{H_i}\vartheta_i^\ast}.
\end{align}
Then
\begin{align}
m_i(\rho)=\langle\Psi_i,\chi_{\rho}\rangle_G.
\end{align}
Thus each physical probe position defines one linear functional on the character table of the little group.

\subsection{The visible class-function space}

Let
\begin{align}
\mathcal V_{\mathrm{vis}}(\bm k_\ast)
=
\operatorname{span}
\{\Psi_{\bm r}:\text{all physical probe positions }\bm r\}
\subseteq
\operatorname{Class}(G)
\end{align}
be the visible class-function space.  Two characters have identical local indices at every probe position precisely when their difference is orthogonal to this space:
\begin{align}
\chi_{\rho}-\chi_{\rho'}
\in
\mathcal V_{\mathrm{vis}}(\bm k_\ast)^\perp.
\end{align}
Equivalently, the complete family of one-point probes determines exactly the projection of $\chi_\rho$ onto $\mathcal V_{\mathrm{vis}}(\bm k_\ast)$.

\subsection{Reduction of the probe family}

For fixed $\bm k_\ast$, every probe position $\bm r$ determines a subgroup-character pair $(H_{\bm r},\vartheta_{\bm r})$, and $\Psi_{\bm r}$ depends only on this pair.  Since $G$ is finite, it has finitely many subgroups, and each subgroup has finitely many one-dimensional characters.  Hence only finitely many distinct probe-induced functions $\Psi_{\bm r}$ can occur.

Choose a maximal linearly independent subfamily
\begin{align}
\Psi_1,\ldots,\Psi_N.
\end{align}
It is a basis of $\mathcal V_{\mathrm{vis}}(\bm k_\ast)$: if some $\Psi_{\bm r}$ lay outside its span, adjoining it would contradict maximality.  Therefore a finite set of physical probe positions captures all information accessible to one-point probes. The substantive task below is to show that, for symmorphic wallpaper groups, physically realised rotation-centre and mirror-line probes span this space.

\subsection{Matrix form and the uniqueness criterion}

Choose probe positions $\bm r_1,\ldots,\bm r_N$ such that $\Psi_1,\ldots,\Psi_N$ form a basis of $\mathcal V_{\mathrm{vis}}(\bm k_\ast)$.  Write the visible part of an irrep character as
\begin{align}
\chi_\rho^{\mathrm{vis}}
=
\sum_{j=1}^{N}c_{\rho,j}\Psi_j.
\end{align}
Then
\begin{align}
m_i(\rho)
=
\left\langle\Psi_i,\chi_\rho^{\mathrm{vis}}\right\rangle_G
=
\sum_{j=1}^{N}P_{ij}c_{\rho,j},
&&
P_{ij}
\equiv
\langle\Psi_i,\Psi_j\rangle_G.
\end{align}
Equivalently,
\begin{align}
\bm m_\rho=P\bm c_\rho,
\end{align}
where
\begin{align}
\bm m_\rho
=
\left(m_1(\rho),\ldots,m_N(\rho)\right)^T.
\end{align}
For a basis $\{\Psi_i\}$, the Gram matrix $P$ has full rank, so the local indices determine $\chi_\rho^{\mathrm{vis}}$.  Consequently, to reconstruct the physical character it is enough to prove
\begin{align}
\mathcal V_{\mathrm{vis}}(\bm k_\ast)
=
\operatorname{Class}_{\mathrm{phys}}(G).
\end{align}
The following sections establish this equality by constructing the required physical probes.

\subsection{Physical character space at a time-reversal invariant momentum}

At a time-reversal invariant momentum, the local phases of a symmorphic stabiliser are real:
\begin{align}
\vartheta_i(h)=\pm1.
\end{align}
Consequently,
\begin{align}
m_i(\bar\rho)=m_i(\rho).
\end{align}
A density measurement therefore cannot distinguish an ordinary complex irrep $\rho$ from its conjugate $\bar\rho$.

Define the corresponding physical character by
\begin{align}
\chi_{[\rho]}
=
\begin{cases}
\chi_\rho,
&
\rho\simeq\bar\rho,
\\[1mm]
\chi_\rho+\chi_{\bar\rho},
&
\rho\not\simeq\bar\rho.
\end{cases}
\end{align}
The span of these characters is denoted
\begin{align}
\operatorname{Class}_{\mathrm{phys}}(G).
\end{align}
In a time-reversal-symmetric system, a distinct pair $\rho,\bar\rho$ forms one irreducible corepresentation of the magnetic little group.  Away from a time-reversal invariant momentum, no such pairing is imposed and we take
\begin{align}
\operatorname{Class}_{\mathrm{phys}}(G)
=
\operatorname{Class}(G).
\end{align}

\subsection{Rotation centres and their local phases}

Let $\mathcal R$ be an $n$-fold lattice rotation.  A point $\bm r$ is a centre of an affine rotation with linear part $\mathcal R$ precisely when
\begin{align}
(1-\mathcal R)\bm r=\bm t\in\Lambda,
\end{align}
where $\Lambda$ is the Bravais lattice.  Translating the centre by $\bm\lambda\in\Lambda$ changes its label by
\begin{align}
\bm t
\longmapsto
\bm t+(1-\mathcal R)\bm\lambda.
\end{align}
The translation-inequivalent centres are therefore classified by
\begin{align}
Q_{\mathcal R}
\equiv
\Lambda/(1-\mathcal R)\Lambda.
\end{align}
For a two-dimensional rotation,
\begin{align}
|Q_{\mathcal R}|
=
|\det(1-\mathcal R)|
=
2-2\cos\left(\tfrac{2\pi}{n}\right).
\end{align}
For the crystallographic orders,
\begin{align}
n=2,3,4,6
\qquad\Longrightarrow\qquad
|Q_{\mathcal R}|=4,3,2,1.
\end{align}
These are the four twofold, three threefold, two fourfold, and one sixfold centre classes.

At a momentum satisfying
\begin{align}
\mathcal R\bm k_\ast
=
\bm k_\ast+\bm G,
\end{align}
the local phase associated with the centre labelled by $\bm t$ is
\begin{align}
\vartheta_{\bm t}(\mathcal R)
=
 e^{-i\bm k_\ast\cdot\bm t}.
\end{align}
This phase is well defined on $Q_{\mathcal R}$, since
\begin{align}
e^{-i\bm k_\ast\cdot(1-\mathcal R)\bm\lambda}
=
 e^{-i(\bm k_\ast-\mathcal R^{-1}\bm k_\ast)\cdot\bm\lambda}
=
1.
\end{align}
Thus the quotient determines both the available rotation centres and the local characters realised by them.

\subsection{Physical realisation of the separating probes}

\paragraph{Lemma.}
For every symmorphic wallpaper little co-group, the cyclic subgroups, local phases, and reflection classes used in the reconstruction below are realised by physical probe positions.

\paragraph{Proof.}
First consider a point group with maximal rotation order $n$.  Applying the preceding quotient count to the powers of the maximal rotation, and removing centres already fixed by a higher-order rotation, gives
\begin{equation}
\begin{array}{c|c}
 n & \text{Rotational stabilisers}\\
\hline
 2 & 4\,C_2\\
 3 & 3\,C_3\\
 4 & 2\,C_4\ \text{and}\ 2\,C_2\\
 6 & 1\,C_6,\ 2\,C_3,\ \text{and}\ 3\,C_2.
\end{array}
\end{equation}
For $n=4$, the two $C_4$ centres are among the four centres of the order-two power, leaving two exact $C_2$ centres.  For $n=6$, the unique $C_6$ centre is among both the three $C_3$ centres and the four $C_2$ centres.  A point fixed by both a threefold and a twofold generator is fixed by the $C_6$ group they generate, so there is no further overlap.  Hence every divisor subgroup required below is represented by an actual rotation centre.  In a mirror-containing group, its full stabiliser may be $D_d$ rather than $C_d$; this case is treated explicitly below.

The pairing
\begin{align}
([\bm k],[\bm t])
\longmapsto
 e^{-i\bm k\cdot\bm t}
\end{align}
between $\mathcal R$-invariant momentum classes and
$Q_{\mathcal R}$ is perfect, meaning that it identifies the
group of $\mathcal R$-invariant momentum classes with the
character group $\widehat{Q_{\mathcal R}}$.  Triviality on every $[\bm t]$ forces $\bm k$ to be reciprocal-lattice equivalent to zero, and the two finite groups have the same order $|\det(1-\mathcal R)|$.  Therefore the nonzero invariant momentum classes realise the nontrivial characters of $Q_{\mathcal R}$.  In the crystallographic cases this gives
\begin{equation}
\begin{array}{c|c}
\text{Case} & \text{Phases realised by the centre classes}\\
\hline
\text{Nonzero $C_2$-invariant momentum} & +1,-1\\
C_4\text{-invariant }M & +1,-1\\
C_3\text{-invariant corner} & 1,\omega,\omega^2.
\end{array}
\end{equation}

Finally, choose the symmorphic origin to be fixed by the point group.  Every point-group reflection $\sigma$ then has a representative $\{\sigma|0\}$ with a fixed mirror line.  A generic point on that line, away from its intersections with other symmetry loci, has stabiliser $C_s$.  For even $n$, representatives of the two reflection conjugacy classes give the two required mirror families.  The origin itself has the full $D_n$ stabiliser.  This proves the claim. \hfill$\square$

\subsection{Cyclic little co-groups at \texorpdfstring{$\Gamma$}{Gamma}}

Let
\begin{align}
G=C_n=\langle c\rangle,
\qquad
n\in\{1,2,3,4,6\}.
\end{align}
The ordinary complex irreps are
\begin{align}
\rho_j(c)=e^{2\pi i j/n},
\qquad
j\in\mathbb Z_n.
\end{align}
At a time-reversal invariant momentum, $j$ and $-j$ belong to the same physical conjugation class.  Define
\begin{align}
e(j)
=
\operatorname{ord}\left(e^{2\pi i j/n}\right)
=
\tfrac{n}{\gcd(j,n)}.
\end{align}
For $n=1,2,3,4,6$, the order $e(j)$ uniquely determines the physical conjugation class.

For every divisor $d$ of $n$, let
\begin{align}
C_d=\langle c^{n/d}\rangle.
\end{align}
At $\Gamma$, the local phase is trivial.  Let $\rho_e$ denote the physical irrep whose rotation eigenvalue has order $e$, and let $d_e$ be its real dimension.  The $C_d$ projector gives
\begin{align}
m_d(\rho_e)
=
d_e\,\mathbf 1_{e\mid n/d}.
\end{align}
Writing
\begin{align}
q=\tfrac{n}{d},
\end{align}
the normalised probe matrix is
\begin{align}
Z_{eq}=\mathbf 1_{e\mid q},
\qquad e,q\mid n.
\end{align}
This is the incidence matrix of the divisor lattice.  Its inverse is given by number-theoretic Möbius inversion.  If
\begin{align}
F(q)=\sum_{e\mid q}a(e),
\end{align}
then
\begin{align}
a(e)=\sum_{q\mid e}\mu(e/q)F(q),
\end{align}
where $\mu$ is the Möbius function.  Hence the divisor probes reconstruct every coefficient in $\operatorname{Class}_{\mathrm{phys}}(C_n)$.

The divisor set includes $d=1$.  The corresponding generic probe gives the dimension $d_e$.  For example, at $\Gamma$ in $C_6$, the probes with stabilisers $C_1,C_2,C_3,C_6$ give
\begin{align}
\begin{array}{c|cccc}
 & C_1 & C_2 & C_3 & C_6\\
\hline
A   & 1 & 1 & 1 & 1\\
B   & 1 & 0 & 1 & 0\\
E_1 & 2 & 0 & 0 & 0\\
E_2 & 2 & 2 & 0 & 0
\end{array}.
\end{align}
This matrix has full rank.  Omitting the $C_1$ column removes the dimension functional but does not create a collision between the four zero patterns.

\subsection{Cyclic little co-groups at nonzero momentum}

Let a probe have stabiliser $C_d$ and local character
\begin{align}
\vartheta(c^{n/d})
=
 e^{-2\pi i q/d}.
\end{align}
For the complex irrep $\rho_j$,
\begin{align}
m_{d,q}(j)
&=
\tfrac1d
\sum_{\ell=0}^{d-1}
 e^{-2\pi i q\ell/d}
 e^{2\pi i j\ell/d}
\\
&=
\begin{cases}
1,
&j=q\pmod d,
\\
0,
&j\neq q\pmod d.
\end{cases}
\end{align}
Thus a shifted rotation centre is a finite Fourier projector for the angular-momentum label modulo $d$.

The physical-realisation lemma exhausts the crystallographic possibilities.  A nonzero $C_2$-invariant momentum has centre phases $\pm1$, which separate the two $C_2$ irreps.  At the square-lattice $M$ point, the two $C_4$ centres carry phases $+1$ and $-1$, whilst the twofold centres supply the remaining divisor test.  At a hexagonal corner, the three $C_3$ centres carry the phases $1,\omega,\omega^2$, and hence
\begin{align}
m_s(\rho_j)
=
\tfrac13
\sum_{\ell=0}^{2}
\omega^{(s+j)\ell}
=
\delta_{s+j,0\ {\rm mod}\ 3}.
\end{align}
Every $C_3$ irrep is therefore allowed at exactly one member-resolved centre.  A sixfold little co-group occurs only at $\Gamma$, which was covered by the divisor argument.

It follows that all cyclic symmorphic wallpaper little co-groups are separated, with only the unavoidable conjugation pairing at a time-reversal invariant momentum.

\subsection{Mirror-containing little co-groups}

Let
\begin{align}
G=C_{nv}\simeq D_n.
\end{align}
A character of $D_n$ is determined by its restriction to the rotation subgroup $C_n$ and by its values on the reflection classes.  There is one reflection class for odd $n$ and two reflection classes for even $n$.

A generic point on a mirror line has stabiliser
\begin{align}
C_s=\{E,\sigma\}.
\end{align}
Allowing for the known local sign $\epsilon_\sigma=\vartheta(\sigma)=\pm1$, its projector rank is
\begin{align}
m_{\sigma,\epsilon_\sigma}(\rho)
=
\tfrac12
\left[d_\rho+\epsilon_\sigma\chi_\rho(\sigma)\right].
\end{align}
This determines the reflection character:
\begin{align}
\chi_\rho(\sigma)
=
\epsilon_\sigma
\left[2m_{\sigma,\epsilon_\sigma}(\rho)-d_\rho\right].
\end{align}
One mirror family therefore determines the reflection class for odd $n$, whilst two inequivalent mirror families determine the two reflection classes for even $n$.

A rotation centre in a mirror group may have stabiliser $D_d$ rather than $C_d$.  Splitting the projector into rotation and reflection parts gives
\begin{align}
m_{D_d,\vartheta}(\rho)
=
\tfrac12m_{C_d,\vartheta|_{C_d}}(\rho)
+
\tfrac{1}{2d}
\sum_{\tau\in D_d\setminus C_d}
\vartheta(\tau)\chi_\rho(\tau).
\end{align}
The reflection characters in the second term have already been determined by the mirror-line probes.  Therefore
\begin{align}
m_{C_d,\vartheta|_{C_d}}(\rho)
=
2m_{D_d,\vartheta}(\rho)
-
\tfrac1d
\sum_{\tau\in D_d\setminus C_d}
\vartheta(\tau)\chi_\rho(\tau).
\end{align}
The cyclic divisor and Fourier arguments then determine the rotational part of the character.  Hence rotation-centre and mirror-line probes together span $\operatorname{Class}_{\mathrm{phys}}(D_n)$.

\subsection{Uniqueness of the dark set}

The preceding subsections prove that the complete index vector determines the physical character.  We now show that the labelled zero pattern is itself sufficient for the symmorphic wallpaper little groups.

For a cyclic group at $\Gamma$, the nonzero divisor probes are exactly those satisfying $e\mid\tfrac{n}{d}.$ This set determines the order $e$ of the rotation eigenvalue.  At a shifted centre, the Fourier projector determines the angular-momentum label directly.

For a one-dimensional irrep $\lambda$ of $D_n$, a pure mirror probe has
\begin{align}
m_\sigma(\lambda)
=
\tfrac12\left[1+\lambda(\sigma)\right].
\end{align}
For odd $n$, the parity on the single reflection class determines $\lambda$.  For even $n$, choose representatives $s$ and $cs$ of the two reflection classes.  Their parities also determine the rotational sign, since
\begin{align}
\lambda(c)=\lambda(cs)\lambda(s).
\end{align}
Thus the mirror zero pattern determines every one-dimensional irrep.

For a two-dimensional irrep $E_j$, every reflection character vanishes, so
\begin{align}
m_\sigma(E_j)=1
\end{align}
for every mirror line.  Any one-dimensional irrep with a dark mirror is therefore separated immediately from every $E_j$.  The only one-dimensional irrep bright on all the pure mirror families is the trivial irrep.  At the origin-centred probe with full stabiliser $D_n$,
\begin{align}
m_{D_n}(\mathbf 1)=1,
\qquad
m_{D_n}(E_j)=0,
\end{align}
because an irreducible two-dimensional representation contains no trivial $D_n$ subrepresentation.  This separates the remaining one-dimensional case from the two-dimensional irreps.

It remains to distinguish the different $E_j$.  At a centre with stabiliser $D_d$, the reflection characters of $E_j$ vanish, so the preceding decomposition reduces to
\begin{align}
m_{D_d,\vartheta}(E_j)
=
\tfrac12m_{C_d,\vartheta|_{C_d}}(E_j).
\end{align}
Consequently,
\begin{align}
m_{D_d,\vartheta}(E_j)=0
\quad\Longleftrightarrow\quad
m_{C_d,\vartheta|_{C_d}}(E_j)=0.
\end{align}
The physically realised $D_d$ centre therefore carries exactly the same zero-pattern information as the corresponding cyclic test on $E_j$.  The cyclic argument distinguishes the different $E_j$.

Combining the cyclic and dihedral cases, we have proved the following:

\noindent\fbox{\parbox{\dimexpr\columnwidth-2\fboxsep-2\fboxrule\relax}{%
    \textbf{Theorem.} Let $G$ be the little co-group of a symmorphic wallpaper group, and let $\rho$ be a single isolated irrep. Probing all member-resolved rotation centres and mirror-line families determines $\rho$ exactly at a non-TRIM and determines the pair $\{\rho,\bar{\rho}\}$ at a TRIM.}}

If time-reversal symmetry is present, a distinct pair $\rho,\bar{\rho}$ forms one irreducible corepresentation of the magnetic little group. In that formulation, the physical corepresentation is determined exactly.

\subsection{Nonsymmorphic groups}

A fixed-point-free glide cannot belong to a one-point real-space stabiliser and therefore never appears in any local projector sum.  One-point probes can consequently resolve only character data carried by symmetries with fixed points.  The symmorphic construction above succeeds because the required character space is spanned by probes associated with physical rotation centres and mirror lines.  No analogous complete one-point basis can recover character directions carried only by fixed-point-free nonsymmorphic operations --- for every nonsymmorphic symmetry, one fewer independent class function is available.

The preceding argument therefore gives a complete uniform proof for symmorphic wallpaper groups and identifies the structural obstruction in the nonsymmorphic case.  The collisions between nonsymmorphic irreps are explicitly observed in Sec.~\ref{supp-wallpaper}.

\end{document}